\documentclass{jfm}
\usepackage{graphicx}
\usepackage{epstopdf, epsfig}
\usepackage{amsmath}
\usepackage{mathrsfs}
\usepackage{float}
\usepackage[utf8]{inputenc}
\usepackage{listings}
\usepackage{color}
\usepackage{amsfonts}
\usepackage{amssymb}
\usepackage{array}
\usepackage{cellspace}
\usepackage{algpseudocode}
\usepackage{algorithm}
\usepackage[normalem]{ulem}

\DeclareMathOperator{\grad}{grad}
\DeclareMathOperator{\sym}{sym}
\DeclareMathOperator{\skw}{skw}

\DeclareMathOperator{\tr}{tr}
\DeclareMathOperator{\divr}{div}

\DeclareMathOperator{\Reyn}{Re}
\DeclareMathOperator{\St}{St}

\shorttitle{A general fluid-sediment mixture model and constitutive theory}
\shortauthor{A. S. Baumgarten and K. Kamrin}

\title{A general fluid-sediment mixture model and constitutive theory validated in many flow regimes}

\author{Aaron S. Baumgarten\aff{1} \and Ken Kamrin\aff{2}\corresp{\email{kkamrin@mit.edu}}}

\affiliation{\aff{1}Department of Aeronautics and Astronautics, Massachusetts Institute of Technology,
Cambridge, MA 02139, USA
\aff{2}Department of Mechanical Engineering, Massachusetts Institute of Technology, Cambridge, MA 02139, USA}

\begin{document}

\maketitle

\begin{abstract}
We present a thermodynamically consistent constitutive model for fluid-saturated sediments, spanning dense to dilute regimes, developed from the basic balance laws for two phase-mixtures. The model can represent various limiting cases, such as pure fluid and dry grains. It is formulated to capture a number of key behaviors such as: (i) viscous inertial rheology of submerged wet grains under steady shearing flows, (ii) the critical state behavior of grains, which causes granular Reynolds dilation/contraction due to shear, (iii) the change in the effective viscosity of the fluid due to the presence of suspended grains, and (iv) the Darcy-like drag interaction observed in both dense and dilute mixtures, which gives rise to complex fluid-grain interactions under dilation and flow. The full constitutive model is combined with the basic equations of motion for each mixture phase and implemented in the material point method (MPM) to accurately model the coupled dynamics of the mixed system. Qualitative results show the breadth of problems which this model can address. Quantitative results demonstrate the accuracy of this model as compared with analytical limits and experimental observations of fluid and grain behaviors in inhomogeneous geometries.

\end{abstract}


\section{Introduction}
Mixtures of fluids and sediments play an important role in many industrial and geotechnical engineering problems, from transporting large volumes of industrial wastes to building earthen levees and dams. To solve these problems, engineers have traditionally relied on the myriad of empirical models developed in the last century. These empirical models are derived by coupling relevant experimental observations to an understanding of the underlying physics governing the behavior of these mixtures. The model reported in \cite{einstein} describes the increase in effective fluid viscosity due to dilute suspensions of grains. The Darcy-like drag law given in \cite{carman} describes the pressure drop in a fluid as it flows through a bed of densely packed grains. The work by \cite{turian} characterizes the flow of slurries in pipelines. Other models (such as in \cite{pailha}) describe more complex problems (such as the initiation of submerged granular avalanches); however, each of these models can only provide a description of a specific regime of mixture and flows.

To address an engineering problem that involves complex interactions of fluids and sediments spanning many flow regimes requires a more general modeling approach. A natural first step is to model the underlying physics directly by solving the coupled fluid grain interactions at the micro-scale (as in the coupled lattice Boltzmann and discrete element method, LBM-DEM, proposed in \cite{cook}). Many problems of interest, however, involve far too much material for a direct approach to be computationally viable. We therefore turn to a continuum modeling approach, where the small scale structures and physics are homogenized into bulk properties and behaviors.

Recent work simulating fluid-sediment mixtures as continua (see \cite{soga}) presents a versatile foundation, but the reported results are highly sensitive to the choice of sediment constitutive model (see \cite{ceccato2016b} and \cite{fern}); even when pore pressure is uniform, no existing dry granular plasticity model correctly predicts the granular part of the rheology of saturated media.  In this work, we carefully formulate a new set of constitutive rules governing the fluid and sediment phases of the continuum mixture. Using these rules, we construct a model that recovers the correct limiting behaviors --- i.e. dry and viscous granular inertial rheologies, change of effective fluid viscosity due to suspended grains, Stokes and Carman-Kozeny drags, and Reynolds dilation --- and smoothly transitions between flow regimes covering the range from dense slurry-like flow to dilute suspensions. We implement our model in MPM and validate this implementation against several dynamic experiments involving submerged glass beads. We also consider the application of our model to the problems of slope collapse and intrusion.

\section{Theory and Formulation}\label{Sec: theory_and_formulation}
Here we lay out the theoretical framework for the two-phase mixture model. In the formulation of this theory, we use the standard notation of continuum mechanics from \cite{gurtin}. In particular, the trace of the tensor $\mathsfbi{A}$ is given by $\tr\mathsfbi{A}$ and the transpose by $\mathsfbi{A}^{\top}$. Every tensor admits the unique decomposition into a deviatoric part $\mathsfbi{A_0}$ and isotropic part by $\mathsfbi{A} = \mathsfbi{A_0} + \tfrac{1}{3}\tr(\mathsfbi{A})\mathsfbi{1}$ with $\mathsfbi{1}$ the identity tensor.

\subsection{Mixture Theory}
To develop the model we start by considering a mixture of grains and fluid. We assume that the grains are rough (i.e. physical, frictional contact can occur between grains; see \cite{zhao2002}), made of incompressible material with \textit{true} density $\rho_s$ (i.e. the density of a grain), and essentially spherical with mean diameter $d$. Additionally, we assume that the grains are quasi-mono-disperse (no size-segregation during flow) and that grains are large enough to neglect Brownian effects (i.e. $d \gtrsim 100 \mu \text{m}$ for common engineering slurries). We also assume that the grains are fully immersed in a barotropic viscous fluid having \textit{true} density $\rho_f$ and viscosity $\eta_0$. We use the term `\textit{true}' to mean those properties of the material in the mixture \textit{before} the mixture is homogenized. A representative volume of material, $\Omega$, can therefore be decomposed into a solid volume, $\Omega_s$, and a fluid volume, $\Omega_f$, such that $\Omega = \Omega_s \cup \Omega_f$.

\begin{figure}
	\centering
	\includegraphics[scale=0.4]{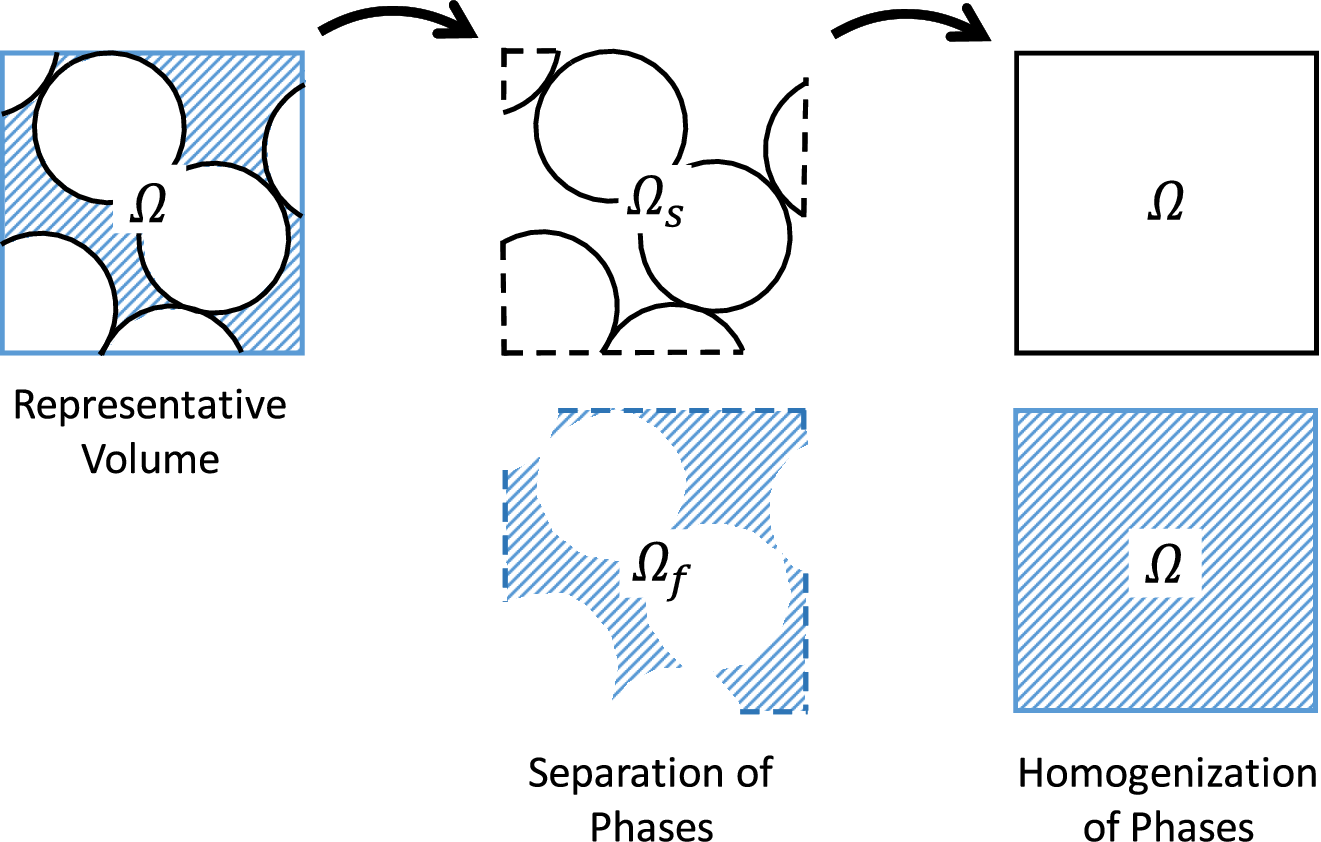}
	\caption{Pictorial description of the representative volume $\Omega$, the decomposition of the domain into fluid and solid volumes, and the homogenization of the two phases.}
	\label{Fig: homogenization}
\end{figure}

Figure \ref{Fig: homogenization} shows how this volume is decomposed and the important step of homogenizing the solid volume and fluid volume into two, overlapping continua. In the analysis that follows, $\psi_s$ will refer to some field $\psi$ defined on the solid phase, and $\psi_f$ will refer to some field $\psi$ defined on the fluid phase. If no subscript is given, then that field is defined on the mixture as a whole.

\subsubsection{Homogenization of Phases} \label{Sec: homogenization}
The effective densities, $\overline{\rho}_{s}$ and $\overline{\rho}_{f}$, and phase velocities, $\mathbi{v_{s}}$ and $\mathbi{v_{f}}$, of the mixture are defined such that conservation of mass and momentum in the continuum correspond to conservation of mass and momentum in the real mixture. For this, we consider a representative volume of material, $\Omega$, that contains a \textit{large} number of individual grains. For the continuum approximation to be valid, \textit{large} is defined such that grain-scale phenomena are smoothed out and bulk behavior is captured. The volume of grains $\Omega_s$ and volume of fluid $\Omega_f$ within $\Omega$ allow us to define the solid phase volume fraction or \textit{packing fraction}, $\phi$, and a fluid phase volume fraction or \textit{porosity}, $n$, as,
\begin{equation}
\overline{\rho}_s = \phi \rho_s, \qquad \overline{\rho}_f = n \rho_f, \qquad \text{with}\qquad \phi = 1 - n.
\label{Eqn: porosity}
\end{equation}
The external body force acting on each homogenized phase (per unit volume), $\mathbi{b_{0s}}$ and $\mathbi{b_{0f}}$, is proportional to the local effective density,
\begin{equation}
\mathbi{b_{0s}} = \overline{\rho}_s\mathbi{g},
\qquad \mathbi{b_{0f}} = \overline{\rho}_f\mathbi{g}
\label{Eqn: bodyforce}
\end{equation}
where $\mathbi{g}$ is the gravitational acceleration vector.

We next define the mixture Cauchy stress, $\boldsymbol{\sigma}$, according to Cauchy's Theorem such that the stress response of the mixture is expressed as the sum of the phase-wise effective Cauchy stresses, $\boldsymbol{\sigma_{s}}$ and $\boldsymbol{\sigma_{f}}$, i.e.
\begin{equation}
\boldsymbol{\sigma} = \boldsymbol{\sigma_s} + \boldsymbol{\sigma_f}.
\label{Eqn: mixture_stress}
\end{equation}

\subsubsection{Overlapping Continuum Bodies}\label{Sec: continuum}
\begin{figure}
	\centering
	\includegraphics[scale=0.35]{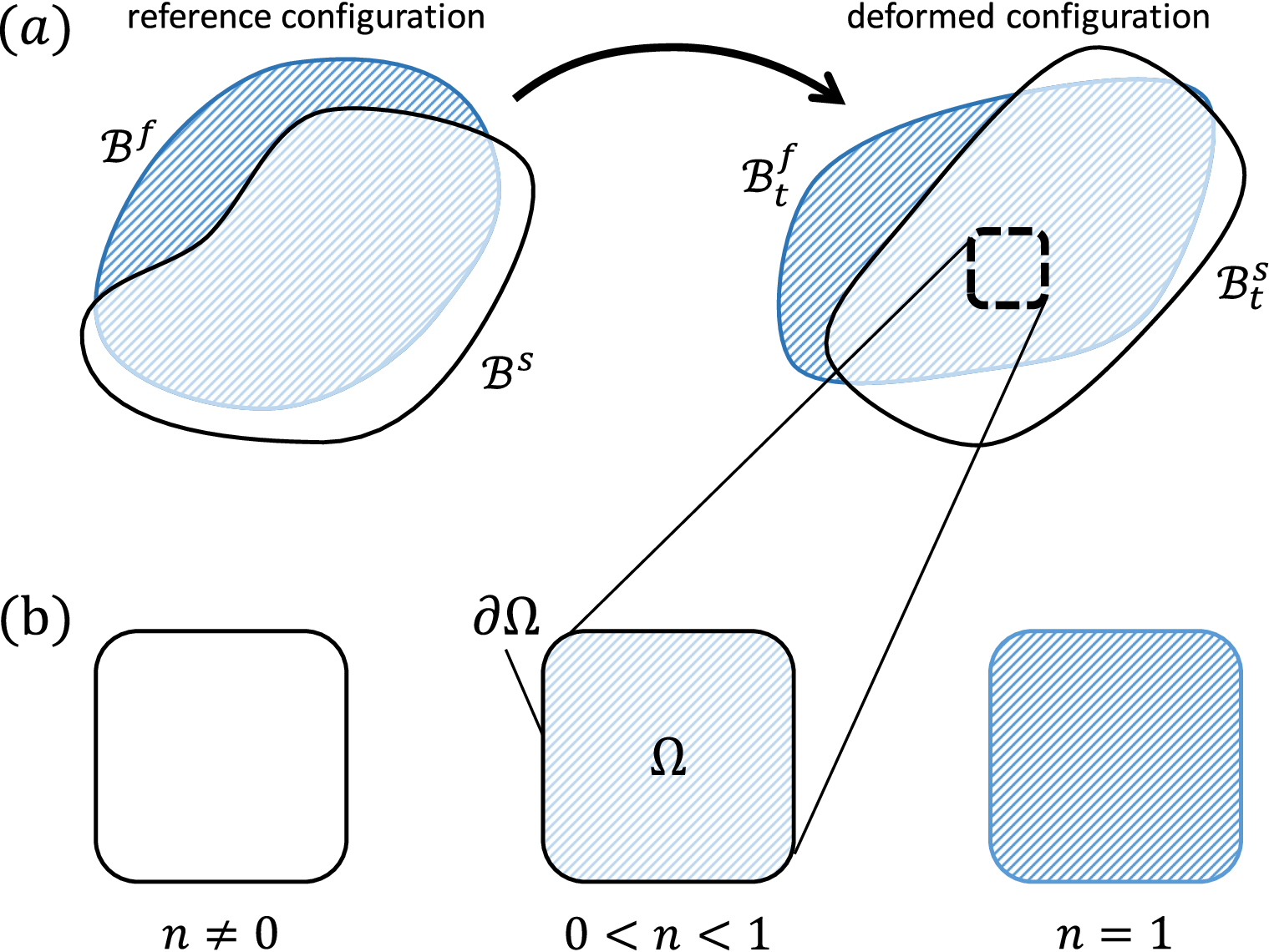}
	\caption{(a) Pictorial definition of the reference bodies, $\mathcal{B}^s$ and $\mathcal{B}^f$, and deformed bodies, $\mathcal{B}^s_t$ and $\mathcal{B}^f_t$. (b) Parts in the deformed body are \textit{always} fully saturated with porosity $n > 0$. In the limit of a fluid-only volume, the porosity $n=1$. In the limit of a solid-only volume, we \textit{do not} let the porosity $n$ go to zero, instead we let the fluid viscosity, bulk modulus, and \textit{true} density go to zero which effectively removes the fluid by making it stress- and density-free.}
	\label{Fig: configuration}
\end{figure}

When considering a mixture problem, we begin by defining each phase as its own continuum body, as shown in figure \ref{Fig: configuration}(a). $\mathcal{B}^s$ defines the initial solid phase body (or reference body) and $\mathcal{B}^f$ defines the fluid phase reference body. At some later time $t$, these bodies are represented by $\mathcal{B}^s_t$ and $\mathcal{B}^f_t$.

To determine the behavior of a volume of mixture $\Omega$, as shown in figure \ref{Fig: configuration}(b), we let that volume define a part in each continuum body. The full mixture is defined by the sum of these parts. If the volume of mixture is composed of fluid only, the porosity $n$ is unity. We also enforce that, in the absence of a solid phase, the local solid phase stress is zero, $\boldsymbol{\sigma_s} = \mathsfbi{0}$. In this limit, we expect the behavior of the mixture to be identical to that of a barotropic viscous fluid on its own. If the volume of mixture is solid only, the porosity $n$ is \textit{not} zero (it would only be zero in the limit of vanishing pore space between grains). In this limit, the behavior of the mixture should be identical to that of a dry granular material. To ensure this, we enforce that the local fluid phase stress is zero, $\boldsymbol{\sigma_f} = \mathsfbi{0}$, and that the \textit{true} fluid density, $\rho_{f}$, vanishes.

\subsubsection{Mass Conservation}
We now define the equations governing the evolution of the \textit{true} fluid density (i.e. the density of the fluid that is between the grains), $\rho_f$, and the effective densities of both phases, $\overline{\rho}_{s}$ and $\overline{\rho}_{f}$. Recalling that the solid grains are assumed to be incompressible, $\rho_s$ is constant. However, $\overline{\rho}_s$ changes when the solid phase compacts or dilates as the structure of the granular skeleton changes. Since we will often have fields which \textit{belong} to one phase or another (e.g. $\rho_s$ \textit{belongs} to the solid phase), it is convenient to define the material derivatives on each phase as follows,
\begin{equation}
\frac{D^s \psi}{Dt} = \frac{\partial \psi}{\partial t} + \mathbi{v_s} \cdot \grad \psi, \qquad
\frac{D^f \psi}{Dt} = \frac{\partial \psi}{\partial t} + \mathbi{v_f} \cdot \grad \psi
\label{Eqn: material_deriv}
\end{equation}

Mass conservation in a part of the solid phase continuum (as defined by a volume $\Omega$) is enforced by setting the material derivative of solid mass in the volume to zero. As shown in \cite{bandara}, this requires that,
\begin{equation}
\frac{D^s \overline{\rho}_s}{Dt} + \overline{\rho}_s \divr \mathbi{v_s} = 0
\label{Eqn: solid_density}
\end{equation}
A simple expansion of this expression using the definition of porosity from \eqref{Eqn: porosity} yields an expression for the rate of change of the local measure of porosity,
\begin{equation}
\frac{\partial n}{\partial t} =  (1-n) \divr \mathbi{v_s} - \mathbi{v_s} \cdot \grad n
\label{Eqn: partialn}
\end{equation}
Mass conservation of a fluid part defined by the arbitrary volume $\Omega$ is enforced by,
\begin{equation}
\frac{D^f \overline{\rho}_f}{Dt} + \overline{\rho}_f \divr \mathbi{v_f} = 0
\label{Eqn: fluid_density}
\end{equation}
Combining \eqref{Eqn: fluid_density} with \eqref{Eqn: porosity} and \eqref{Eqn: partialn}, we find the correct form of the evolution law for the \textit{true} fluid density,
\begin{equation}
\frac{n}{\rho_f} \frac{D^f \rho_f}{Dt} = -\divr\big((1-n)\mathbi{v_s} + n\mathbi{v_f}\big)
\label{Eqn: true_fluid_density}
\end{equation}

\subsubsection{Momentum Balance}
Conservation of linear momentum is enforced locally for the each continuum body (see figure \ref{Fig: configuration}) as follows,
\begin{equation}
\begin{aligned}
\overline{\rho}_s \frac{D^s \mathbi{v_s}}{Dt} &= \mathbi{b_{0s}} - \mathbi{f_b} - \mathbi{f_d} + \divr \boldsymbol{\sigma_s}
\\ \overline{\rho}_f \frac{D^f \mathbi{v_f}}{Dt} &= \mathbi{b_{0f}} + \mathbi{f_b} + \mathbi{f_d} + \divr \boldsymbol{\sigma_f}
\label{Eqn: momentum_strong}
\end{aligned}
\end{equation}
where $\mathbi{f_b}$ and $\mathbi{f_d}$ are inter-phase body forces. $\mathbi{f_d}$ is the \textit{inter-phase drag} or Darcy's law force. $\mathbi{f_b}$ has the form of the \textit{buoyant} force described in \cite{drumheller} for immiscible mixtures,
\begin{equation}
\mathbi{f_b} = p_f \grad (n).
\label{Eqn: buoyant}
\end{equation}

We let the solid phase stress $\boldsymbol{\sigma_s}$ take the classic form,
\begin{equation}
\boldsymbol{\sigma_s} = \boldsymbol{\tilde{\sigma}} - (1-n) p_f \mathsfbi{1}.
\label{Eqn: solid_stress}
\end{equation}
The effective granular stress $\boldsymbol{\tilde{\sigma}}$ is the portion of the solid phase stress resulting from granular contact forces and from microscopic viscous stresses on grains from the fluid medium; it excludes the pressurization of the grains due to the pressure of the pore fluid. When the solid phase is \textit{dense}, this also describes the Terzaghi effective stress that governs plastic flow of the solid phase. The term $p_f$ is the \textit{true} fluid phase pore pressure. Since the fluid is barotropic, this is determined by the \textit{true} fluid density $\rho_f$.

The expression for the fluid phase stress $\boldsymbol{\sigma_f}$ is,
\begin{equation}
\boldsymbol{\sigma_f} = \boldsymbol{\tau_f} - n p_f \mathsfbi{1}.
\label{Eqn: fluid_stress}
\end{equation}
The fluid phase stress is decomposed into a deviatoric part, $\boldsymbol{\tau_f}$, (i.e. $\tr(\boldsymbol{\tau_f}) = 0$) and a isotropic part, $n p_f \mathsfbi{1}$.
With the expressions for the stresses and the buoyant body force given in \eqref{Eqn: solid_stress}, \eqref{Eqn: fluid_stress}, and \eqref{Eqn: buoyant}, we recover the equations of motion from \cite{jackson}.

The solid phase equation of motion is given as,
\begin{equation}
\overline{\rho}_s \frac{D^s \mathbi{v_s}}{Dt} = \overline{\rho}_s \mathbi{g} - \mathbi{f_d} + \divr (\boldsymbol{\tilde{\sigma}}) - (1-n) \grad(p_f)
\label{Eqn: solid_closed_form}
\end{equation}
and the fluid phase equation of motion is given as,
\begin{equation}
\overline{\rho}_f \frac{D^f \mathbi{v_f}}{Dt} = \overline{\rho}_f \mathbi{g} + \mathbi{f_d} + \divr (\boldsymbol{\tau_f}) - n \grad(p_f).
\label{Eqn: fluid_closed_form}
\end{equation}
The equations in \eqref{Eqn: solid_closed_form} and \eqref{Eqn: fluid_closed_form} fully describe the motion and behavior of the mixture; however, we still need to define the specific rules governing the viscous drag between the phases $\mathbi{f_d}$, the elastic-plastic behavior of the solid phase $\boldsymbol{\tilde{\sigma}}$, the pore fluid pressure $p_f$, and the viscous shear response of the fluid phase $\boldsymbol{\tau_f}$. By carefully defining these four constitutive rules, we capture the rheologically correct behavior for mixtures of fluid and grains.

\subsubsection{First and Second Laws of Thermodynamics}
To formulate the rules for $\mathbi{f_d}$, $\boldsymbol{\tilde{\sigma}}$, $p_f$, and $\boldsymbol{\tau_f}$, we start by defining the thermodynamic laws governing our mixture. When considering a single phase of material, it is often useful to assume that \textit{internal energy} $(\varepsilon)$, \textit{entropy} $(s)$, and \textit{absolute temperature} $(\vartheta)$ are basic properties of a material. That is, they do not need to be defined in terms of other more basic properties. For our mixture model, we assume that analogous continuum fields exist describing the energy, entropy, and temperature of the two continuum phases; however the physical basis of these fields is poorly defined (see  \cite{wilmanski} and \cite{klika}). We therefore rely on the intuition developed in \cite{gurtin} to specialize the thermodynamic analysis from \cite{drumheller} to a mixture of grains (represented by an elastic-plastic porous solid) with a barotropic viscous fluid.

The full thermodynamic analysis is shown in appendix \ref{appA}, and a brief summary of the resulting constitutive rules is given in table \ref{Tab: constitutive_rules}. Through the analysis, we find that the fluid pore pressure $p_f$ must be defined by the fluid phase specific free energy function $\hat{\psi}_f(\rho_f)$ and that the fluid shear stress $\boldsymbol{\tau_f}$ and inter-phase drag $\mathbi{f_d}$ must both be dissipative, that is they must `resist' their driving motion (the symmetric part of the fluid phase velocity gradient $\mathsfbi{D_f}$ and the difference in phase velocities $(\mathbi{v_s-v_f})$ respectively). We also show that the solid phase effective granular stress $\boldsymbol{\tilde{\sigma}}$ can be expressed as that of an elastic-plastic solid with behavior defined by \eqref{Eqn: solid_stress_free_energy}. For these kinds of materials, the stress is determined by the the strain-energy function $\hat{\varphi}_s$, which depends on the \textit{elastic part of the deformation gradient} $\mathsfbi{F^e}$ as defined in \eqref{Eqn: kroner_lee}, which in turn defines the elastic volume Jacobian $J^e$ and the \textit{right elastic Cauchy-Green tensor} $\mathsfbi{C^e}$. The elastic tensor $\mathsfbi{F^e}$ is a history dependent material tensor that evolves through time according to a decomposition of the solid phase strain-rate $\mathsfbi{D_s}$ into an elastic strain-rate $\mathsfbi{D^e}$ and a plastic part $\mathsfbi{\tilde{D}^p}$ according to \eqref{Eqn: additive_decomposition} and subject to the dissipative inequality in \eqref{Eqn: solid_plasticity_free_energy}.  We note that in most common granular materials the bulk elastic deformations are extremely small, especially compared to plastic deformation, however grains do have finite stiffness and proper accounting of granular elasticity is important for thermodynamic consistency of the constitutive relations.

\begin{table}
	\begin{center}
		\begin{tabular}{ lcc } 
			\textbf{Rule}& \centering \textbf{Expression} & \textbf{Number} \\ 
			Fluid Pore Pressure Equality & $\displaystyle
			p_f - \rho_f^2 \frac{\partial \hat{\psi}_f(\rho_f)}{\partial \rho_f} = 0$ & \eqref{Eqn: fluid_pressure_free_energy}\\[10pt]
			Fluid Shear Stress Inequality & $\displaystyle\boldsymbol{\tau_f}:\mathsfbi{D_{f}} \geq 0$ & \eqref{Eqn: fluid_stress_free_energy}\\[5pt]
			Effective Granular Stress Equality & $\displaystyle\boldsymbol{\tilde{\sigma}} - 2 J^{e-1} \mathsfbi{F^e} \frac{\partial \hat{\varphi}_s(\mathsfbi{C^e})}{\partial \mathsfbi{C^e}} \mathsfbi{F^{e\top}} = 0$ & \eqref{Eqn: solid_stress_free_energy}\\[8pt]
			Effective Granular Stress Inequality & $\displaystyle \boldsymbol{\tilde{\sigma}}:\mathsfbi{\tilde{D}^p} - J^{e-1} \varphi_s \tr(\mathsfbi{\tilde{D}^p}) \geq 0$ & \eqref{Eqn: solid_plasticity_free_energy}\\[8pt]
			Drag Law Inequality & $\displaystyle\mathbi{f_d} \cdot (\mathbi{v_s} - \mathbi{v_f}) \geq 0$ &\eqref{Eqn: interphase_drag_free_energy}\\
		\end{tabular}
	\end{center}
	\caption{Summary of thermodynamic rules for constitutive laws derived in appendix \ref{appA}.}
	\label{Tab: constitutive_rules}
\end{table}

\subsection{Inter-Phase Drag Law}
The flow of a viscous fluid around and between grains of sediment will result in an inter-phase drag that we represent with the drag force $\mathbi{f_d}$. This drag force can be understood as a body force acting on one phase by the other and has units $N/m^3$. In this work we assume that this force depends only on the relative velocities of the two phases $(\mathbi{v_s} - \mathbi{v_f})$, the porosity of the mixture $n$, the grain diameter $d$, and the fluid viscosity $\eta_0$. We neglect dependence on material orientation or rotation (e.g. a fabric tensor) and neglect the effects of tortuosity (see \cite{coussy2004}) on the apparent inter-phase drag.

\begin{figure}
	\centering
	\includegraphics[scale=0.35]{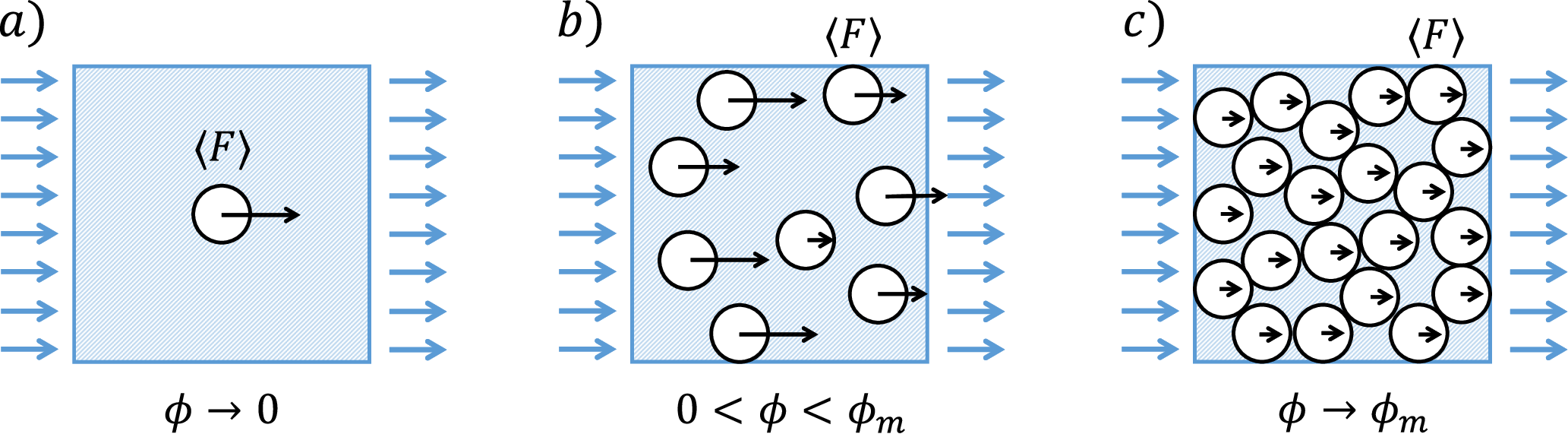}
	\caption{The three regimes over which the inter-phase drag $\mathbi{f_d}$ must be defined. The normalized average drag force $\langle F \rangle$ is taken from \cite{beetstra1} to be the average force on a single grain for a given packing fraction $\phi$ at a given flow rate.}
	\label{Fig: drag_law}
\end{figure}

For \textit{small} flow velocities, the drag interaction between the fluid and the solid grains in the dilute limit shown in figure \ref{Fig: drag_law}a) is given analytically by the Stokes-Einstein equation. 
\begin{equation}
\langle F \rangle = 3 \pi \eta_0 d u\label{Eqn: stokes_einstein}
\end{equation}
where $u$ is the free-steam flow speed. \textit{Small} here is taken to mean $\Reyn \to 0$ with,
\begin{equation}
\Reyn \equiv \frac{n \rho_f d \|\mathbi{v_s-v_f}\|}{\eta_0}
\label{Eqn: reynolds}
\end{equation}
The leading $n$ in \eqref{Eqn:  reynolds} is taken from \cite{dupuit} and relates the free-stream velocity to the average pore velocity. Normalizing the inter-phase drag by the volume average of the Stokes-Einstein drag on a single grain suggests the following functional form of $\mathbi{f_d}$,
\begin{equation}
\mathbi{f_d} = \frac{18 \phi (1-\phi) \eta_0}{d^2} \ \hat{F}(\phi,\Reyn) \  (\mathbi{v_s} - \mathbi{v_f})
\label{Eqn: beetstra_final}
\end{equation}
with $\hat{F}(\phi, \Reyn)$ a function of non-dimensional parameters only. The \textit{only} thermodynamic requirement on $\mathbi{f_d}$ (see \eqref{Eqn: interphase_drag_free_energy}) is satisfied if $\hat{F}(\phi, \Reyn) \geq 0$ for all values of $\phi$ and $\Reyn$.

Determining the expression for $\hat{F}(\phi, \Reyn)$ for the full range of potential packing fractions $(0 \leq \phi \lesssim 0.65)$ has historically been an intractable challenge. Analytical methods \textit{cannot} be used for high Reynolds number flows $(\Reyn > 1)$ and flows with non-negligible packing fractions $(\phi > 0)$ (see \cite{clift}). Experimentally, any \textit{loose} packing $(\phi \lesssim 0.58)$ without sustained granular contacts will quickly compact, making the collection of accurate measurements near impossible.

Recent work by \cite{beetstra1} and \cite{beetstra2} make use of the lattice-Boltzman method to simulate the flow of fluid around mono- and bi-disperse packings of spheres for $0.10 <\phi < 0.6$ and $\Reyn < 1000$. These simulations give the following form of $\hat{F}$ at low Reynolds numbers $(\Reyn \to 0)$,
\begin{equation}
\hat{F}(\phi,0) = \frac{10 \phi}{(1-\phi)^2} + (1-\phi)^2(1 + 1.5 \sqrt{\phi})
\label{Eqn: beetstra1}
\end{equation}
with the following high Reynolds correction,
\begin{equation}
\hat{F}(\phi,\Reyn) = \hat{F}(\phi,0) + \frac{0.413 \Reyn}{24 (1-\phi)^2} \bigg( \frac{(1-\phi)^{-1} + 3 \phi (1-\phi) + 8.4 \Reyn^{-0.343}}{1 + 10^{3\phi} \Reyn^{-(1+4\phi)/2}} \bigg)
\label{Eqn: beetstra2}
\end{equation}
In the dilute, low Reynolds limit, \eqref{Eqn: beetstra1} and \eqref{Eqn: beetstra2} recover the Stokes-Einstein inter-phase drag. In the dense, low Reynolds limit, \eqref{Eqn: beetstra1} and \eqref{Eqn: beetstra2} recover the Carman-Kozeny inter-phase drag from \cite{carman} as used in \cite{bandara},
$$\lim_{\phi \to 0} \hat{F}(\phi,0) = 1, \qquad \lim_{\phi \to 1} \hat{F}(\phi,0) = \frac{10 \phi}{(1-\phi)^2} .$$

\subsection{Fluid Phase Pore Pressure}
The fluid phase pore pressure is governed by the constitutive relation given in \eqref{Eqn: fluid_pressure_free_energy}. We let the fluid phase free energy function, $\hat{\psi}(\rho_f)$, be given by,
\begin{equation}
\hat{\psi}_f(\rho_f) = \kappa \bigg(\frac{\ln(\rho_{0f}) - \ln(\rho_f) - 1}{\rho_f^2}\bigg), \qquad \text{s.t.} \qquad
p_f = \kappa \ln\bigg( \frac{\rho_f}{\rho_{0f}} \bigg)
\label{Eqn: fluid_phase_pore_pressure}
\end{equation}
where $\rho_{0f}$ is the true fluid density for which $p_f = 0$ and $\kappa$ is the fluid bulk modulus with units of Pa.

\subsection{Fluid Phase Shear Stress}
We assume that the functional form of $\boldsymbol{\tau_f}$ is given by, $\boldsymbol{\tau_f} = \boldsymbol{\hat{\tau}_f}(\mathsfbi{D_{f}},\phi)$ with $\boldsymbol{\hat{\tau}_f}$ isotropic and linear in $\mathsfbi{D_{f}}$, the symmetric part of the fluid strain-rate tensor (see \eqref{Eqn: spin_and_strainrate_tensors}). From \cite{truesdell}, the representation theorem for isotropic linear tensor functions requires that
$$\boldsymbol{\hat{\tau}_f}(\mathsfbi{D_{f}},\phi) = 2 \mu(\phi) \mathsfbi{D_{f}} + \lambda(\phi) \tr(\mathsfbi{D_{f}}) \mathbf{1}$$
We assume $\boldsymbol{\tau_f}$ is deviatoric, which requires $\lambda(\phi)=-\tfrac{2}{3}\mu(\phi)$. And the thermodynamic restriction on $\boldsymbol{\tau_f}$ in \eqref{Eqn: fluid_stress_free_energy} yields $\mu(\phi) \geq 0$. We let the effective fluid phase viscosity, $\mu(\phi)$, be given by the linear relation from \cite{einstein} such that,
\begin{equation}
\boldsymbol{\tau_f} = 2 \eta_0 \big(1 + \tfrac{5}{2}\phi\big) \mathsfbi{D_{0f}}
\label{Eqn: fluid_shear_stress}
\end{equation}
with $\eta_0$ defined previously as the \textit{true} fluid viscosity.

\subsection{Solid Phase Stress Evolution}
The solid phase effective granular stress is a function of the accumulated elastic deformation in the solid phase, $\mathsfbi{F^e}$, as defined in \eqref{Eqn: solid_stress_free_energy} (see Table 1). In appendix \ref{appB}, we show that for \textit{stiff} elastic materials, \eqref{Eqn: solid_stress_free_energy} is satisfied if the effective granular stress is evolved according to the following approximation using the Jaumann objective rate of $\boldsymbol{\tilde{\sigma}}$,
$$\overset{\Delta}{\boldsymbol{\tilde{\sigma}}} \equiv \frac{D^s \boldsymbol{\tilde{\sigma}}}{Dt} - \mathsfbi{W_s}\boldsymbol{\tilde{\sigma}} + \boldsymbol{\tilde{\sigma}}\mathsfbi{W_s} \approx \mathscr{C}[\mathsfbi{D^e}]$$
with $\mathscr{C}$ an elastic stiffness tensor defined in \eqref{Eqn: elastic_stress}, $\mathsfbi{W_s}$ the skew part of the solid phase velocity gradient, and $\mathsfbi{D^e} = \mathsfbi{D_s} - \mathsfbi{\tilde{D}^p}$ (see \eqref{Eqn: additive_decomposition}). The material derivative of the effective granular stress is therefore given by,
\begin{equation} 
\frac{D^s \boldsymbol{\tilde{\sigma}}}{Dt} = 2G \mathsfbi{D^e_0} + K \tr(\mathsfbi{D^e})\mathsfbi{1} + \mathsfbi{W_s}\boldsymbol{\tilde{\sigma}} - \boldsymbol{\tilde{\sigma}}\mathsfbi{W_s}
\label{Eqn: solid_phase_granular_stress_evolution}
\end{equation}
In this way, we evolve the effective granular stress according to how the solid phase is straining $(\mathsfbi{D_s})$ minus how much of that strain-rate is plastic $(\mathsfbi{\tilde{D}^p})$. Physically, when the solid phase is flowing, most of the strain is accumulated plastically; when the solid phase is static (or resisting flow), strain is accumulated elastically.  In this sense, through careful selection of the plastic flow rule $\mathsfbi{\tilde{D}^p}$, the model can represent both flowing and static (sub-yield) behaviors of the solid phase.

\subsection{Solid Phase Plastic Flow Rules}\label{Sec: plastic_flow_rules}
We let $\mathsfbi{\tilde{D}^p}$ have the following form,
\begin{equation}
\mathsfbi{\tilde{D}^p} = \frac{\dot{\bar{\gamma}}^p}{\sqrt{2}} \frac{\boldsymbol{\tilde{\sigma}_0}}{\|\boldsymbol{\tilde{\sigma}_0}\|} + \frac{1}{3} \big( \beta \dot{\bar{\gamma}}^p + \dot{\xi}_1 + \dot{\xi}_2 \big) \mathsfbi{1}
\label{Eqn: plastic_flow_rule}
\end{equation}
where the `over-dot' operator $\dot{\psi}$ is equivalent to the material derivative $\frac{D^s \psi}{Dt}$. The only thermodynamic constraint on this plastic flow relation is given by \eqref{Eqn: cauchy_stress_dissipation}. It is a simple exercise to show that the following formulation obeys that inequality.

\begin{figure}
	\centering
	\includegraphics[scale=0.4]{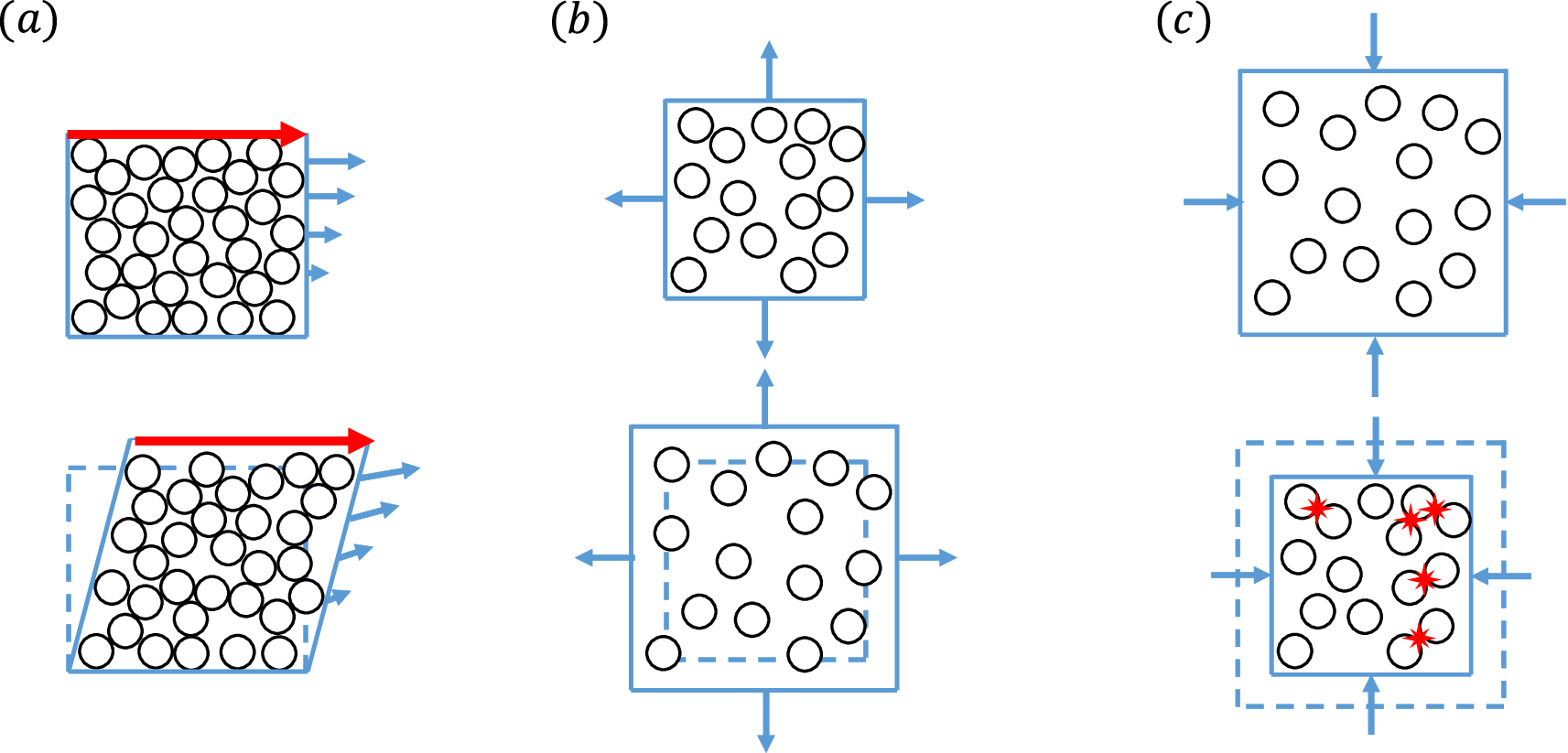}
	\caption{(a) In shear, the granular phase will obey critical state behavior and `open-up'. This phenomena is called Reynolds' dilation and is captured by the rate of \textit{plastic dilation}, $\beta\dot{\bar{\gamma}}^p$. (b) In expansion, the granular phase will `open' freely. This phenomena is stress-free and is captured by the rate of \textit{plastic expansion}, $\dot{\xi}_1$. (c) In compaction, granular collisions will result in a macroscopic pressure. This phenomena is governed by the rate of \textit{plastic compaction}, $-\dot{\xi}_2$.}
	\label{Fig: yield_surfaces}
\end{figure}
The \textit{equivalent plastic shear strain-rate} $\dot{\bar{\gamma}}^p$, the rate of \textit{plastic expansion} $\dot{\xi}_1$ (figure \ref{Fig: yield_surfaces}b), the rate of \textit{plastic compaction} $-\dot{\xi}_2$ (figure \ref{Fig: yield_surfaces}c), and the rate of \textit{Reynolds dilation} $\beta \dot{\bar{\gamma}}^p$ (figure \ref{Fig: yield_surfaces}a) are the scalar measures that give the solid phase plastic flow. These flow measures are uniquely determined by the solid phase strain-rate $\mathsfbi{D_s}$, the solid phase effective stress $\boldsymbol{\tilde{\sigma}}$, and the current state of the mixture. 

The \textit{dilation angle}, $\beta$, governs the rate of Reynolds dilation during plastic shear (see \cite{roux1998}, \cite{roux2001}, and \cite{rudnicki}) and allows the material to dilate when shearing over-compacted grains and contract when shearing under-compacted grains.
We use the functional form of $\beta$ given in \cite{pailha},
\begin{equation}
\beta = K_3 (\phi - \phi_{eq})
\label{Eqn: beta}
\end{equation}
where $K_3$ is a unit-less material parameter and $\phi_{eq}$ is the rate-dependent equilibrium packing fraction achieved in steady-state shearing, given by \cite{amarsid} as,
\begin{equation}
\phi_{eq} = \frac{\phi_m}{1 + aI_m}
\label{Eqn: critical_state_packing}
\end{equation}
with $a$ a material parameter, $I_m$ the \textit{mixed inertial number}, and $\phi_m$ a material parameter describing the maximum possible packing fraction for a granular material in steady-state shearing flow. The non-dimensional inertial numbers (including the \textit{inertial number}, $I$, and the \textit{viscous inertial number}, $I_v$) are defined as,
\begin{equation}
I = \dot{\bar{\gamma}}^p d \sqrt{\frac{\rho_s}{\tilde{p}}}, \qquad I_v = \frac{\eta_0 \dot{\bar{\gamma}}^p}{\tilde{p}}, \qquad 
I_m = \sqrt{I^2 + 2 I_v}
\label{Eqn: inertial_numbers}
\end{equation}
%
with $\tilde{p}$ the granular pressure described in \cite{boyer}. The specific form of the \textit{mixed inertial number}, $I_m$, was determined in \cite{amarsid} by analyzing numerous 2D shearing flows spanning the inertial and viscous regimes ($0 \lesssim I \lesssim 0.1$ and $0 \lesssim I_v \lesssim 0.2$).

To determine $\dot{\bar{\gamma}}^p$, $\dot{\xi}_1$, and $\dot{\xi}_2$ it is convenient to express their functional dependences implicitly in terms of the yield conditions given below. First, we uniquely define the equivalent plastic shear rate $\dot{\bar{\gamma}}^p$ by solving
\begin{equation}
\begin{aligned}
	f_1 &= \bar{\tau} - \max\big((\mu_p + \beta)\tilde{p},\ 0\big)\\[1pt]
	f_1 &\leq 0, \qquad \dot{\bar{\gamma}}^p \geq 0, \qquad f_1 \dot{\bar{\gamma}}^p = 0
\end{aligned}
\label{Eqn: f1_yield_surface}
\end{equation}
with,
\begin{equation}
\bar{\tau} = \frac{\|\boldsymbol{\tilde{\sigma}_0}\|}{\sqrt{2}}, \qquad \tilde{p} = -\frac{1}{3} \tr(\boldsymbol{\tilde{\sigma}}).
\label{Eqn: tau_and_p}
\end{equation}
Solutions to this system have non-zero plastic shearing only when the yield condition, $f_1=0$, is met, and vanishing plastic shear-rate when below yield, $f_1<0$. We let $\mu_p = \hat{\mu}_p(\phi, I_v, I_m)$, which is formulated to capture both the $\mu(I)$ dry granular rheology from \cite{jop} and the $\mu(I_v)$ low Stokes mixture rheology from \cite{boyer}, as will be shown in section \ref{Sec: verification}. The functional form of $\hat{\mu}_p$ is defined as,
\begin{equation}
\hat{\mu}_p(\phi,I_v,I_m) = \mu_1 + \frac{\mu_2 - \mu_1}{1 + (b/I_m)} + \tfrac{5}{2} \bigg(\frac{\phi I_v}{a I_m}\bigg).
\label{Eqn: internal_friction_coefficient}
\end{equation}
Note that in steady-state shearing, $\phi = \phi_{eq}$ and $\hat{\mu}_p$ reduces to a function of $I_v$ and $I_m$ only. 

Granular separation, represented by the rate of plastic expansion, $\dot{\xi_1}$, is obtained from the conditions
\begin{equation}
\begin{aligned}
f_2 &= -\tilde{p}\\[1pt]
f_2 &\leq 0, \qquad \dot{\xi}_1 \geq 0, \qquad f_2 \dot{\xi}_1 = 0.
\label{Eqn: f2_yield_surface}
\end{aligned}
\end{equation}
These conditions enforce the assumption that non-cohesive grains cannot support tension.  Hence, the granular media undergoes plastic expansion $\dot{\xi}_1$, representing grain separation, in lieu of developing tensile granular stress states.

The flow rule governing plastic compaction, $-\dot{\xi}_2$, arises from solving the system below:
\begin{equation}
\begin{aligned}
f_3 &= g(\phi) \tilde{p} - (a\phi)^2 \big[ (\dot{\bar{\gamma}}^p - K_4 \dot{\xi}_2)^2 d^2 \rho_s + 2 \eta_0 (\dot{\bar{\gamma}}^p - K_4 \dot{\xi}_2) \big]
\\[1pt]
f_3 &\leq 0, \qquad \dot{\xi}_2 \leq 0, \qquad f_3 \dot{\xi}_2 = 0
\label{Eqn: f3_yield_surface}
\end{aligned}
\end{equation}
with,
\begin{equation}
g(\phi) = \bigg\{
\begin{matrix}
(\phi_m - \phi)^2& \text{if \quad} \phi < \phi_m\\
0& \text{if \quad} \phi \geq \phi_m.
\end{matrix}
\label{Eqn: f3_g_expression}
\end{equation}
The form of $g(\phi)$ and the $f_3$ yield surface is chosen such that when the material is being compacted or sheared but is less dense than the critical packing, $\phi < \phi_m$, there is an upper bound on the admissible effective pressure $\tilde{p}$. However, in the compacted regime, $\phi \geq \phi_m$, any pressure is admissible, as the grains are assumed to always be touching. The upper bound on the value of $\tilde{p}$ is determined by inverting the expression for $\phi_{eq}$ defined in \eqref{Eqn: critical_state_packing}. The unit-less $K_4$ coefficient defines the relative importance of the plastic compaction rate in determining this upper bound compared to the plastic shear-rate.

Together \eqref{Eqn: solid_phase_granular_stress_evolution}, \eqref{Eqn: plastic_flow_rule}, \eqref{Eqn: additive_decomposition}, \eqref{Eqn: f1_yield_surface}, \eqref{Eqn: f2_yield_surface}, and \eqref{Eqn: f3_yield_surface} uniquely determine the plastic flow rates $\dot{\bar{\gamma}}^p$, $\dot{\xi}_1$, and $\dot{\xi}_2$. It is also important to note that the specific forms of \eqref{Eqn: f2_yield_surface} and \eqref{Eqn: f3_yield_surface} restrict $\dot{\xi}_1$ to be zero when plastic compaction occurs and restrict $\dot{\xi}_2$ to be zero when plastic expansion occurs.

\subsection{Summary of Model Assumptions}
The model presented in this section is formulated to capture several key phenomena observed in mixtures of fluids and grains. In the development of this model we have assumed that the granular material is quasi-mono-disperse, composed of incompressible cohesion-less solid grains, and fully-saturated with an isothermal Newtonian liquid. We have neglected Brownian effects on the mixture, limiting the applicability of our model to the study of granular mixtures which are dominated by gravitational energy or by shearing time-scales ($d \gtrsim 100 \mu m$ for common engineering slurries). In addition, the evolution law for the effective granular stress (including the plastic flow rule) is only applicable in the limit of \textit{stiff} elasticity ($G,\ K \gg \tilde{p}$).

We have derived our inter-phase drag law (see \eqref{Eqn: beetstra_final}) from the empirical relations given in \cite{beetstra1}, \cite{beetstra2}, and \cite{carman}. The simulations and experiments that underpin these empirical relations suggest that this drag law is applicable for $0 \leq \phi \lesssim 0.65$ and $\Reyn \leq 1000$. In addition the internal friction coefficient for the solid phase (see \eqref{Eqn: internal_friction_coefficient}) is developed through consideration of models presented in \cite{boyer}, \cite{amarsid}, and \cite{stickel}. The data which underpins these empirical models suggest that the internal friction model is applicable for $I \lesssim 0.2,$ $I_v \lesssim 0.1$, $I_m \lesssim 0.6$, and $0 \leq \St < \infty$.

\section{Analytical Verification of Model}\label{Sec: verification}
In this section we verify that the model laid out in section \ref{Sec: theory_and_formulation} has the correct limiting behavior in a simple shearing flow. In particular we are interested in showing that under the appropriate conditions, the following rheologies are captured.

\begin{itemize}
	\item $\mu(I)$, $\phi(I)$ \quad steady-state dry granular inertial rheology.
	\item $\mu(I_v)$, $\phi(I_v)$ \quad steady-state viscous inertial rheology.
	\item $\eta_r(\phi)$ \quad slurry/suspension effective viscosity.
\end{itemize}

These phenomena should arise as different cases of steady shearing flow wherein the mixture is co-moving such that, $\mathbi{v_s} = \mathbi{v_f}$ and,
$$\grad(\mathbi{v_s}) = \grad(\mathbi{v_f}) = \mathsfbi{L} =
\begin{bmatrix}
0& \dot{\gamma}& 0\\
0& 0& 0\\
0& 0& 0
\end{bmatrix}
$$
where $\dot{\gamma}$ is the applied \textit{steady shear-rate}. Since the mixture is uniform and $\tr(\mathsfbi{L}) = 0$, \eqref{Eqn: true_fluid_density} tells us that the \textit{true} fluid density, $\rho_f$, is constant. By \eqref{Eqn: fluid_phase_pore_pressure}, this means that the fluid phase pore pressure remains constant, $p_f = p_{eq}$, with $p_{eq}$ some constant equilibrium pressure.

The fluid phase shear stress, $\boldsymbol{\tau_f}$, is determined by \eqref{Eqn: fluid_shear_stress},
\begin{equation}
\boldsymbol{\tau_f} = \eta_0(1 + \tfrac{5}{2}\phi)\begin{bmatrix}
0& \dot{\gamma} & 0\\
\dot{\gamma} & 0& 0\\
0& 0& 0
\end{bmatrix}.
\label{Eqn: steady_shear_fluid_shear_stress}
\end{equation}
In the solid phase, there are two regimes of interest, the compacted regime with $\phi \geq \phi_m$ and the non-compacted regime with $\phi < \phi_m$. In the compacted regime, the sustained granular contacts result in non-steady behavior (the positivity of the dilatation angle $\beta$ from \eqref{Eqn: beta} results in continuous growth of the pressure $\tilde{p}$). For this reason, we will be more interested in the behavior of the non-compacted regime, where steady-state flow is possible.

Assuming that the solid phase begins in a stress-free state and that the \textit{shear modulus}, $G$, is much greater than the characteristic shear stress, it can be shown that \eqref{Eqn: solid_phase_granular_stress_evolution} and \eqref{Eqn: plastic_flow_rule} together imply that $\boldsymbol{\tilde{\sigma}}$ will reach a steady value with the \textit{equivalent plastic shear rate} $\dot{\bar{\gamma}}^p$ non-zero and equivalent to total \textit{steady shear rate} $\dot{\gamma}$. The solid phase effective granular stress then satisfies,
\begin{equation} 
\boldsymbol{\tilde{\sigma}} \approx
\begin{bmatrix}
\tilde{p}& \mu_p \tilde{p}& 0\\
\mu_p \tilde{p}& \tilde{p}& 0\\
0& 0& \tilde{p}
\end{bmatrix}
\label{Eqn: steady_shear_solid_stress}
\end{equation}

The total mixture stress as defined in \eqref{Eqn: mixture_stress} is characterized by the mixture pressure $p = -\tfrac{1}{3}\tr(\boldsymbol{\sigma})$ and the mixture shear stress $\tau = \tfrac{1}{\sqrt{2}}\|\boldsymbol{\sigma_0}\|$. In the case of steady shearing flow we find,
\begin{equation} 
\begin{aligned}
\tau &= \eta_0 \big(1 + \tfrac{5}{2}\phi\big)\dot{\gamma} + \mu_p \tilde{p}\\
p &= p_{eq} + \tilde{p}
\end{aligned}
\label{Eqn: steady_shear_traction}
\end{equation}
with the steady state packing fraction, $\phi = \phi_{eq}$, given by \eqref{Eqn: critical_state_packing}.

\subsection{Dry Granular Flow}
In steady simple shear flow, dry granular materials have been shown to obey the $\mu(I)$ and $\phi(I)$ rheology as given in \cite{jop} and \cite{dacruz},
$$\mu(I) = \mu_1 + \frac{\mu_2 - \mu_1}{1+(I_0/I)}, \qquad \text{and} \qquad \phi(I) = \phi_m - cI$$
with $c$ some material constant and where $\mu(I)$ is the ratio between the measured shear stress $\tau$ and the measured granular pressure $\tilde{p}$. This behavior is captured by our model in the limit that $\eta_0 \to 0$. By the definitions of the inertial numbers in \eqref{Eqn: inertial_numbers}, if $\eta_0 = 0$, then $I_m = I$. Additionally, for $\eta_0 = 0$, $\mu = \mu_p$ as given in \eqref{Eqn: internal_friction_coefficient}. Expanding the expression for $\phi_{eq}$ from \eqref{Eqn: critical_state_packing} around $I=0$ (where existing data has been collected), we find that our model predicts the following steady shear behavior,
$$\mu = \mu_1 + \frac{\mu_2 - \mu_1}{1+(b/I)}, \qquad \text{and} \qquad \phi \approx \phi_m - a\phi_mI + O(I^2)$$
Which is a reasonable approximation to known fits of the $\mu(I)$, $\phi(I)$ rheology if $a = (c / \phi_m)$ and $b = I_0$.

\subsection{Viscous Granular Mixtures} \label{Sec: boyer_verification}
\cite{boyer} experimentally investigate the steady-state rheology of mixtures undergoing steady, quasi-2D shear flow at low Stokes numbers. The Stokes number of interest in this context is defined in \cite{amarsid} as,
\begin{equation}
\St = \frac{\rho_s d^2 \dot{\gamma}}{\eta_0} = \frac{I^2}{I_v}
\label{Eqn: stokes_number}
\end{equation}
In the limit that $\St \to 0$, the mixed inertial number $I_m$ is dominated by the viscous inertial number $I_v$, such that $I_m = \sqrt{2I_v}$.

\cite{boyer} defines the $\mu(I_v)$ and $\phi(I_v)$ viscous granular rheologies as follows,
$$\mu(I_v) = \mu_1 + \frac{\mu_2 - \mu_1}{1 + (I_0/I_v)} + I_v + \tfrac{5}{2} \phi_m \sqrt{I_v} \qquad \text{and} \qquad \phi(I_v) = \frac{\phi_m}{1+\sqrt{I_v}}$$

\begin{figure}
	\centering
	\includegraphics[scale=0.45]{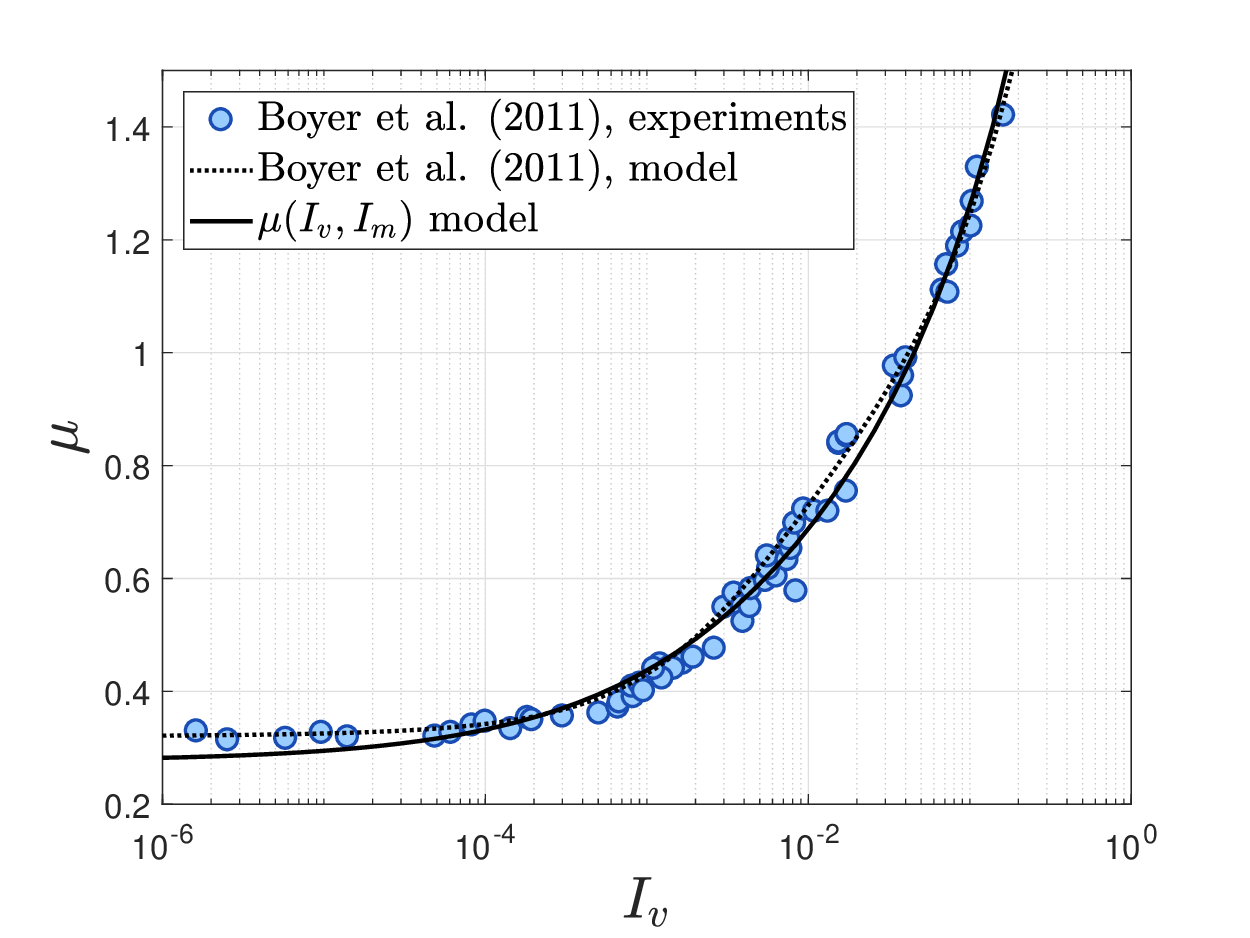}
	\caption{Plot of the ratio between the shear stress and effective granular pressure ($\mu$) against the inertial number $I_v$. Data collected by \cite{boyer} is shown as the shaded blue circles. The $\mu(I_v)$ rheology from that work is represented by the dotted line. The combined response of the mixture model presented in this work (see \eqref{Eqn: steady_shear_boyer_mu}) is represented by the solid line.}
	\label{Fig: boyer_comparison}
\end{figure}

It can be shown from \eqref{Eqn: critical_state_packing}, \eqref{Eqn: inertial_numbers}, and \eqref{Eqn: internal_friction_coefficient} that the steady shear response of our mixture is given by,
\begin{equation} 
\mu(I_v,I_m) = \mu_1 + \frac{\mu_2 - \mu_1}{1 + (b/\sqrt{2I_v})} + I_v  + \tfrac{5}{2} \phi_m \sqrt{\frac{I_v}{2a^2}}, \qquad \text{and} \qquad \phi = \frac{\phi_m}{1+a\sqrt{2I_v}}
\label{Eqn: steady_shear_boyer_mu}
\end{equation}
The expression for the steady-state packing fraction $\phi$ in \eqref{Eqn: steady_shear_boyer_mu} identically recovers the $\phi(I_v)$ fit from \cite{boyer} when $a = \tfrac{1}{\sqrt{2}}$. The $\mu(I_v)$ function of \cite{boyer} is not reproduced exactly with our model; however, as shown in figure \ref{Fig: boyer_comparison}, we can fit our form $\mu(I_v,I_m)$ to their data directly. Strong agreement is found between our model fit, the model fit in \cite{boyer}, and the data collected in that work. The fit parameters for the plot in figure \ref{Fig: boyer_comparison} are given in table \ref{Tab: boyer_fit}

\begin{table}
	\begin{center}
		\setlength{\tabcolsep}{1em} 
		\begin{tabular}{  c  c  c  } 
			Parameter & $\mu(I_v)$ & $\mu(I_v,I_m)$ \\ 
			$\mu_1$ & 0.32 & 0.2764 \\ 
			$\mu_2$ & 0.7 & 0.8797 \\ 
			$I_0$ & 0.005 & - \\ 
			$\phi_m$ & 0.585 & 0.585 \\ 
			$a$ & - & 0.7071 \\ 
			$b$ & - & 0.1931 \\ 
		\end{tabular}
	\end{center}
	\caption{Parameters for model fit to data in figure \ref{Fig: boyer_comparison}.}\label{Tab: boyer_fit}
\end{table}

\subsection{Suspension Effective Viscosity}
Significant work has been done on understanding the behavior of co-moving suspensions of granular material in fluids. We are particularly interested in the change in effective fluid viscosity of suspensions due to the solid phase volume fraction as reviewed and summarized in \cite{stickel} with $\eta_r$, the \textit{relative viscosity},
$$\eta_r = \frac{\tau}{\eta_0 \dot{\gamma}}, \qquad \lim_{\phi \to \phi_m}{\eta_r} = \infty, \qquad \lim_{\phi \to 0}{\frac{\eta_r-1}{\phi}} = [\eta]$$
In the dense limit ($\phi \to \phi_m$), the viscosity of the suspension approaches infinity and in the dilute limit ($\phi \to 0$), the viscosity of the mixture should vary linearly with $[\eta]$ where $[\eta] = \tfrac{5}{2}$ for hard spheres (\cite{stickel}).

As in section \ref{Sec: boyer_verification}, we are concerned with the behavior of our mixture model in the low Stokes limit such that $I_m = \sqrt{2I_v}$. Therefore we find, $\eta_r = 1 + \tfrac{5}{2}\phi + \tfrac{\mu_p}{I_v}$, which by \eqref{Eqn: critical_state_packing} and \eqref{Eqn: internal_friction_coefficient} is equivalently,
\begin{equation} 
\eta_r(\phi) = 1 + \frac{5}{2}\phi\bigg(\frac{\phi_m}{\phi_m - \phi}\bigg) + 2\bigg(\frac{a\phi}{\phi_m-\phi}\bigg)^2\bigg(\mu_1 + \frac{\mu_2 - \mu_1}{1 + ab\phi/(\phi_m-\phi)}\bigg)
\label{Eqn: steady_shear_relative_viscosity}
\end{equation}
It can be shown that this relation achieves both limiting behaviors required of effective viscosity models.

\begin{figure}
	\centering
	\includegraphics[scale=0.45]{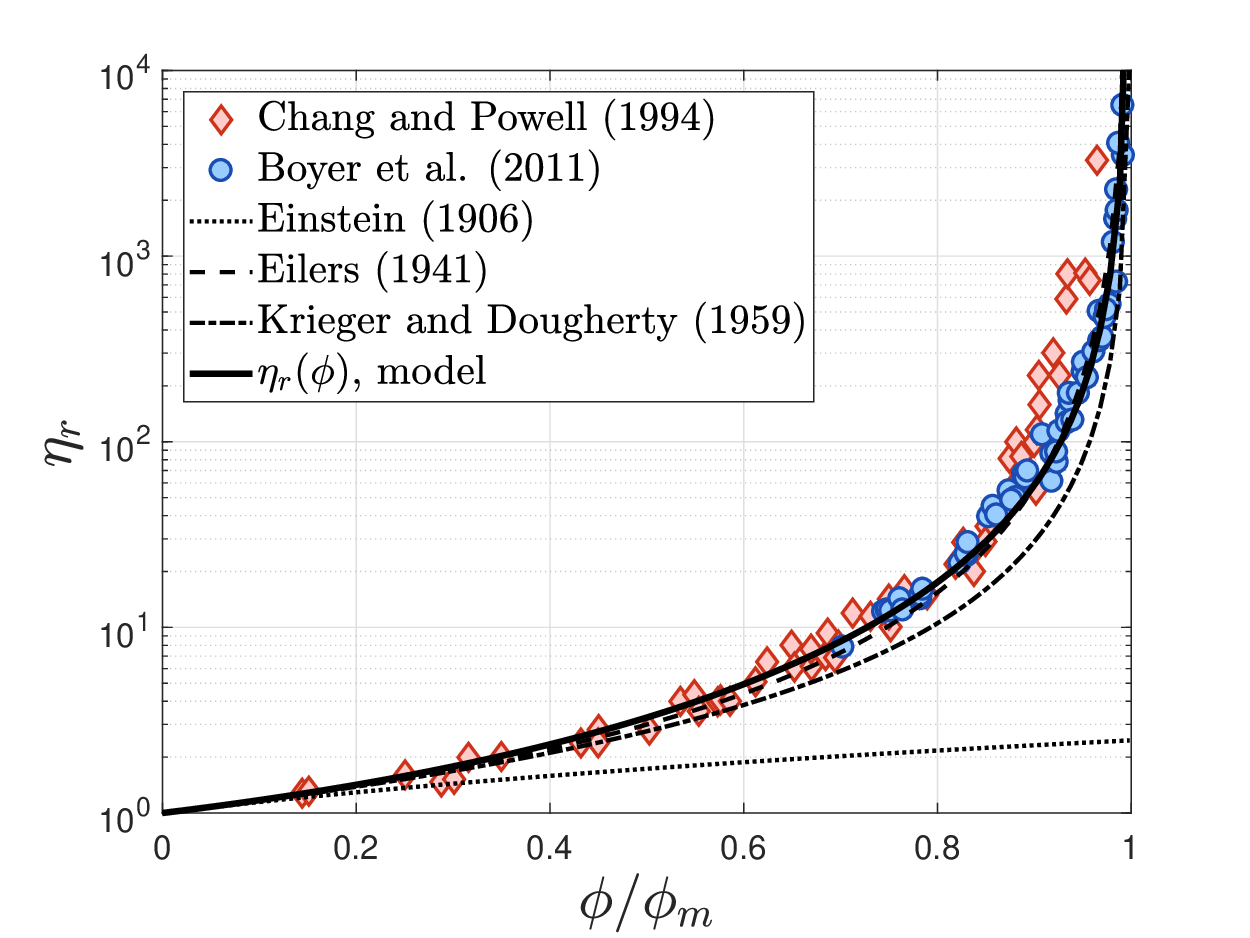}
	\caption{Plot of various models for effective viscosity $\eta_r$ versus the relative packing fraction $\phi/\phi_m$. The model given in this work by \eqref{Eqn: steady_shear_relative_viscosity} is shown by the solid line. The red diamonds represent the experimental results reported in \cite{chang1994} (from \cite{chong}, \cite{poslinski}, \cite{storms}, \cite{shapiro}, \cite{chang1993}, and \cite{chang1994}). The blue circles represent the experimental results reported in \cite{boyer}.}
	\label{Fig: powell_comparison}
\end{figure}

By noting the similarity of the materials used by \cite{chang1994} (PS and PMMA), to that used in \cite{boyer}, we use the coefficients determined in section \ref{Sec: boyer_verification} and given in table \ref{Tab: boyer_fit} to compare \eqref{Eqn: steady_shear_relative_viscosity} against the experimental measurements reported in \cite{chang1994} and \cite{boyer} (see figure \ref{Fig: powell_comparison}). 

\section{Numerical Implementation}\label{Sec: implementation}
We are interested in time-accurate simulations of fluid-sediment mixtures undergoing arbitrarily large deformations. To do this, we use a material point method (MPM) framework capable of simultaneously solving all of the governing equations shown in table \ref{Tab: governing_equations}. This MPM framework is a derivative of that shown in \cite{dunatunga} and borrows heavily from methods described in \cite{abe} and \cite{bandara}.

\begin{table}
\begin{center}
\def\arraystretch{2}
\cellspacebottomlimit=5pt
\cellspacetoplimit=5pt
\begin{tabular}{ l  c  c } 
	\textbf{Rule}&  \textbf{Expression} & \textbf{Number} \\ 
	Solid Phase Mass Conservation & $\displaystyle\frac{D^s \bar{\rho}_s}{Dt} + \bar{\rho}_s \divr \mathbi{v_s} = 0$ & \eqref{Eqn: solid_density} \\[5pt]
	Fluid Phase Mass Conservation &  $\displaystyle\frac{D^f \bar{\rho}_f}{Dt} + \bar{\rho}_f \divr \mathbi{v_f} = 0$ & \eqref{Eqn: fluid_density}\\[5pt]
	Fluid Phase True Density &  $\displaystyle\frac{n}{\rho_f}\frac{D^f \rho_f}{Dt} = -\divr \big((1-n)\mathbi{v_s} + n\mathbi{v_f}\big)$ & \eqref{Eqn: true_fluid_density}\\[5pt]
	Solid Phase Momentum Balance &  $\displaystyle\bar{\rho}_s\frac{D^s \mathbi{v_s}}{Dt} = \bar{\rho}_s\mathbi{g} - \mathbi{f_d} + \divr(\boldsymbol{\tilde{\sigma}}) - (1-n)\grad(p_f)$ & \eqref{Eqn: solid_closed_form}\\[5pt]
	Fluid Phase Momentum Balance & $\displaystyle\bar{\rho}_f\frac{D^f \mathbi{v_f}}{Dt} = \bar{\rho}_f \mathbi{g} + \mathbi{f_d} + \divr(\boldsymbol{\tau_f}) - n \grad(p_f)$ & \eqref{Eqn: fluid_closed_form}\\[5pt]
	Darcy's Drag Law & $\displaystyle\mathbi{f_d} = \frac{18\phi(1-\phi)\eta_0}{d^2} \ \hat{F}(\phi,\Reyn) \ (\mathbi{v_s-v_f})$ & \eqref{Eqn: beetstra_final} \\[5pt]
	Fluid Phase Pore Pressure & $\displaystyle p_f = \kappa \ln\bigg(\frac{\rho_f}{\rho_{0f}}\bigg)$ & \eqref{Eqn: fluid_phase_pore_pressure}\\[5pt]
	Fluid Phase Shear Stress & $\displaystyle \boldsymbol{\tau_f} = 2 \eta_0 \big(1 + \tfrac{5}{2}\phi\big) \mathsfbi{D_{0f}}$ & \eqref{Eqn: fluid_shear_stress}\\[5pt]
	Solid Phase Effective Stress & $\displaystyle \frac{D^s \boldsymbol{\tilde{\sigma}}}{Dt} = 2G\mathsfbi{D^e_0} + K \tr(\mathsfbi{D^e})\mathsfbi{1} + \mathsfbi{W_s}\boldsymbol{\tilde{\sigma}} - \boldsymbol{\tilde{\sigma}}\mathsfbi{W_s}$ & \eqref{Eqn: solid_phase_granular_stress_evolution}\\[5pt]
	Additive Flow Rate Decomposition & $\displaystyle \mathsfbi{D_s} = \mathsfbi{D^e} + \mathsfbi{\tilde{D}^p}$ & \eqref{Eqn: additive_decomposition}\\[5pt]
	Solid Phase Plastic Flow Rate & $\displaystyle \mathsfbi{\tilde{D}^p} = \frac{\dot{\bar{\gamma}}^p}{\sqrt{2}} \frac{\boldsymbol{\tilde{\sigma}_0}}{\|\boldsymbol{\tilde{\sigma}_0}\|} + \tfrac{1}{3}\big(\beta \dot{\bar{\gamma}}^p + \dot{\xi}_1 + \dot{\xi}_2\big)\mathsfbi{1}$ & \eqref{Eqn: plastic_flow_rule} \\[5pt]
	Dilation Angle & 
	$\beta = K_3\, (\phi-\phi_{eq})$ & \eqref{Eqn: beta}\\[5pt]
	Critical State Packing Fraction & $\displaystyle \phi_{eq} = \frac{\phi_m}{1+aI_m}$ & \eqref{Eqn: critical_state_packing}\\[5pt]
	Internal Friction Coefficient & $\displaystyle \mu_p = \mu_1 + \frac{\mu_2 - \mu_1}{1 + (b/I_m)} + \tfrac{5}{2} \bigg(\frac{\phi I_v}{aI_m}\bigg)$ & \eqref{Eqn: internal_friction_coefficient}\\[15pt]
	Granular Shear Flow Rule & $\displaystyle
	\begin{aligned}
	&f_1 = \bar{\tau} - \max\big((\mu_p + \beta)\tilde{p},\ 0\big)\\[1pt]
	&f_1 \leq 0, \qquad \dot{\bar{\gamma}}^p \geq 0, \qquad f_1 \dot{\bar{\gamma}}^p = 0
	\end{aligned}$ &\eqref{Eqn: f1_yield_surface}\\[15pt]
	Granular Separation Rule & $\displaystyle
	\begin{aligned}
	&f_2 = -\tilde{p}\\[1pt]
	&f_2 \leq 0, \qquad \dot{\xi}_1 \geq 0, \qquad f_2 \dot{\xi}_1 = 0
	\end{aligned}$ &\eqref{Eqn: f2_yield_surface}\\[15pt]
	Granular Compaction Rule & $\displaystyle
	\begin{aligned}
	&f_3 = g(\phi) \tilde{p} - (a\phi)^2 \big[ \zeta^2 d^2 \rho_s + 2 \eta_0 \zeta \big]\\[1pt]
	&f_3 \leq 0, \qquad \dot{\xi}_2 \leq 0, \qquad f_3 \dot{\xi}_2 = 0\\[3pt]
	&\zeta = \dot{\bar{\gamma}}^p - K_4 \dot{\xi}_2
	\end{aligned}$ &\eqref{Eqn: f3_yield_surface}\\[17pt]
	&$\displaystyle g(\phi) = \bigg\{\def\arraystretch{1}\begin{matrix}
	(\phi_m-\phi)^2 &\text{if} \quad \phi < \phi_m\\
	0 & \text{if} \quad \phi \geq \phi_m
	\end{matrix}$&\eqref{Eqn: f3_g_expression} \\
	\end{tabular}
	\end{center}
\caption{Summary of governing equations derived in section \ref{Sec: theory_and_formulation}.}\label{Tab: governing_equations}
\end{table}

\begin{figure}
	\centering
	\includegraphics[scale=0.4]{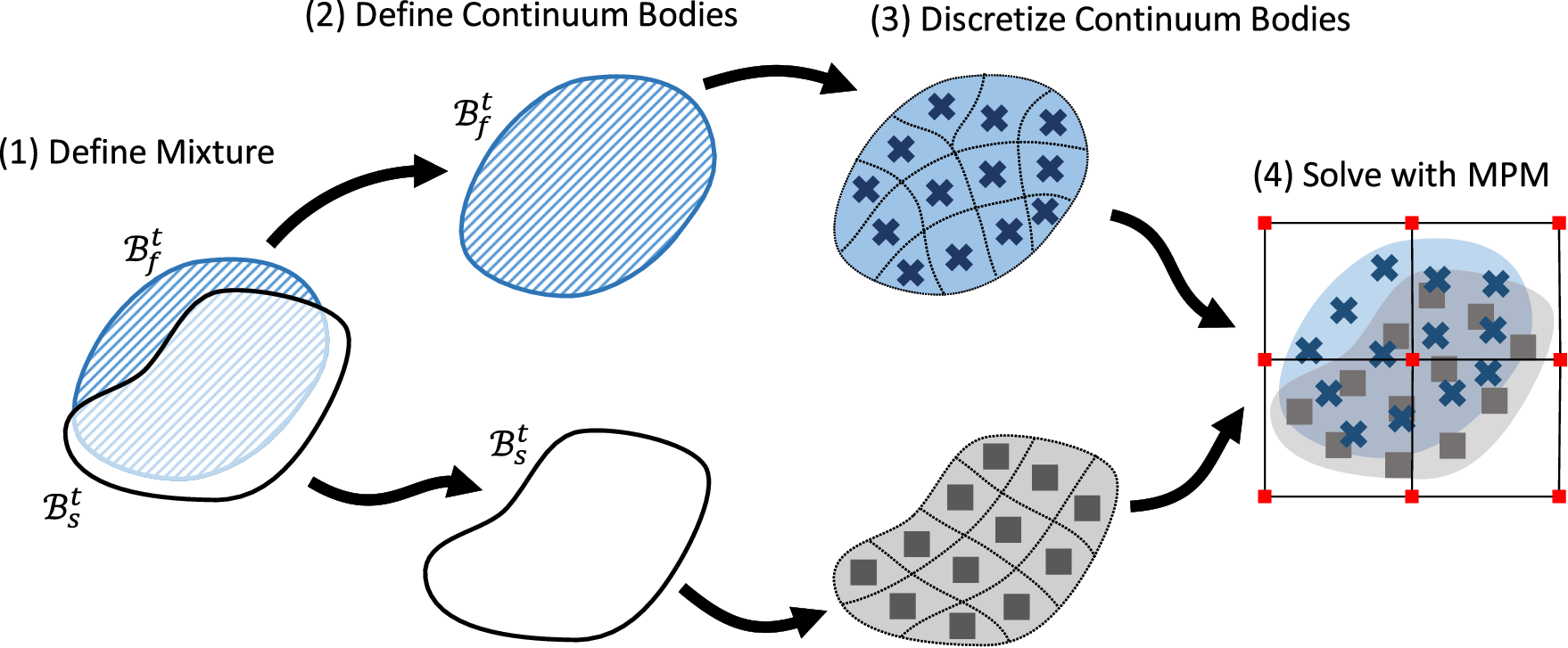}
	\caption{Solving mixture problems using the material point method. (1) Define the mixture and initial configuration including densities, porosities, stresses. (2) Define the solid and fluid phase continuum bodies. (3) Break the continuum bodies into piecewise-defined blocks of material represented by discrete material points. (4) Solve the equations of motion for the mixture on a background grid according to the material point method algorithm described in section \ref{Sec: time_marching}}
	\label{Fig: discretization}
\end{figure}

Figure \ref{Fig: discretization} shows the basic method we implement. First, the mixture problem is defined and the material configurations are given (figure \ref{Fig: discretization}(1)). The two phases are then separated into the continuum bodies described in figure \ref{Fig: configuration} (figure \ref{Fig: discretization}(2)). These continuum bodies are discretized into continuum `chunks' defined by two sets of Largrangian material point tracers. These tracers carry the full description of the continuum bodies (e.g. stress, density, velocity) and advect material information through space (figure \ref{Fig: discretization}(3)). These two sets of tracers are then placed into a simulation domain which is discretized into a background grid. The background grid is where the equation of motion are solved in the weak form (figure \ref{Fig: discretization}(4)).

Time integration of the mixture problem is achieved by using an explicit time-marching algorithm. During each discrete time-step, the mixture state (stored on the two sets of material points) is projected to the nodes which define the background Cartesian grid. A finite-element-like step is performed which solves the system of equations in table \ref{Tab: governing_equations} and updates the nodal representation of the mixture velocities and acceleration. These accelerations and velocities are then used to update the mixture state (as stored on the two sets of material points). At the end of the time-step, the grid is reset, and the procedure is repeated. In this way, we track the state of the mixture on a moving set of material point tracers and solve the equations of motion of a background finite-element-like grid. Specific details about our implementation of this framework, boundary conditions, and novel numerical corrections can be found in appendix \ref{appC}.

\section{Results}\label{Sec: results}
To validate our model, we use the numerical method described in section \ref{Sec: implementation} to simulate underwater column collapses and quasi-2D erosion flows for comparison with experimental data reported by \cite{rondon} and \cite{allen}. We also explore two applications of our method for potential use in impact/penetration problems (as explored in \cite{ceccato}) or for loaded slope failures (see summary of numerical work in this area by \cite{soga}).

\subsection{Numerical Validation of Model and Method}
In this section, we show that our model parameters can be fit to a particular class of fluid-sediment mixtures (in this case glass beads immersed in oil/water mixtures, see \cite{pailha}) and that these fit parameters can be used to accurately simulate an underwater column collapse (from \cite{rondon}) and quasi-2D erosion flows (from \cite{allen}).

\subsubsection{Model Fit to Glass Beads}\label{Sec: model_fit}
\begin{figure}
	\centering
	\includegraphics[scale=0.3]{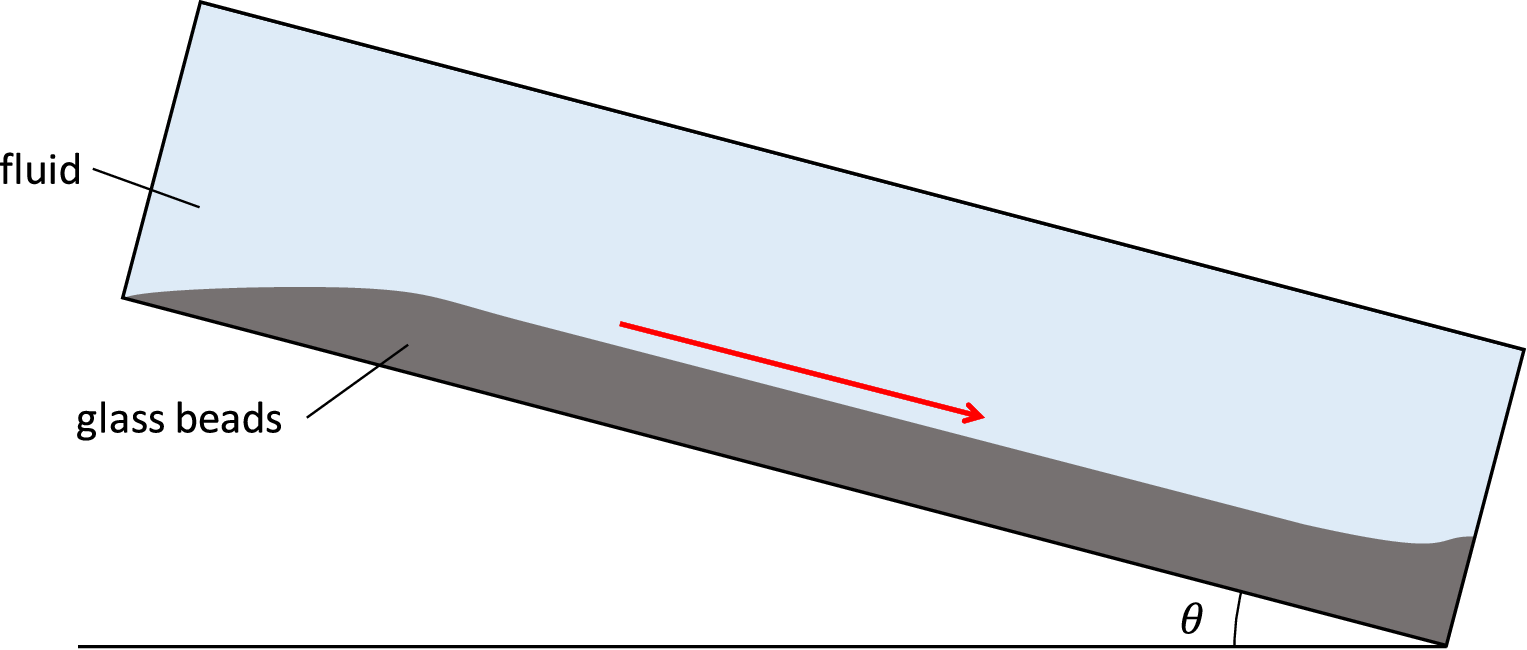}
	\caption{Experimental geometry used by \cite{pailha}. A bed of glass beads is immersed in a tank of viscous fluid. The incline of the base of the tank, $\theta$, is changed to induce submerged slope avalanches.}
	\label{Fig: avalanche_setup}
\end{figure}

\cite{pailha} characterize the behavior of glass beads flowing down a chute while immersed in a viscous fluid (setup shown in figure \ref{Fig: avalanche_setup}). The glass beads have density $\rho_s = 2500 \tfrac{\text{kg}}{\text{m}^3}$ and diameter $d = 160 \mu \text{m}$. Two mixtures of water/oil are reported and have viscosities $\eta_0 = 9.8 \times 10^{-3} \text{ Pa}\cdot\text{s}$ and $\eta_0 = 96 \times 10^{-3} \text{ Pa}\cdot\text{s}$.

\begin{figure}
	\centering
	\includegraphics[scale=0.4]{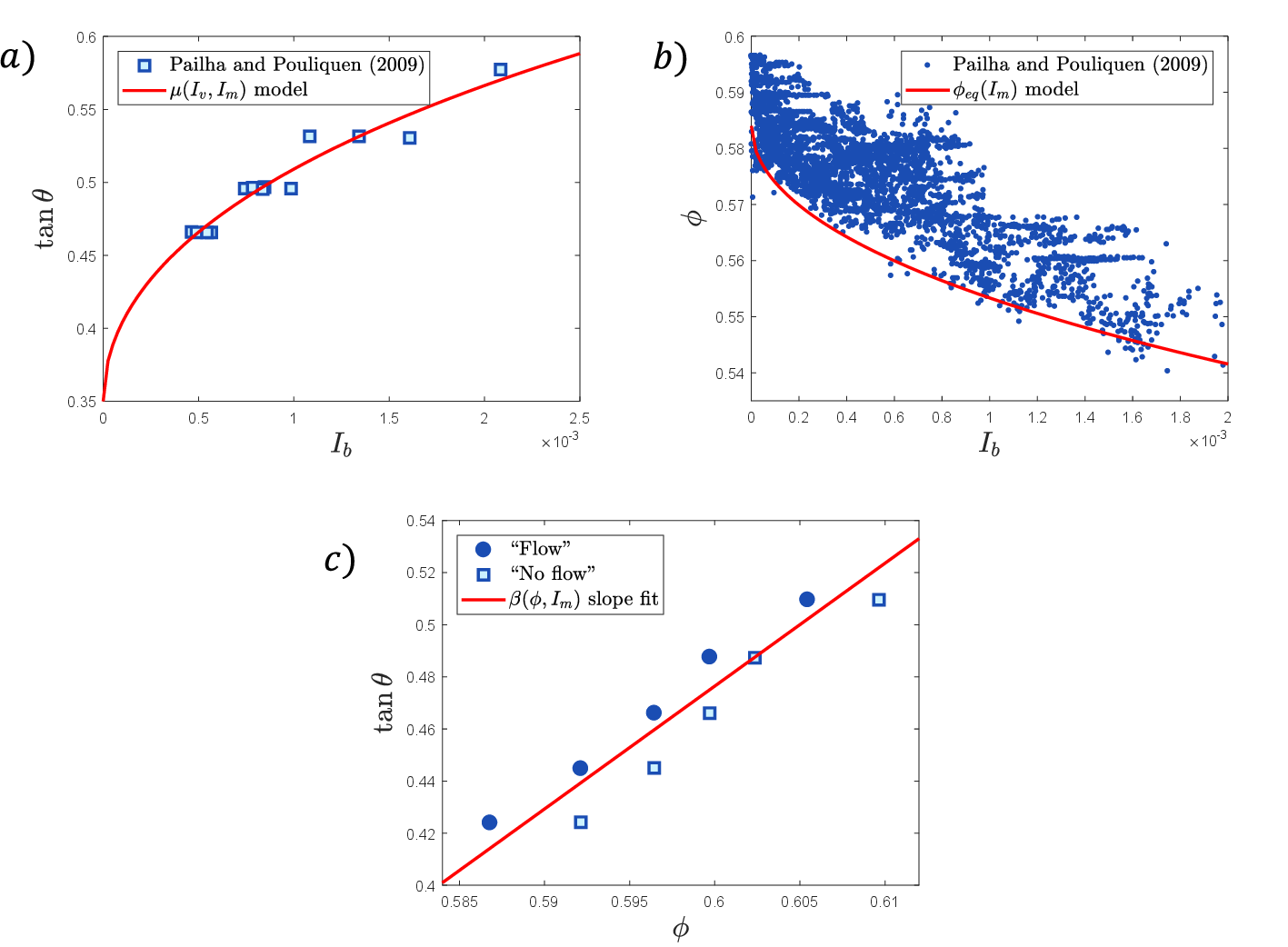}
	\caption{Model fit to the experimental data presented in figure 5 of \cite{pailha}. (a) Plot of internal friction coefficient against inertial number. (b) Critical state packing fraction fit to extreme measurements of $\phi$ at various flow rates. (c) $\beta$ slope coefficient $K_3$ and $K_4$ fit to the critical angle between flowing and static slopes.}
	\label{Fig: pailha_plots}
\end{figure}

In order to fit our model to the characterization of this mixture, we have focused on a subset of the reported data shown in figure \ref{Fig: pailha_plots}. Figure \ref{Fig: pailha_plots}b) shows the measured packing fraction of numerous flows/times plotted against the inertial number $I_b$ (defined in \cite{pailha}). We assume that the chute flow profile is parabolic (as is proven in \cite{cassar}) such that $I_b \approx I_v$. We further assume that all of the reported flows are in the low Stokes limit $(\St \to 0)$ such that $I_m \approx \sqrt{2I_b}$. Fitting \eqref{Eqn: critical_state_packing} to the lower extrema of the data, we find the following material parameters,
$$\phi_m = 0.584, \quad a = 1.23$$

Figure \ref{Fig: pailha_plots}a) shows the measured internal friction angle $\tan(\theta)$ plotted against the experimental inertial number $I_b$. Assuming that all measurements were taken when the flows had reached steady state, $\tan(\theta) \approx \mu$ with $\mu$ given in \eqref{Eqn: steady_shear_boyer_mu}. Fitting this equation to the data, we find the following material parameters,
$$\mu_1 = 0.35,\quad \mu_2 = 1.387,\quad b = 0.3085$$

Figure \ref{Fig: pailha_plots}c) shows a set of flow onset measurements. At the transition from the ``No flow" state to the ``Flow" state, $\tan(\theta) \approx \mu_1 + \beta$. We assume that near the onset of flow $I_m = 0$.
Therefore, the slope of the transition line between flowing and non-flowing behavior will be given by $K_3$, and since the rate of compaction in these flows is small (the $K_4$ term), we let,
$$
K_3 = 4.715, \quad K_4 = 0
$$

With the parameters above determined for glass beads, we can now simulate other experiments which use similar mixtures. The remaining parameters ($\rho_s$, $\rho_{0f}$, $\eta_0$, and $d$) are determined by the specific materials used in the relevant experiments.

\subsubsection{Granular Column Collapse of Glass Beads}
\begin{figure}
	\centering
	\includegraphics[scale=0.3]{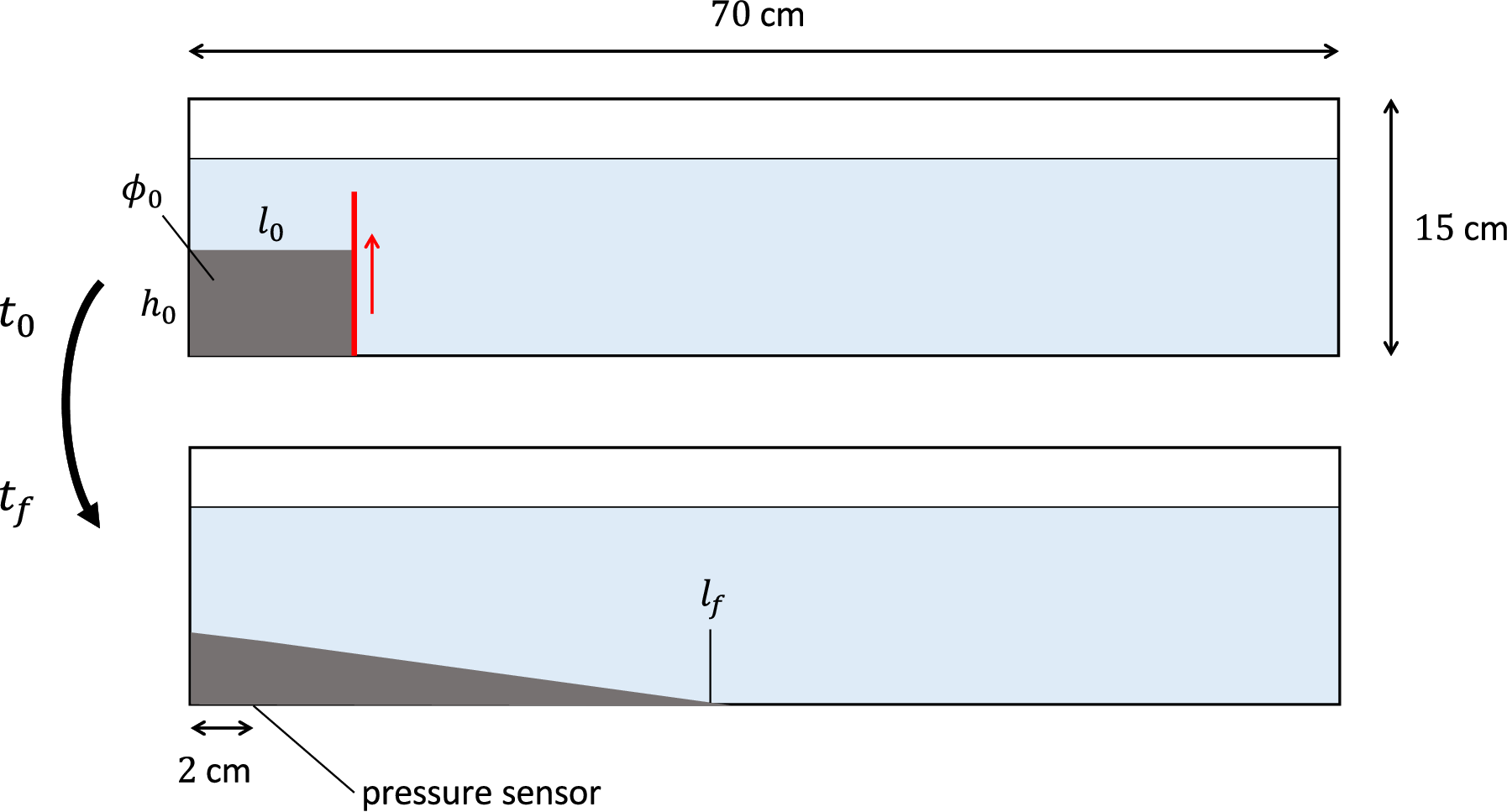}
	\caption{The experimental setup used by \cite{rondon}. A column of small glass spheres with initial packing fraction $\phi_0$ is held in place by a retaining wall and immersed in a long tank filled with a viscous fluid. At time $t_0$, the wall is removed and the column is allowed to collapse. A pressure sensor at the base of the column (2 cm from the edge of the tank) collects pore pressure data during the collapse. The run-out profiles of the column are captured with a camera.}
	\label{Fig: column_collapse_setup}
\end{figure}

\cite{rondon} explore the behavior of collapsing granular columns submerged in a fluid with viscosity $\eta_0 = 12 \text{ or } 23$ cP and density $\rho_{0f} \approx 1000 \tfrac{\text{kg}}{\text{m}^3}$. A bed of glass beads with diameter $d = 225 \mu\text{m}$ and density $\rho_s = 2500 \tfrac{\text{kg}}{\text{m}^3}$ was held at some initial packing fraction behind a retaining wall (see figure \ref{Fig: column_collapse_setup}). Once the wall was removed, the dynamics of the column were measured and reported.

In this work, we are interested in the behavior of two of the columns reported in that work. The two columns are made of the same mass of glass beads and formed into a \textit{loose} column and a \textit{dense} column. The \textit{loose} column has initial height $h_0 = 4.8$ cm, initial width $l_0 = 6.0$ cm, and initial packing fraction $\phi_0 = 0.55$. The \textit{dense} column has initial height $h_0 = 4.2$ cm, initial width $l_0 = 6.0$ cm, and initial packing fraction $\phi_0 = 0.60$. Both columns are immersed in a fluid tank measuring 70 cm $\times$ 15 cm $\times$ 15 cm. It was observed that the initially \textit{loose} column collapsed much faster with much longer run-out than the initially \textit{dense} column. 

To simulate these two column collapses, we consider a reduced computational domain by assuming that the flow is approximately plane-strain (quasi-2D) and that the fluid tank can be shortened to 30 cm in length and 10 cm in height without significantly affecting the dynamics of the column collapse (we let the fluid partially fill the tank to a height of 8 cm). We then run our model with the same initial conditions as described in \cite{rondon}, computational parameters given in table \ref{Tab: pailha_simulation}, and remaining material parameters given in section \ref{Sec: model_fit}. The fluid-wall interaction is governed by a simple frictionless boundary condition while the grain-wall interaction is governed by the frictional boundary rule described in section \ref{Sec: boundary_friction}.

\begin{table}
	\begin{center}
		\setlength{\tabcolsep}{1em} 
		\begin{tabular}{ l  l  l } 
			Parameter & 300$\times$100 Simulations & 120$\times$40 Simulations \\ 
			Elements & 300 $\times$ 100 & 120 $\times$ 40\\
			Points per Cell & 4 & 4\\
			$\Delta t$ & $5\cdot 10^{-5}$ s & $2\cdot 10^{-5}$ s\\ 
			$\Delta x$ & 1.0 mm & 2.5 mm\\
			$t_0$ & 0s & 0s\\
			$t_f$ & 20 s & 60 s\\
			$G$ & $3.8 \cdot 10^{4}$ Pa & $3.8 \cdot 10^{5}$ Pa \\
			$K$ & $8.3 \cdot 10^{4}$ Pa & $8.3 \cdot 10^{5}$ Pa \\
			$\eta_0$ & $1.2 \cdot 10^{-2}$ Pa$\cdot$s & $1.2 \cdot 10^{-2}$ Pa$\cdot$s \\
			$\kappa$ & $1.0 \cdot 10^{5}$ Pa & $1.0 \cdot 10^6$ Pa\\
		\end{tabular}
	\end{center}
\caption{Simulation parameters for column collapses run on different grids.}\label{Tab: pailha_simulation}
\end{table}

\begin{figure}
	\centering
	\includegraphics[scale=0.33]{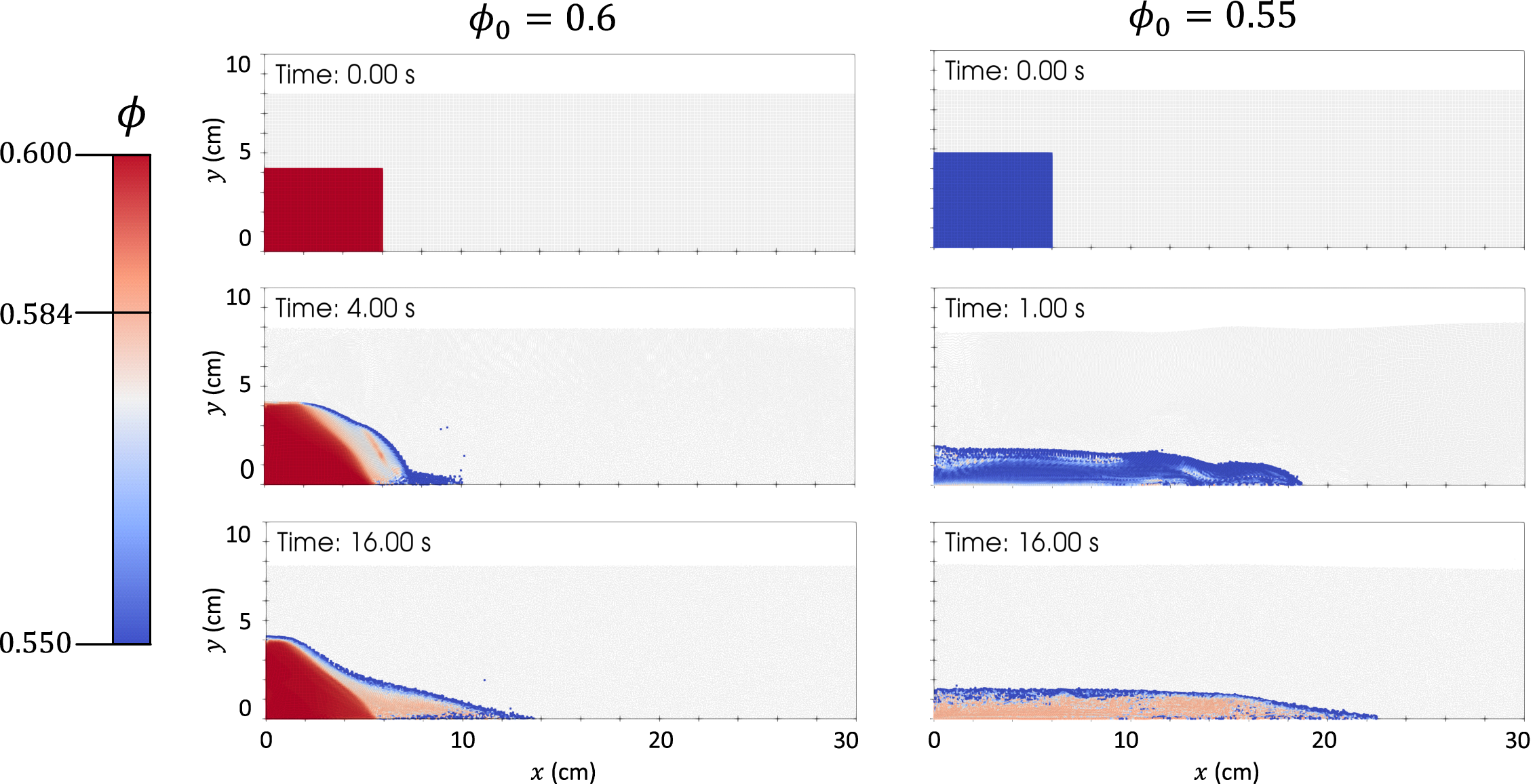}
	\caption{Comparison between simulated collapses for the \textit{loose} initial packing (right) and the \textit{dense} initial packing (left) using the 300$\times$100 element grid described in table \ref{Tab: pailha_simulation}. Solid phase material points are colored by packing fraction according to the scale at the left. Fluid phase material points are colored light gray.}
	\label{Fig: column_collapse_time_series}
\end{figure}

In both the experiments and simulations, the only differences between the \textit{dense} and \textit{loose} columns are the initial packing fraction, the initial column height, and the initial hydrostatic stress state. The resulting differences in the simulated flow dynamics are due to the different solutions picked out by the governing equations given these initial conditions. A series of snapshots taken from these two simulations (as run on the 300$\times$100 grid) are shown in figure \ref{Fig: column_collapse_time_series}.

\begin{figure}
	\centering
	\includegraphics[scale=0.31]{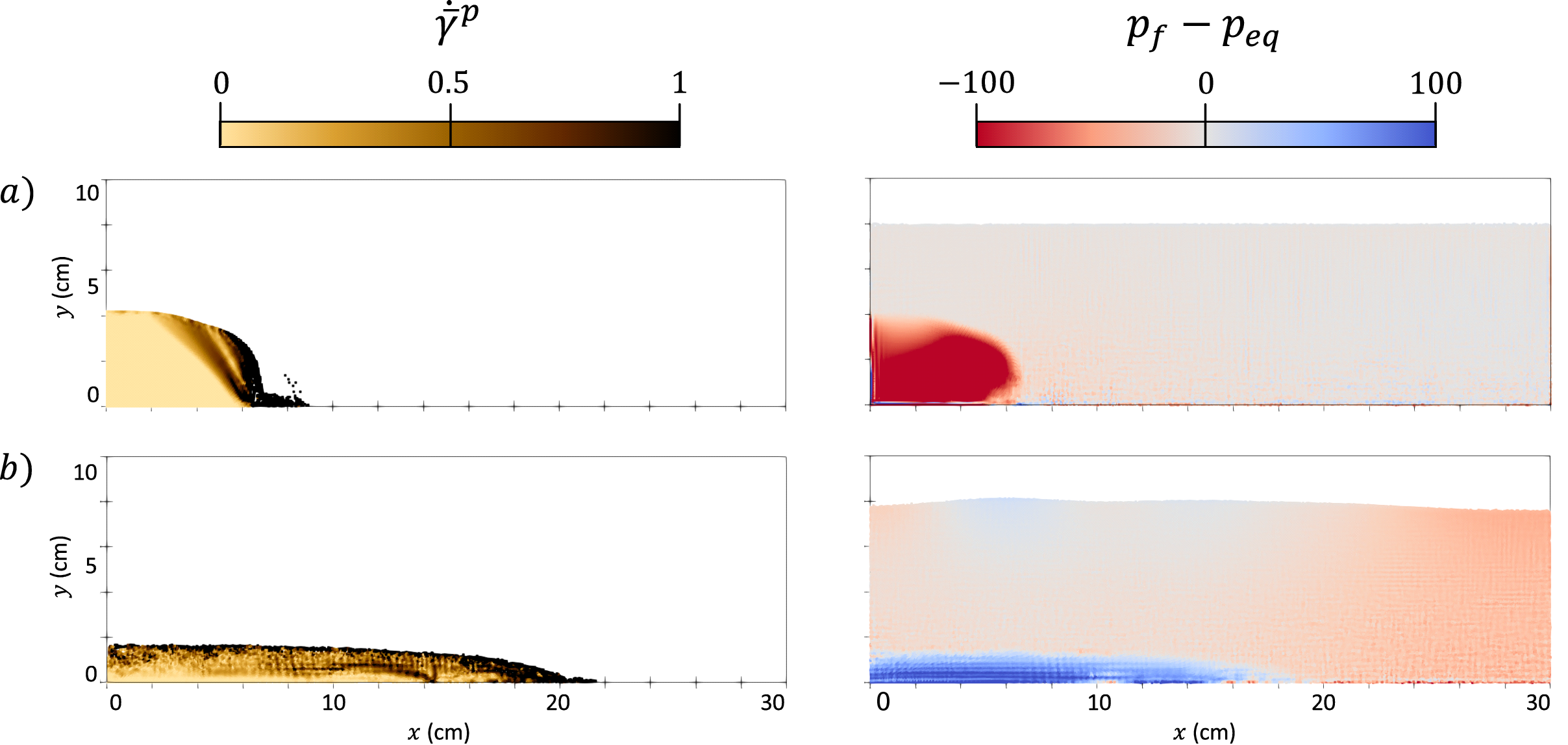}
	\caption{Snapshot of simulated solid phase equivalent plastic shear rate (left) and fluid phase excess pore pressure (right) at $t=4$s for (a) the initially \textit{dense} column and (b) the initially \textit{loose} column. The plastic shearing rate is visualized at the material point centroids of the solid phase. The post-processed excess pore pressure (see \cite{dunatunga}) as compared to a hydrostatic baseline $p_{eq}$ is visualized at the fluid phase centroids.}
	\label{Fig: column_snapshot}
\end{figure}

In addition to visualizing the solid phase dilation and compaction as in figure \ref{Fig: column_collapse_time_series}, we can also examine the differences in shearing rate and fluid pore pressure as shown in figure \ref{Fig: column_snapshot}. As the initially \textit{dense} column collapses, the solid phase experiences \textit{shear dilation}, increasing the porosity of the mixture. This results in pore tension in the fluid phase as fluid is drawn into the increased pore space (see figure \ref{Fig: column_snapshot}a). This increased pore tension within the collapsing column (as compared to the surrounding fluid) increases the effective granular pressure given by $\tilde{p}$ in \eqref{Eqn: tau_and_p} and therefore strengthens the solid phase resulting in a slower collapse process. On the other hand, as the initially \textit{loose} column collapses the solid phase experiences \textit{plastic compaction}, reducing the porosity of the mixture. This has the opposite effect, causing an excess positive pore pressure (see figure \ref{Fig: column_snapshot}b) which reduces the strength of the solid phase. It is this coupling of solid phase flow to fluid phase pressure to solid phase strength that results in these two completely different collapse behaviors.

\begin{figure}
	\centering
	\includegraphics[scale=0.45]{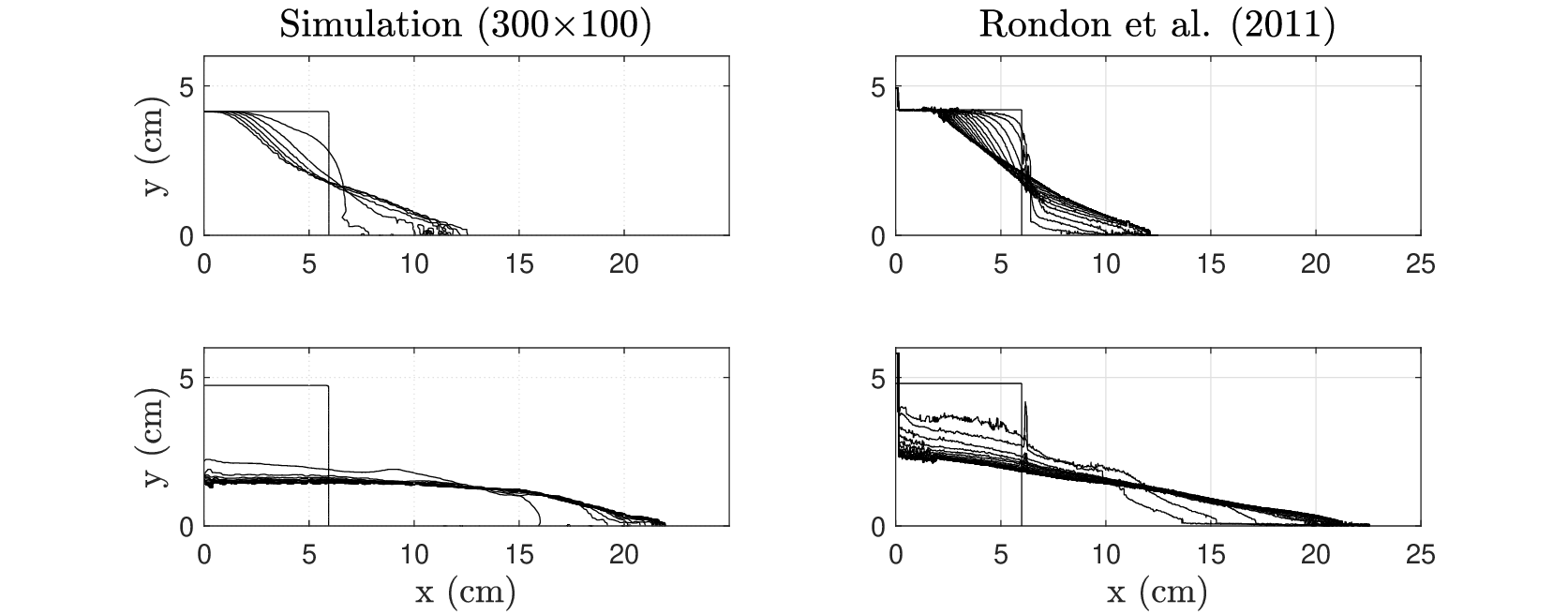}
	\caption{Contours of the collapsing columns from the \textit{dense} simulation (top, left) taken at 3s intervals and the \textit{loose} simulation (bottom, left) taken at 0.66s intervals. The corresponding contours for the \textit{dense} experiment (top, right) and \textit{loose} experiment (bottom, right) from \cite{rondon} are also shown. The simulated profiles are generated by plotting the contour of the nodal porosity field (given by the coefficients $\{n_i\}$) at $n = 0.45$.}
	\label{Fig: matlab_profiles}
\end{figure}

\begin{figure}
	\centering
	\includegraphics[scale=0.35]{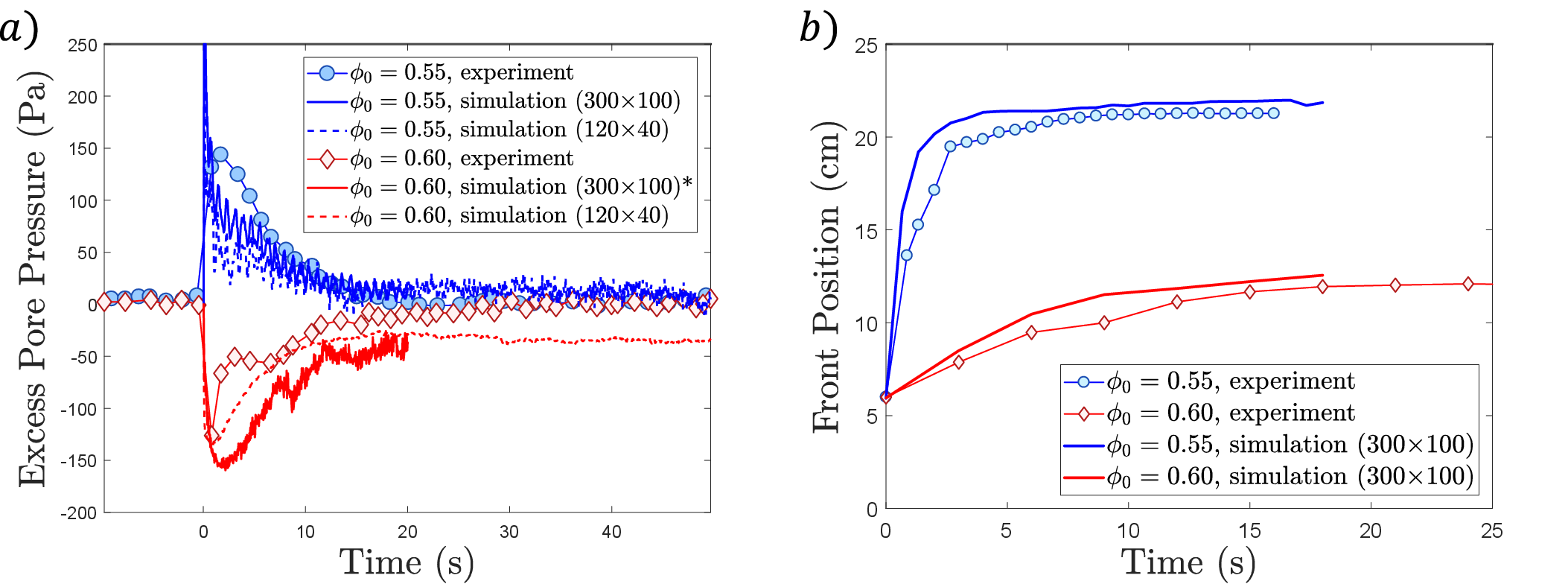}
	\caption{(a) Comparison of the simulated excess pore pressure for the \textit{loose} initial packing (top, blue) and the \textit{dense} initial packing (bottom, red). The base pore pressure for all simulations is approximately 800 Pa. (b) Comparison between simulated front positions for the \textit{loose} initial packing (top, blue) and the \textit{dense} initial packing (bottom, red).}
	\label{Fig: rondon_plots}
\end{figure}

By accurately modeling these complex interactions, we are able to capture the vastly different collapse profiles (see figure \ref{Fig: matlab_profiles}), predict the measured excess pore pressure (see figure \ref{Fig: rondon_plots}a), and match the time-accurate front motion (see figure \ref{Fig: rondon_plots}b) reported in \cite{rondon}. The collapse profiles shown in figure \ref{Fig: matlab_profiles} are the $n=0.45$ contours of the nodal porosity field $n(\mathbf{x})$ and show reasonable similarity to the experimental profiles, though there are some artifacts of the finite grid spacing visible near the front of the collapsing column.

The pore pressure in figure \ref{Fig: rondon_plots}(b) shows the weighted average nodal representation of pressure (as defined in \cite{dunatunga}) near (but not at) the lower domain boundary and 2 cm from the left wall. This value is compared to a hydrostatic reference value to find the excess pore pressure. In the \textit{dense} 300$\times$100 simulation, the fluid phase material points exhibited excessive clumping (see section \ref{Sec: delta_position}), so a second nodal sample was taken at the same height, 2 cm from the right wall and used as the reference value. The front positions shown in figure \ref{Fig: rondon_plots}(b) are determined by taking the maximum $x$-position of the collapse profiles shown in figure \ref{Fig: matlab_profiles}. 

The time-history of the simulated pore pressures in figure \ref{Fig: rondon_plots} show close agreement to the experimental measurements; however, the \textit{dense} simulations appear to saturate at a negative excess pore pressure. This discrepancy is likely due to the high frequency error observed in the MPM stress field before nodal averaging (see \cite{dunatunga}). \cite{mast2012} propose several methods of mitigating these errors and the associated kinematic locking, but we do not implement them here. All together, the results shown in figures \ref{Fig: matlab_profiles} and \ref{Fig: rondon_plots} indicate that our model is capable of accurately predicting the dynamics of submerged granular column collapses and captures the sensitivity of the problem to small changes in initial conditions.

\subsubsection{Quasi-2D Flow of Glass Beads}
\begin{figure}
	\centering
	\includegraphics[scale=0.3]{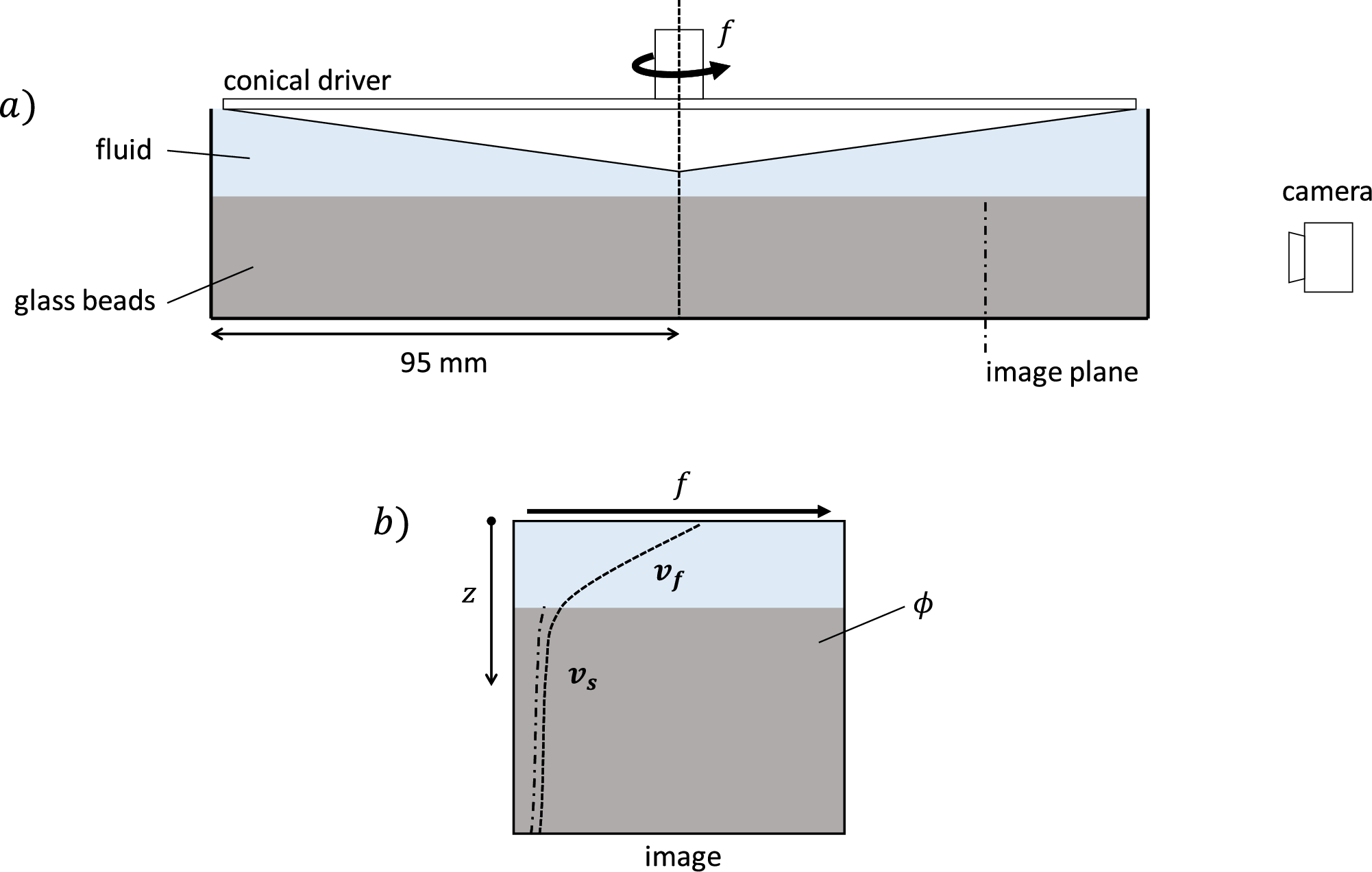}
	\caption{(a) Experimental setup of \cite{allen}. An approximately 9 mm bed of grains is immersed in a cylindrical tank filled with fluid. A conical driver is submerged to the granular surface and driven by a motor at a specified rotation rate $f$. (b) The resulting flow is imaged at a plane near the edge of the tank. Measurements are taken of phase velocities and packing fractions as a function of distance $z$ from the driving surface.}
	\label{Fig: allen_setup}
\end{figure}

In addition to sudden collapses of granular columns, we are also interested in using our model to simulate steady erosion processes. To gage the accuracy of our method for such problems, we simulate the experiments performed by \cite{allen}. As shown in figure \ref{Fig: allen_setup}, the experimental setup approximates a 2D erosion flow by driving a conical motor at a prescribed rotation rate, $f$, above an immersed granular bed of glass beads. The fields reported, obtained using index-matching, are a function of vertical depth below the driving surface, $z$.

The mixture of fluid and grains used in \cite{allen} is similar to that used in \cite{pailha}, suggesting that we can use the same material parameters determined in Section \ref{Sec: model_fit}. The remaining material parameters are given by the specific materials used in the experiment: $\rho_{0f} = 1002 \tfrac{\text{kg}}{\text{m}^3}$, $\eta_0 = 0.021$ Pa$\cdot$s, $\rho_s = 2500 \tfrac{\text{kg}}{\text{m}^3}$, and $d = 1.05$ mm.

We simulate four of the reported flows in that work, $f/f_c = \{0.37,1.04,1.26,1.33\}$, where $f$ is the assigned driving frequency and $f_c$ is the reported critical driving frequency around which grains become suspended in the fluid flow. We set up an $x$-periodic domain measuring 15.5 mm by 15.5 mm and drive the upper surface at a velocity determined by the ratios the driving frequency $f$. We let the lower wall be governed by a no-slip boundary condition. The resulting fluid flow is allowed to reach steady-state and the flow properties are averaged over a 12 s time window. A summary of the simulation setup is given in table \ref{Tab: allen_simulation}.

\begin{table}
	\begin{center}
		\setlength{\tabcolsep}{1em} 
		\begin{tabular}{  l  l  l  l  l  } 
			Parameter & $f/f_c = 0.37$ & $f/f_c = 1.04$ & $f/f_c = 1.26$ & $f/f_c = 1.33$ \\ 
			Bed Height & 10.5 mm & 11.0 mm & 11.0 mm & 11.4 mm\\
			Driving Velocity & $0.2325 \tfrac{\text{m}}{\text{s}}$ & $0.6536 \tfrac{\text{m}}{\text{s}}$ & $0.7919 \tfrac{\text{m}}{\text{s}}$ & $0.8359 \tfrac{\text{m}}{\text{s}}$\\
			Elements & 20 $\times$ 20 & 20 $\times$ 20 & 20 $\times$ 20 & 20 $\times$ 20\\
			Points per Cell & 9 & 9 & 9 & 9\\
			$\Delta t$ & $2\cdot 10^{-5}$ s & $2\cdot 10^{-5}$ s & $2\cdot 10^{-5}$ s& $2\cdot 10^{-5}$ s\\ 
			$\Delta x$ & 775 $\mu$m & 775 $\mu$m & 775 $\mu$m & 775 $\mu$m \\
			$t_0$ & 0s & 0s & 0s & 0s\\
			$t_f$ & 30 s & 30 s & 30s & 30s\\
			$G$ & $3.8 \cdot 10^{4}$ Pa & $3.8 \cdot 10^{4}$ Pa & $3.8 \cdot 10^{4}$ Pa & $3.8 \cdot 10^{4}$ Pa \\
			$K$ & $8.3 \cdot 10^{4}$ Pa & $8.3 \cdot 10^{4}$ Pa & $8.3 \cdot 10^{4}$ Pa & $8.3 \cdot 10^{4}$ Pa \\
			$\eta_0$ & $2.1 \cdot 10^{-2}$ Pa$\cdot$s & $2.1 \cdot 10^{-2}$ Pa$\cdot$s & $2.1 \cdot 10^{-2}$ Pa$\cdot$s & $2.1 \cdot 10^{-2}$ Pa$\cdot$s\\
			$\kappa$ & $1.0 \cdot 10^{5}$ Pa & $1.0 \cdot 10^{5}$ Pa & $1.0 \cdot 10^{5}$ Pa & $1.0 \cdot 10^{5}$ Pa\\
			$\phi_0$ & 0.585 & 0.585 & 0.585 & 0.585\\
		\end{tabular}
	\end{center}
\caption{Simulation parameters for four erosion flows run at different driving velocities.}\label{Tab: allen_simulation}
\end{table}

\begin{figure}
	\centering
	\includegraphics[scale=0.32]{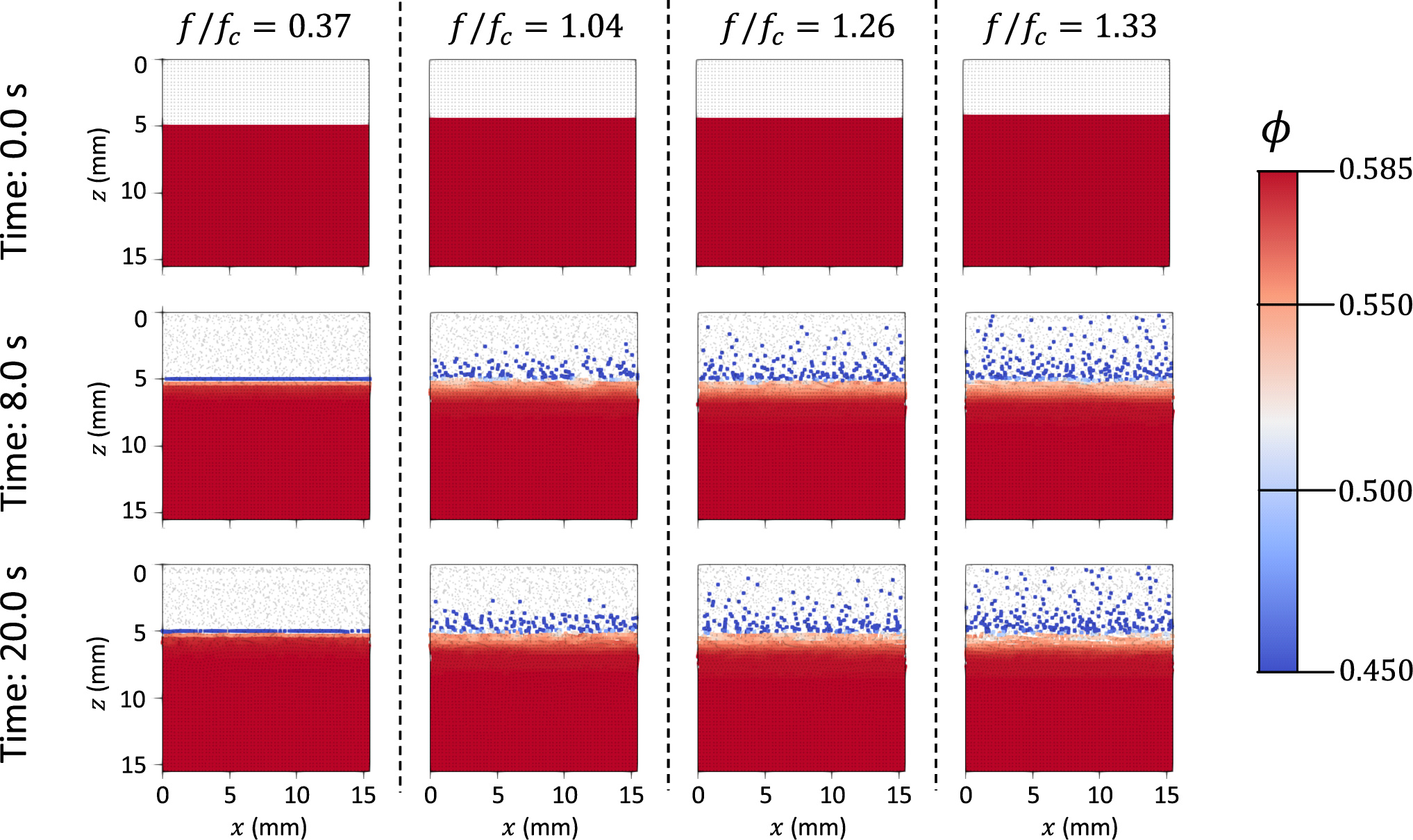}
	\caption{Comparison between the simulated erosion flows described in table \ref{Tab: allen_simulation}. Solid phase material points are colored by packing fraction according to the scale at the right. Fluid phase material points are colored light gray. In all cases, the shearing of the fluid phase induces motion in the solid phase. As the driving frequency $f$ increases above the critical $f_c$ (as reported in \cite{allen}) solid phase material is `picked up' and becomes suspended in the fluid.}
	\label{Fig: allen_time_series}
\end{figure}

\begin{figure}
	\centering
	\includegraphics[scale=0.4]{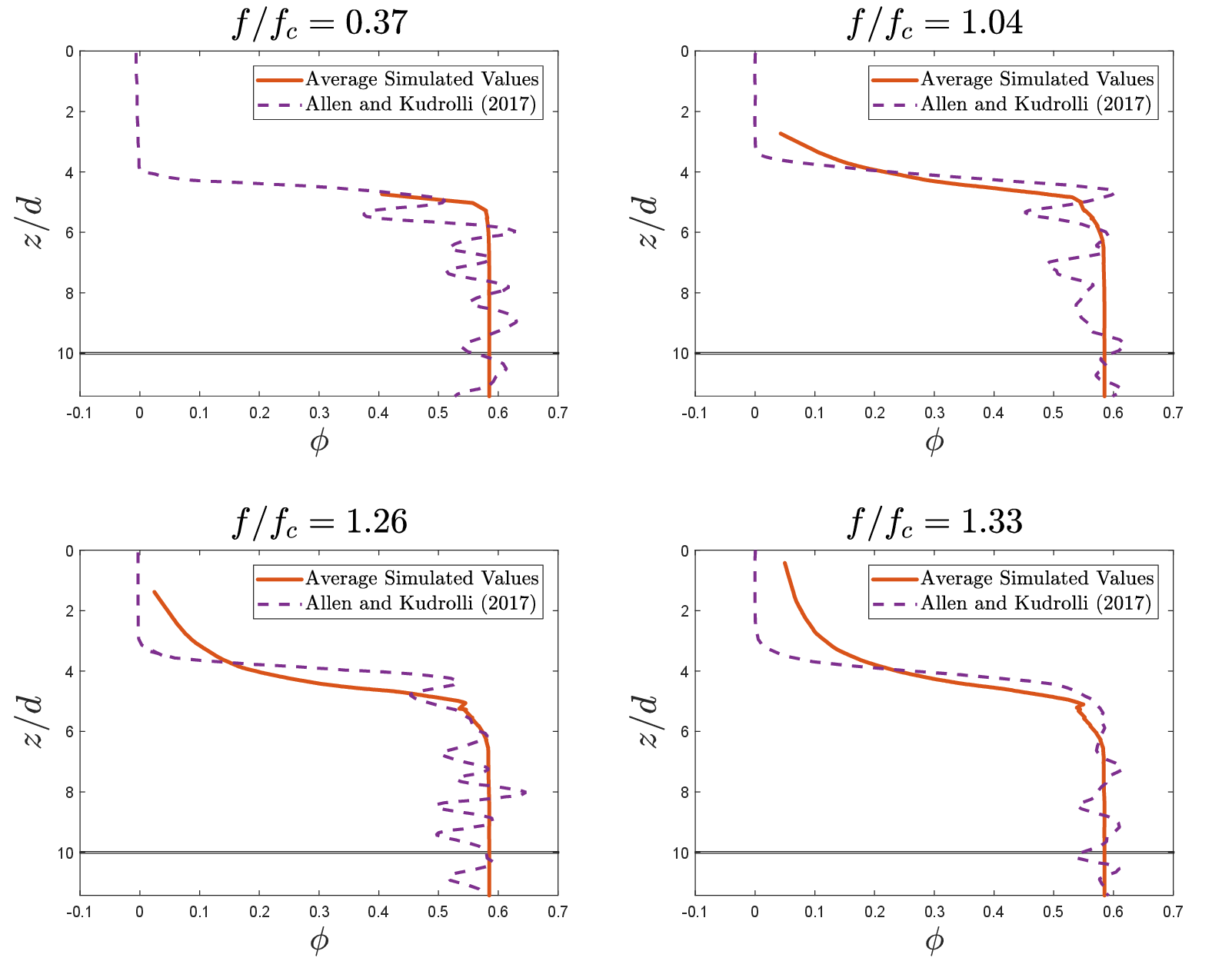}
	\caption{Plots comparing the time-averaged steady state packing fractions as a function of normalized depth reported in \cite{allen} to those found by running the simulations described in table \ref{Tab: allen_simulation}. Very close matching is observed when the solid phase material is \textit{dense}; however the simulated data has a heavy tail in the \textit{dilute} regime. This is likely due to the large empty spaces between the solid phase material points when they become suspended in the fluid flow.}
	\label{Fig: allen_packing}
\end{figure}

\begin{figure}
	\centering
	\includegraphics[scale=0.4]{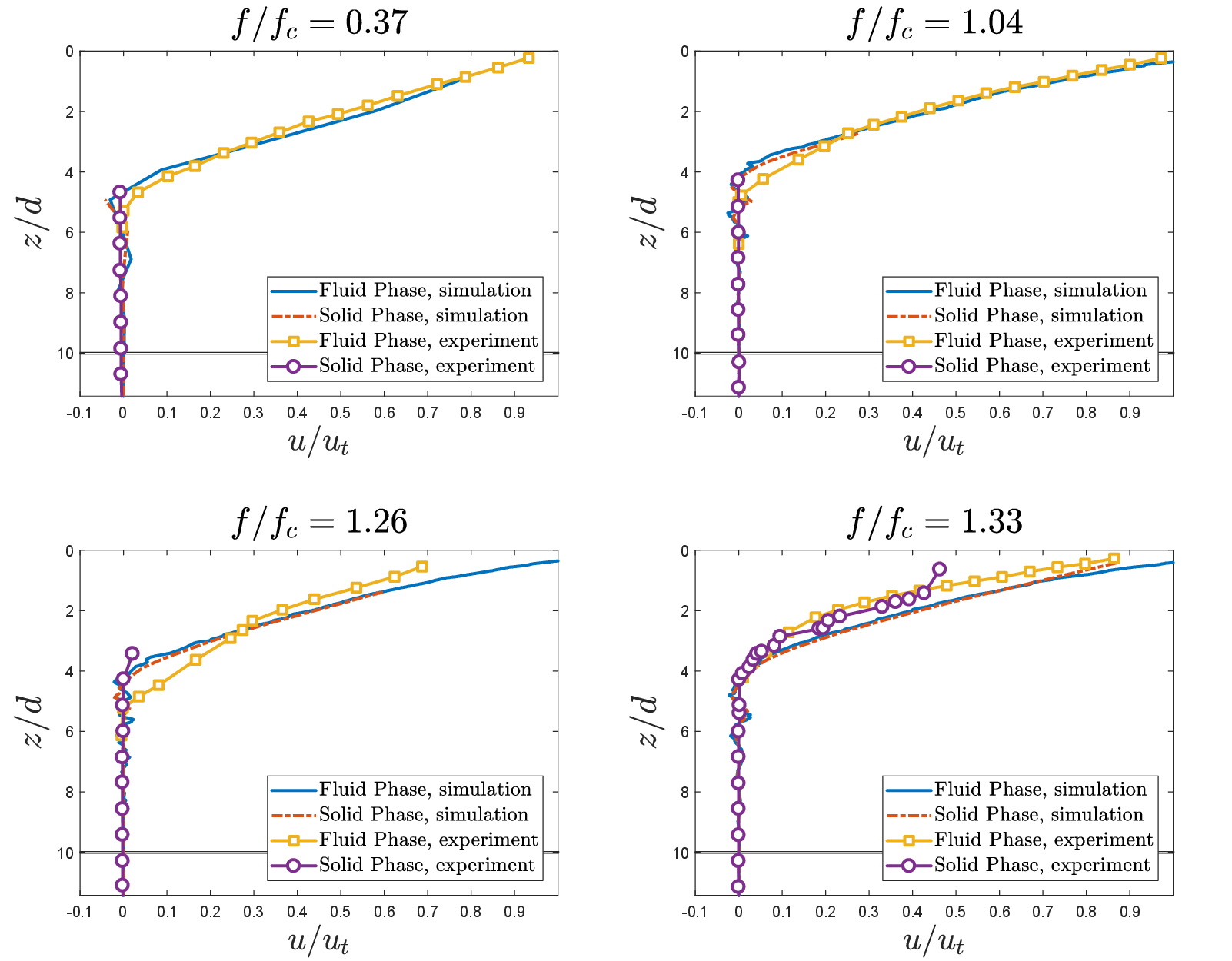}
	\caption{Plots comparing the time-averaged steady state phase velocities $u$ (normalized by the velocity of the driving surface $u_t$) as a function of normalized depth reported in \cite{allen} to those found by running the simulations described in table \ref{Tab: allen_simulation}. The simulated values show strong similarity to the experimental values; however, there are oscillations visible in the simulated profiles. These oscillations are due to well known errors in the material point velocity fields.}
	\label{Fig: allen_velocity}
\end{figure}

A series of simulation snapshots is shown in figure \ref{Fig: allen_time_series}. 
As was observed in \cite{allen}, below the critical driving frequency $f_c$ there is essentially no flow of grains; however, once the driving frequency $f$ is increased above $f_c$, solid phase material is `picked up' by the shearing of the fluid phase and enters into suspension. The steady-state flow predicted by our simulations shows strong similarity to the experimentally measured packing fraction (see figure \ref{Fig: allen_packing}) and phase velocities (see figure \ref{Fig: allen_velocity}).

The simulated packing fractions and velocities are plotted by averaging the material point coefficients 
over a 12s window. The resulting phase velocity and packing fraction averages are then sorted by the average material point centroid position and filtered using the MATLAB \texttt{smooth()} function. It is important to note that as the solid phase dilates, the solid phase material points will separate. After the material points separate by more than 1 element (around $\phi\approx0.2$), the material point value $\phi_p$ will no longer be representative of the true mixture packing fraction.

\subsection{Qualitative Results}
In this section we consider two potential applications of our model and method. The first shows the behavior of a fluid-grain mixture as an intruding body is pressed into its surface. The second shows the effect of water level on the failure of a loaded slope. 

\subsubsection{2D Circular Intruder}\label{Sec: intruder}
The use of the material point method for intrusion into a saturated soil is explored at length in \cite{ceccato}. In that work, the mixture model developed in \cite{bandara} is adjusted to use the Modified Cam Clay model to model the solid phase behavior.

\begin{figure}
	\centering
	\includegraphics[scale=0.3]{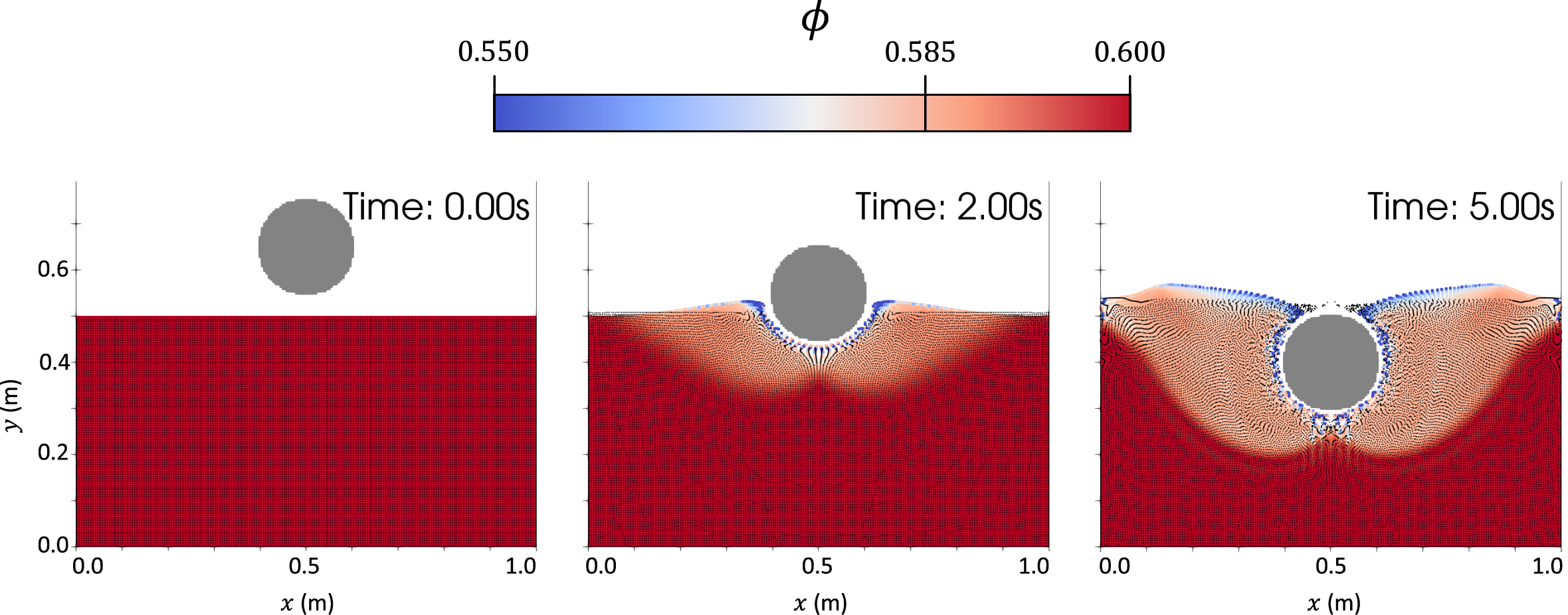}
	\caption{Series of snapshots taken from simulation described in section \ref{Sec: intruder}. Solid phase material points are colored according to packing fraction. Fluid material points are represented by small black dots. Intruder material points are colored light gray. As the intruder enters the mixture, the shearing of the solid phase results in noticeable dilation.}
	\label{Fig: intruder}
\end{figure}

Here we show that our model may be extended to explore similar problems by simulating the intrusion of a disk into a submerged bed of acrylic beads. As an exploratory problem, we use the material parameters given in table \ref{Tab: boyer_fit} and let $d = 1.0$ cm, $\rho_s = 2500 \tfrac{\text{kg}}{\text{m}^3}$, and $\rho_f = 1000 \tfrac{\text{kg}}{\text{m}^3}$. A 1m$\times$1m domain is simulated on a 100$\times$100 element grid with 4 material points per cell. The domain is initially half-filled by a mixture of fluid and grains with packing fraction $\phi_0 = 0.60$. The resulting behavior is shown in figure \ref{Fig: intruder}. 
As the intruder enters the mixture, we observe shear dilation of the granular material and independent motion of the two phases of material as fluid fills in the opening pore space under the intruder, revealing dry granular media at the free surface.

\subsubsection{2D Slope Collapse}\label{Sec: slope}
\begin{figure}
	\centering
	\includegraphics[scale=0.35]{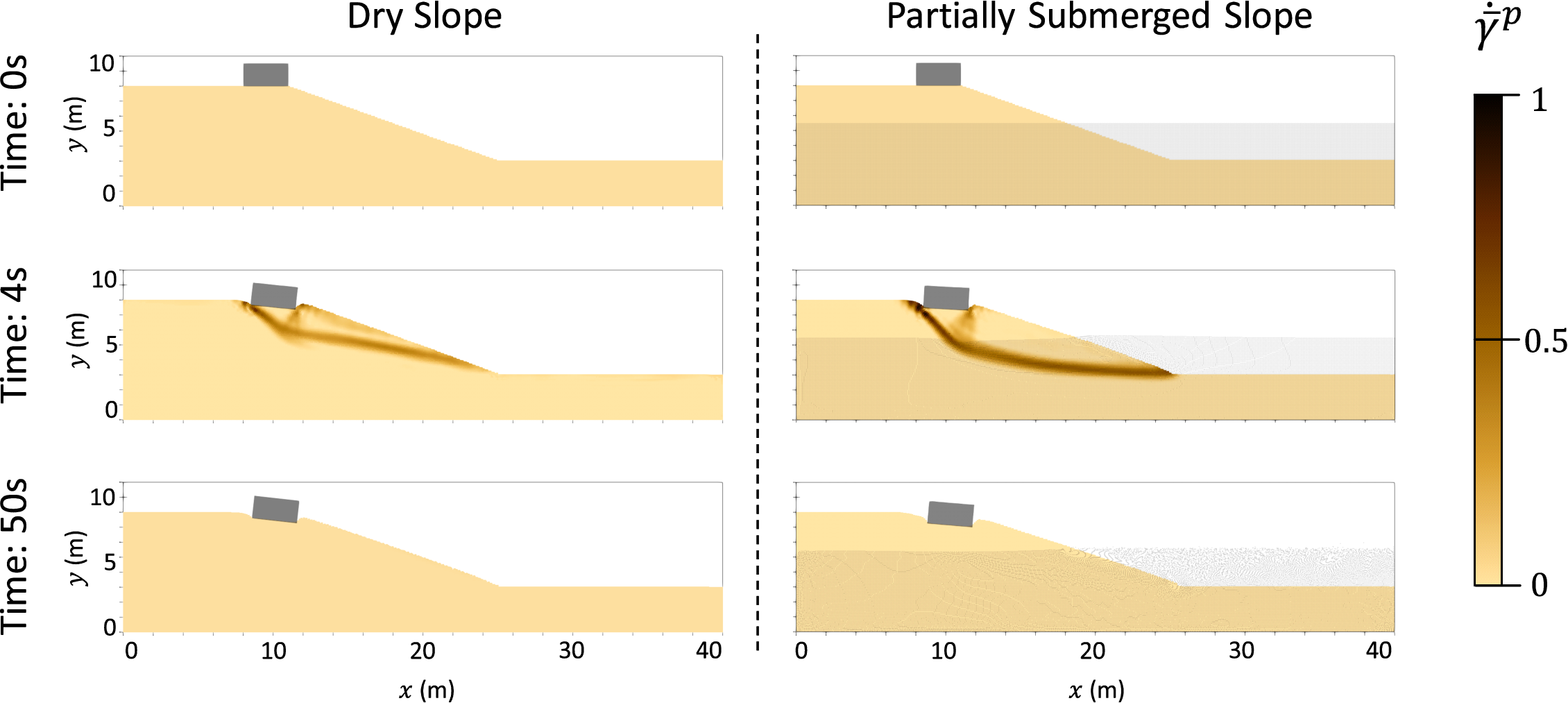}
	\caption{Series of snapshots taken from simulations described in section \ref{Sec: slope}. Solid phase material points are colored according to the \textit{equivalent plastic shearing rate}, $\dot{\bar{\gamma}}^p$. Fluid material points are represented by light gray dots. Block material points are colored light gray.}
	\label{Fig: slope_time_series}
\end{figure}

Another application of interest for our model is the complex interactions between structures and saturated soils. To demonstrate the application of this model to the problem of a loaded slope, we consider two simple cases. In the first case, a dry slope with length 14m and height 5m is loaded with a cement block at the top (see figure \ref{Fig: slope_time_series}). The slope is composed of 2mm diameter grains with density $\rho_s = 2500 \tfrac{\text{kg}}{\text{m}^3}$. In the second case, an identical slope with identical loading and material composition is partially submerged in water (approximating a shoreline).

\begin{figure}
	\centering
	\includegraphics[scale=0.35]{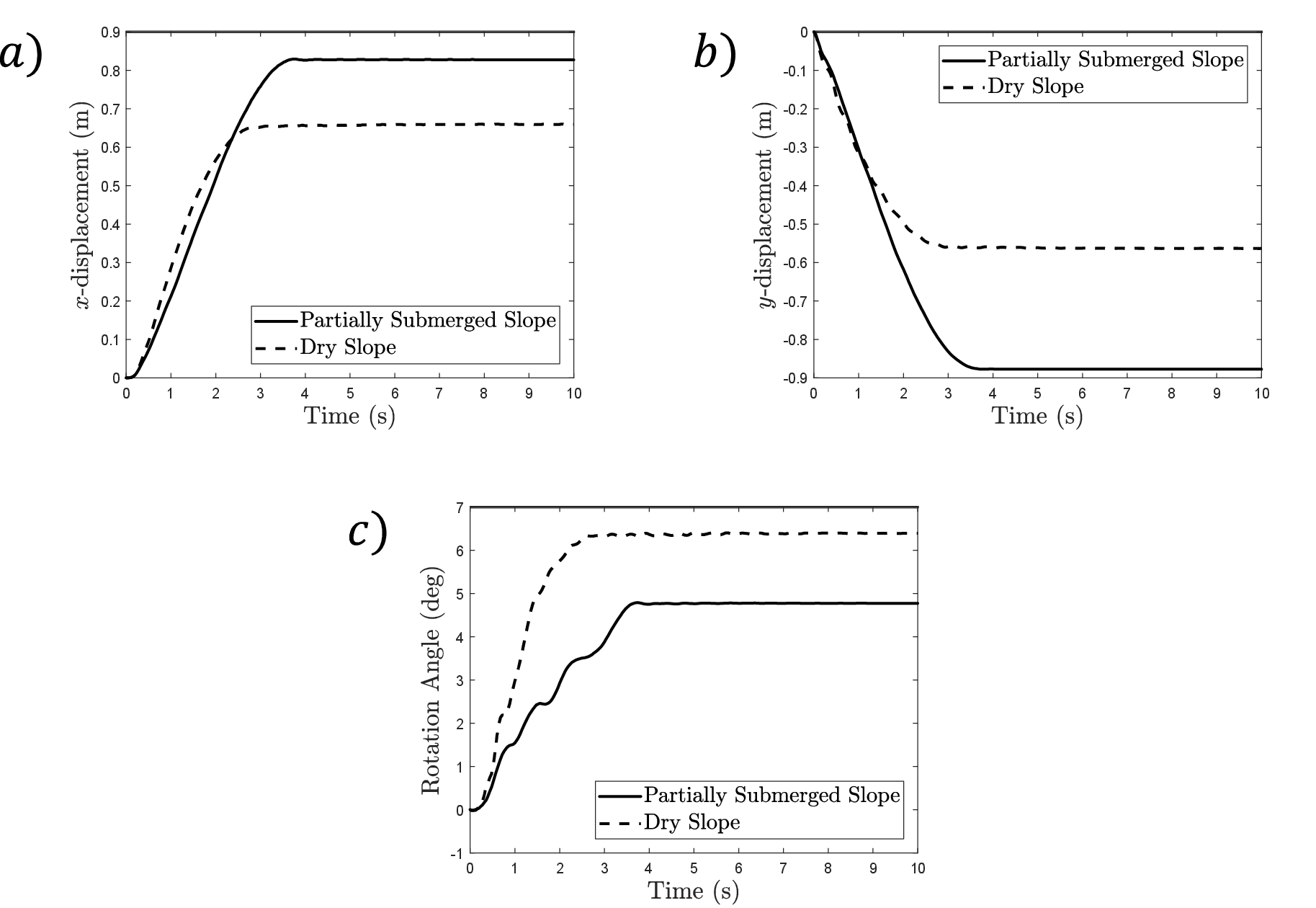}
	\caption{Plots of block motion for simulations described in section \ref{Sec: slope}. a) The $x$-displacement of the block's center of mass. b) The $y$-displacement of the block's center of mass. c) The rotation of the block about its center of mass.}
	\label{Fig: slope_plots}
\end{figure}

The simulations are performed in a 40m$\times$10m domain discretized into 160$\times$40 elements. The material points for the three bodies are seeded with 9 material points per grid cell. The initial packing of the granular slope is $\phi_0 = 0.585$. The resulting collapses are shown in figure \ref{Fig: slope_time_series}. 
We let the material properties be identical to those given in section \ref{Sec: model_fit}. As shown in figure \ref{Fig: slope_plots}, the resulting motion of the block (approximating a structure) on top of the slope has a strong dependence on the water lever in the slope. Over the course of 5 simulated seconds, the block on the partially submerged slope moves 20\% more in the $x$-direction, 36\% more in the $y$-direction, and rotates 34\% less.

\section{Conclusion}\label{Sec: conclusion}
We have developed a full set of constitutive relations for fluid-sediment mixtures which is capable of accurately and robustly modeling both dense and dilute flows of material. Our model is derived from a thermodynamically consistent set of rules and formulated to capture the dry and viscous inertial rheologies of granular materials, the critical state behavior of grains under shear, the change in the effective viscosity of the fluid due to suspended sediments, and a robust Darcy-like inter-phase drag. This model is implemented in MPM and validated against experiment. We characterize mixtures of glass beads immersed in a Newtonian fluid by fitting our model to the experimental data reported in \cite{pailha}. We then take these material parameters and show that our model is able to accurately predict the behavior of both collapsing granular columns (see \cite{rondon}) and shearing of fluid above granular beds (see \cite{allen}) \textit{without re-fitting material properties}. In addition, we also look at the application of this model and method to the problems of intrusion and slope stability.

The model we have presented in this work may be extensible to more general fluid-sediment mixtures such as those involving air (especially for examining the kick-up of dust for vertical take-off and landing vehicles). Other extensions of this model may look at adding cohesion (redefining the $f_1$ and $f_2$ yield conditions), introducing a fabric tensor to the rules governing dilation, or adding non-local effects \citep{kamrin2012,henann2013,kamrin2015} to capture, for example, the exponential-type decay of the granular velocity field deep in fluid-driven beds \citep{houssais2015,allen}.

\begin{acknowledgments}
This work was supported by Army Research Office Grant W911NF-16-1-0440 and National Science Foundation Grant CBET-1253228.  We thank Pascale Aussilous for access to experimental data from \cite{rondon}.
\end{acknowledgments}

\appendix
\section{}\label{appA}
The constitutive rules for our material model given in table \ref{Tab: constitutive_rules} are derived in the following specialization of the two-phase thermodynamic analysis from \cite{drumheller}. The basic rules for our model are similar to those described in \cite{coussy2004}; however, our derivation allows for two mixture temperatures, $\vartheta_s$ and $\vartheta_f$, and (as mentioned previously) we do not explicitly account for tortuosity.

\subsection{First Law of Thermodynamics}
The first law of thermodynamics states that the rate of change of the \textit{total energy} stored within a volume must be equal to the rate of \textit{heat flow} into the volume plus the \textit{external power} exerted on the volume. The total energy stored within a volume is the sum of \textit{internal energy} and \textit{kinetic energy}. We define a local expression for the energy conservation in the mixture in terms of the \textit{specific internal energies} $\varepsilon_s$ and $\varepsilon_f$, the phase-wise \textit{external heat fluxes} $\mathbi{q_s}$ and $\mathbi{q_f}$, the phase-wise \textit{internal heat generation} $q_s$ and $q_f$, and the basic homogenized continuum fields from section \ref{Sec: homogenization},
$$
\begin{aligned}
\bar{\rho}_s \frac{D^s \varepsilon_s}{Dt} + \bar{\rho}_f \frac{D^f \varepsilon_f}{Dt} + \bar{\rho}_s \frac{D^s \mathbi{v_s}}{Dt} \cdot \mathbi{v_s} + \bar{\rho}_f \frac{D^f \mathbi{v_f}}{Dt} \cdot \mathbi{v_f} =&\  (\divr (\boldsymbol{\sigma_s}) + \bar{\rho}_s \mathbi{g})\cdot\mathbi{v_s}
\\
&+ (\divr (\boldsymbol{\sigma_f}) + \bar{\rho}_f \mathbf{g})\cdot\mathbi{v_f}
\\
&+ \boldsymbol{\sigma_s}:\grad(\mathbi{v_s}) + \boldsymbol{\sigma_f}:\grad(\mathbi{v_f})
\\
&+ q_s + q_f - \divr(\mathbi{q_s} + \mathbf{q_f})
\end{aligned}
$$
which, with the momentum balance expressions in \eqref{Eqn: momentum_strong}, the \textit{buoyant} force from \eqref{Eqn: buoyant}, the specific form of the phase stresses in \eqref{Eqn: solid_stress} and \eqref{Eqn: fluid_stress}, and the evolution law for the \textit{true} fluid density from \eqref{Eqn: true_fluid_density}, becomes,
\begin{equation}
\begin{aligned}
\bar{\rho}_s \frac{D^s \varepsilon_s}{Dt} + \bar{\rho}_f \frac{D^f \varepsilon_f}{Dt} =&\ p_f \bigg(\frac{n}{\rho_f}\frac{D^f \rho_f}{Dt}\bigg)
+ \mathbi{f_d} \cdot (\mathbi{v_s} - \mathbi{v_f})
\\
&+ \boldsymbol{\tilde{\sigma}}:\grad(\mathbi{v_s}) + \boldsymbol{\tau_f}:\grad(\mathbi{v_f})
\\
&+ q_s + q_f - \divr(\mathbi{q_s} + \mathbi{q_f})
\label{Eqn: first_law_strong}
\end{aligned}
\end{equation}

\subsection{Second Law of Thermodynamics}
The second law of thermodynamics states that the rate of change of the \textit{total entropy} within a volume $\Omega$ must always be greater than or equal to the \textit{combined entropy flow} into the volume. \cite{drumheller} gives the following \textit{necessary} condition for entropy imbalance of a mixture,
\begin{equation} 
\bar{\rho}_s \frac{D^s s_s}{Dt} + \bar{\rho}_f \frac{D^f s_f}{Dt} + \divr \bigg( \frac{\mathbi{q_s}}{\vartheta_s} + \frac{\mathbi{q_f}}{\vartheta_f} \bigg) - \frac{q_s}{\vartheta_s} - \frac{q_f}{\vartheta_f} \geq 0
\label{Eqn: entropy_strong}
\end{equation}
We add two additional conditions by considering the entropy flow into each phase separately including the entropy flow due to the \textit{inter-phase heat flow} $q_i$,
\begin{equation}
\begin{aligned}
\bar{\rho}_s\frac{D^s s_s}{Dt} + \divr\bigg(\frac{\mathbi{q_s}}{\vartheta_s}\bigg) - \frac{q_s-q_i}{\vartheta_s} &\geq 0
\\
\bar{\rho}_f\frac{D^f s_f}{Dt} + \divr\bigg(\frac{\mathbi{q_f}}{\vartheta_f}\bigg) - \frac{q_f+q_i}{\vartheta_f} &\geq 0
\label{Eqn: two_phase_entropy_strong}
\end{aligned}
\end{equation}

In the absence of \textit{inter-phase heat flow}, satisfying the conditions in \eqref{Eqn: two_phase_entropy_strong} also necessarily satisfies \eqref{Eqn: entropy_strong}. Combining these last two expressions we find a second condition for entropy balance which does not depend on the inter-phase heat flow, $q_i$.
\begin{equation} 
\bar{\rho}_s \vartheta_s \frac{D^s s_s}{Dt} + \bar{\rho}_f \vartheta_f \frac{D^f s_f}{Dt} - \bigg(\frac{\mathbi{q_s} \cdot \grad(\vartheta_s)}{\vartheta_s} +  \frac{\mathbi{q_f} \cdot \grad(\vartheta_f)}{\vartheta_f}\bigg) + \divr(\mathbi{q_s + q_f})- (q_s + q_f)\geq 0
\label{Eqn: second_law_strong}
\end{equation}

\subsection{Helmholtz Free Energy}
We now introduce the definition for the phase-wise Helmholtz free energies, $\psi_s$ and $\psi_f$, such that,
\begin{equation}
\begin{aligned}
\psi_s =& \varepsilon_s - s_s \vartheta_s
\\
\psi_f =& \varepsilon_f - s_f \vartheta_f
\label{Eqn: helmholtz}
\end{aligned}
\end{equation}
Substituting into the first law expression in \eqref{Eqn: first_law_strong} and combining with the second law expression from \eqref{Eqn: second_law_strong}, the following free energy inequality is found,
\begin{equation}
\begin{aligned}
0 \leq&\ - \bar{\rho}_s \frac{D^s \psi_s}{Dt} - \bar{\rho}_f \frac{D^f \psi_f}{Dt}
- \bar{\rho}_s s_s \frac{D^s \vartheta_s}{Dt} - \bar{\rho}_f s_f \frac{D^f \vartheta_f}{Dt}
\\
&+ p_f \bigg(\frac{n}{\rho_f}\frac{D^f \rho_f}{Dt}\bigg)
+ \mathbi{f_d} \cdot (\mathbi{v_s} - \mathbi{v_f})
+ \boldsymbol{\tilde{\sigma}}:\grad(\mathbi{v_s})\\
&+ \boldsymbol{\tau_f}:\grad(\mathbi{v_f})
- \frac{\mathbi{q_s} \cdot \grad(\vartheta_s)}{\vartheta_s} -  \frac{\mathbi{q_f} \cdot \grad(\vartheta_f)}{\vartheta_f}
\label{Eqn: second_law_helmholtz}
\end{aligned}
\end{equation}

We let the spatial solid phase and fluid phase velocity gradients be expressed in matrix form as,
\begin{equation}
\mathsfbi{L_s} \equiv \grad(\mathbi{v_s}), \qquad \mathsfbi{L_f} \equiv \grad(\mathbi{v_s})
\label{Eqn: phase_velocity_gradients}
\end{equation}
which have unique decompositions into a phase spin tensor, $\mathsfbi{W}$, and a phase strain-rate tensor, $\mathsfbi{D}$, 
\begin{equation}
\mathsfbi{D} = \sym(\mathsfbi{L}) = \tfrac{1}{2}(\mathsfbi{L + L^\top}), \qquad\text{and} \qquad \mathsfbi{W} = \skw(\mathsfbi{L}) = \tfrac{1}{2}(\mathsfbi{L - L^\top})
\label{Eqn: spin_and_strainrate_tensors}
\end{equation}
We further assume that the solid and fluid phases have uniform and constant temperatures, $\vartheta_s$ and $\vartheta_f$, such that \eqref{Eqn: second_law_helmholtz} becomes,
\begin{equation}
- \bar{\rho}_s \frac{D^s \psi_s}{Dt} - \bar{\rho}_f \frac{D^f \psi_f}{Dt} + p_f \bigg(\frac{n}{\rho_f}\frac{D^f \rho_f}{Dt}\bigg) + (\boldsymbol{\tilde{\sigma}}:\mathsfbi{D_s}) + (\boldsymbol{\tau_f}:\mathsfbi{D_{0f}}) + \mathbi{f_d} \cdot (\mathbi{v_s} - \mathbi{v_f}) \geq 0
\label{Eqn: kinematic_free_energy_imbalance}
\end{equation}

\subsection{Fluid Phase Free Energy Function}
The conservative constitutive behavior of the fluid phase is governed by the fluid phase specific free energy, $\psi_f$. We assume that the functional form of the free energy only depends on the true fluid density, $\psi_f = \hat{\psi}_f(\rho_f)$. Substituting into the expression for free energy imbalance in \eqref{Eqn: kinematic_free_energy_imbalance},
\begin{equation}
- \bar{\rho}_s \frac{D^s \psi_s}{Dt}  + (\boldsymbol{\tilde{\sigma}}:\mathsfbi{D_s}) - \frac{n}{\rho_f} \frac{D^f \rho_f}{Dt} \bigg(p_f - \rho_f^2 \frac{\partial \hat{\psi}_f(\rho_f)}{\partial \rho_f}\bigg) + (\boldsymbol{\tau_f}:\mathsfbi{D_{0f}}) + \mathbi{f_d} \cdot (\mathbi{v_s} - \mathbi{v_f}) \geq 0
\label{Eqn: fluid_free_energy_imbalance}
\end{equation}

\subsection{Solid Phase Free Energy Function}\label{Sec: appendix_solid_elasticity}
The solid phase behavior will be governed by an elastic-plastic constitutive relation derived from that given in \cite{anand}. We begin with the definition of the solid phase deformation gradient,
$$\mathsfbi{F} = \frac{\partial\boldsymbol{\chi_s}(\mathbi{X},t)}{\partial \mathbi{X}}, \qquad \frac{D^s \mathsfbi{F}}{Dt} = \mathsfbi{L_s F}$$
where $\boldsymbol{\chi_s}(\mathbi{X},t)$ is the motion function mapping from a position, $\mathbi{X}$, in the solid reference configuration to a position in the solid deformed (current) configuration at time $t$.

We assume the Kroner-Lee decomposition of the deformation gradient,
\begin{equation} 
\mathsfbi{F = F^eF^p}
\label{Eqn: kroner_lee}
\end{equation}
with $\mathsfbi{F^e}$ the \textit{elastic} deformation and $\mathsfbi{F^p}$ the \textit{plastic} deformation. With this, the velocity gradient can be separated into an \textit{elastic} and \textit{plastic} flow,
\begin{equation}
\mathsfbi{L_s} = \mathsfbi{L^e} + \mathsfbi{F^eL^pF^{e-1}}, \qquad \text{s.t.} \qquad \frac{D^s \mathsfbi{F^e}}{Dt} = \mathsfbi{L^eF^e},\quad \frac{D^s \mathsfbi{F^p}}{Dt} = \mathsfbi{L^pF^p}
\label{Eqn: solid_velocity_gradient}
\end{equation}
We assume that the plastic flow, $\mathsfbi{L^p}$,  is symmetric such that,
\begin{equation}
\begin{aligned}
\mathsfbi{D^e} = \sym(\mathsfbi{L^e}), \qquad \mathsfbi{W^e} = \mathsfbi{W_s}, \qquad
\mathsfbi{D^p} = \sym(\mathsfbi{L^p}), \qquad \mathsfbi{W^p} = \mathsfbi{0}
\end{aligned}
\label{Eqn: plastic_and_elastic_flow}
\end{equation}

The right polar decomposition of the elastic deformation is defined as $\mathsfbi{F^e} = \mathsfbi{R^e}\mathsfbi{U^e}$, with $\mathsfbi{R^e}$ the orthogonal \textit{rotation} tensor and $\mathsfbi{U^e}$ the symmetric positive definite \textit{elastic stretch} tensor. The \textit{right Cauchy-Green tensor} is then, $\mathsfbi{C^e} = \mathsfbi{U^e}^2 = \mathsfbi{F^{e\top}F^e}$. Since $\mathsfbi{U^e}$ is symmetric and positive definite, it admits a spectral decomposition which we use to define the \textit{logarithmic strain tensor}, $\mathsfbi{E^e}$ ,
\begin{equation} 
\mathsfbi{U^e} = \sum_{i=1}^3{\lambda_i \mathbi{r}_i \otimes \mathbi{r}_i}, \qquad \text{and} \qquad \mathsfbi{E^e} = \ln(\mathsfbi{U^e}) \equiv \sum_{i=1}^3{\ln(\lambda_i) \mathbi{r}_i \otimes \mathbi{r}_i}
\label{Eqn: logarithmic_strain}
\end{equation}
where $\{\lambda_i\}$ are the principal stretches, $\{\mathbi{r}_i\}$ are the right principal directions, and each $\lambda_i > 0$. Further we define the volumetric Jacobians as,
\begin{equation}
J \equiv \det(\mathsfbi{F}) > 0, \qquad J^e \equiv \det(\mathsfbi{F^e}) > 0, \qquad J^p \equiv \det(\mathsfbi{F^p}) > 0
\label{Eqn: J}
\end{equation}

We introduce the solid phase \textit{volumetric free energy}, $\varphi_s$, which is defined as,
\begin{equation}
\varphi_s = J^e \bar{\rho}_s \psi_s, \qquad \text{s.t.} \qquad \varphi_s = \hat{\varphi}_s(\mathsfbi{C^e}) = \tilde{\varphi}_s(\mathsfbi{E^e})
\label{Eqn: volumetric_free_energy}
\end{equation}
Therefore \eqref{Eqn: fluid_free_energy_imbalance} has the following specialized form,
\begin{equation}
\begin{aligned}
\bigg(\boldsymbol{\tilde{\sigma}} - 2 J^{e-1} \mathsfbi{F^e} \frac{\partial \hat{\varphi}_s(\mathsfbi{C^e})}{\partial \mathsfbi{C^e}} \mathsfbi{F^{e\top}} \bigg) : \mathsfbi{D^e}  + \bigg(\boldsymbol{\tilde{\sigma}}:(\mathsfbi{F^eD^pF^{e-1}}) - ( J^{e-1} \varphi_s \mathsfbi{1} ): \mathsfbi{D^p}\bigg)&
\\
- \frac{n}{\rho_f} \frac{D^f \rho_f}{Dt} \bigg(p_f - \rho_f^2 \frac{\partial \hat{\psi}_f(\rho_f)}{\partial \rho_f}\bigg) + (\boldsymbol{\tau_f}:\mathsfbi{D_{0f}}) + \mathbi{f_d} \cdot (\mathbi{v_s} - \mathbi{v_f}) &\geq 0
\label{Eqn: solid_free_energy_imbalance}
\end{aligned}
\end{equation}

\subsection{Rules for Constitutive Relations}
The expression in \eqref{Eqn: solid_free_energy_imbalance} must be true for \textit{all} flows \textit{everywhere}. Since it is possible to conceive of mixture motions with independently varying (and possibly vanishing) values for $\mathsfbi{D^e}$, $\mathsfbi{D^p}$, $D^s\rho_f/Dt$, $\mathsfbi{D_{0f}}$, and $(\mathbi{v_s- v_f})$, the following relations must each individually be satisfied,
\begin{equation}
p_f - \rho_f^2 \frac{\partial \hat{\psi}_f(\rho_f)}{\partial \rho_f} = 0
\label{Eqn: fluid_pressure_free_energy}
\end{equation}
\begin{equation}
\boldsymbol{\tau_f}:\mathsfbi{D_{0f}} \geq 0
\label{Eqn: fluid_stress_free_energy}
\end{equation}
\begin{equation}
\boldsymbol{\tilde{\sigma}} - 2 J^{e-1} \mathsfbi{F^e} \frac{\partial \hat{\varphi}_s(\mathsfbi{C^e})}{\partial \mathsfbi{C^e}} \mathsfbi{F^{e\top}} = 0
\label{Eqn: solid_stress_free_energy}
\end{equation}
\begin{equation}
\boldsymbol{\tilde{\sigma}}:(\mathsfbi{F^eD^pF^{e-1}}) - ( J^{e-1} \varphi_s \mathsfbi{1} ): \mathsfbi{D^p} \geq 0
\label{Eqn: solid_plasticity_free_energy}
\end{equation} 
\begin{equation}
\mathbi{f_d} \cdot (\mathbi{v_s} - \mathbi{v_f}) \geq 0
\label{Eqn: interphase_drag_free_energy}
\end{equation}

\section{}\label{appB}
Following the thermodynamic analysis in appendix \ref{appA}, we let the solid phase effective granular stress be given by a \textit{stiff} elastic specialization of the model derived in \cite{anand}.

\subsection{Solid Phase Effective Granular Stress}
Beginning with the equality in \eqref{Eqn: solid_stress_free_energy}, we define the solid phase effective stress $\boldsymbol{\tilde{\sigma}}$ as,
\begin{equation} 
\boldsymbol{\tilde{\sigma}} = 2 J^{e-1} \mathsfbi{F^e} \frac{\partial \hat{\varphi}_s(\mathsfbi{C^e})}{\partial \mathsfbi{C^e}} \mathsfbi{F^{e\top}}
\label{Eqn: solid_cauchy_stress_free_energy}
\end{equation}
We then define the elastic stress measure, $\mathsfbi{T^e}$, such that,
\begin{equation}
\mathsfbi{T^e} = J^e \mathsfbi{F^{e\top}} \boldsymbol{\tilde{\sigma}} \mathsfbi{F^{e-\top}}, \qquad \text{and} \qquad \boldsymbol{\tilde{\sigma}} = J^{e-1} \mathsfbi{F^{e-\top} T^e F^{e\top}}
\label{Eqn: elastic_to_effective_stress}
\end{equation}
Combining these expressions, we also have,
\begin{equation}
\mathsfbi{T^e} = 2 \mathsfbi{C^e} \frac{\partial \hat{\varphi}_s(\mathsfbi{C^e})}{\partial \mathsfbi{C^e}} = \frac{\partial \tilde{\varphi}_s(\mathsfbi{E^e})}{\partial \mathsfbi{E^e}}
\end{equation}

We choose the volumetric free energy function, $\tilde{\varphi}_s(\mathsfbi{E^e}) = G \|\mathsfbi{E_0^e}\|^2 + \frac{1}{2} K \tr(\mathsfbi{E^e})^2$, with $G$ the solid \textit{shear modulus} and $K$ the solid \textit{bulk modulus} with units of stress. The elastic stress measure is therefore given by,
\begin{equation}
\mathsfbi{T^e} = \mathscr{C}[\mathsfbi{E^e}] \equiv 2 G \mathsfbi{E_0^e} + K \tr(\mathsfbi{E^e}) \mathsfbi{1}
\label{Eqn: elastic_stress}
\end{equation}

\subsection{Solid Phase Plastic Strain-Rate}\label{Sec: plastic_strain_rate}

The solid phase plastic flow rate, $\mathsfbi{D^p}$, must obey the inequality in \eqref{Eqn: solid_plasticity_free_energy}. By substituting the expression from \eqref{Eqn: elastic_to_effective_stress} into this inequality, we find,
$$\mathsfbi{T^e}:\mathsfbi{D^p} - \varphi_s \mathsfbi{1}:\mathsfbi{D^p} \geq 0$$

We let the granular skeleton of the mixture be \textit{elastically stiff}, such that $\mathsfbi{U^e} \approx \mathsfbi{1}$, $J^e \approx 1$, and $\mathsfbi{E^e} \ll \mathsfbi{1}$. In this limit, \eqref{Eqn: solid_plasticity_free_energy} is dominated by the \textit{stiff plastic dissipation},
\begin{equation}
\mathcal{D} \equiv \mathsfbi{T^e} : \mathsfbi{D^p} \geq 0
\label{Eqn: elastic_stress_dissipation}
\end{equation}

We introduce another measure of the plastic strain-rate, $\mathsfbi{\tilde{D}^p}$, defined as follows,
\begin{equation} 
\mathsfbi{\tilde{D}^p} = \mathsfbi{F^eD^pF^{e-1}}, \qquad \text{s.t} \qquad \mathsfbi{D_s} = \mathsfbi{D^e} + \mathsfbi{\tilde{D}^p}
\label{Eqn: additive_decomposition}
\end{equation}
And therefore, \eqref{Eqn: elastic_stress_dissipation} becomes,
\begin{equation}
\mathcal{D} = \boldsymbol{\tilde{\sigma}}:\mathsfbi{\tilde{D}^p} \geq 0
\label{Eqn: cauchy_stress_dissipation}
\end{equation}
To satisfy the dissipation inequality in \eqref{Eqn: elastic_stress_dissipation}, we define the plastic strain-rate $\mathsfbi{D^p}$ implicitly via $\mathsfbi{\tilde{D}^p}$ defined in \eqref{Eqn: plastic_flow_rule} subject to \eqref{Eqn: cauchy_stress_dissipation}.

\section{}\label{appC}
In this section we describe the specific details of the numerical implementation referenced in section \ref{Sec: implementation}.

\subsection{Material Point Method Discretization}
The material point method, as first derived by \cite{sulsky}, is a numerical scheme for solving dynamic problems in solid mechanics where materials undergo large deformations. The basic algorithm defined in \cite{sulsky} and generalized by \cite{bardenhagen} involves discretizing material fields (such as density and stress) on a set of material point tracers and solving the equations of motion on a background grid.

In \cite{abe} and \cite{bandara}, the material point method is extended to solve the equations of mixtures defined in \cite{jackson} (here in \eqref{Eqn: solid_closed_form} and \eqref{Eqn: fluid_closed_form}). The algorithm presented in this work derives directly from the weak formulation of the governing equations in table \ref{Tab: governing_equations} and differs slightly from prior works (due to different simplifying assumptions).

\subsubsection{Definition of Material Point Tracers}\label{Sec: material_point_definition}
As shown in figure \ref{Fig: discretization}(3), the two continuum bodies defined in figure \ref{Fig: configuration} by $\mathcal{B}_s^t$ and $\mathcal{B}_f^t$ are discretized into material blocks represented by discrete material points. We let the continuum representation of the bodies be given by,
\begin{equation}
\sum_{p = 1}^{N_s}U_{sp}(\mathbi{x},t) = \bigg\{ \begin{matrix}
1&\mathbi{x} \in \mathcal{B}_s^t\\
0&\text{else}
\end{matrix} \qquad \qquad
\sum_{p = 1}^{N_f}U_{fp}(\mathbi{x},t) = \bigg\{ \begin{matrix}
1&\mathbi{x} \in \mathcal{B}_f^t\\
0&\text{else}
\end{matrix}
\label{Eqn: point_characteristic_function}
\end{equation}
where $\mathbi{x}$ is the position vector in the domain $\Omega$, $t$ is time, $U_{sp}(\mathbi{x},t)$ and $U_{fp}(\mathbi{x},t)$ are the $p$th material point characteristic functions (as in \cite{bardenhagen}) that are \textit{co-moving with the material}, and $N_s$ and $N_f$ are the number of solid and fluid material point tracers respectively. Intuitively, the sum of the phase-wise characteristic functions defines a spatial field which is equal to 1 within the body and 0 outside.

We construct the solid continuum fields using the $U_{sp}$ functions with $\bar{\rho}_s(\mathbi{x})$ defined at time $t^k$ by the $N_s$ coefficients $\{\bar{\rho}_{sp}^k\}$ and $\boldsymbol{\tilde{\sigma}}(\mathbi{x})$ by the $N_s$ coefficients $\{\boldsymbol{\tilde{\sigma}}_p^k\}$. The fluid continuum fields are constructed using $U_{fp}$ such that the fields $\bar{\rho}_f(\mathbi{x})$, $\rho_f(\mathbi{x})$, $\boldsymbol{\tau_f}(\mathbi{x})$, and $p_f(\mathbi{x})$ are given at time $t^k$ by the $N_f$ coefficients $\{\bar{\rho}_{fp}^k\}$, $\{\rho_{fp}^k\}$, $\{\boldsymbol{\tau_f}_p^k\}$, and $\{p_{fp}^k\}$ respectively.

We also introduce a measure of material point weights, $v_{sp}^k$ and $v_{fp}^k$, with,
\begin{equation}
v_{sp}^k = \int_{\Omega}{ U_{sp}(\mathbi{x},t^k) dv}, \qquad
v_{fp}^k = \int_{\Omega}{ U_{fp}(\mathbi{x},t^k) dv}
\label{Eqn: material_point_volume}
\end{equation}
Each material point has a centroid (center of mass) which maps to a location $\mathbi{x_s}_p$ for the $p$th solid material point and $\mathbi{x_{f}}_p$ for the $p$th fluid material point. This centroid moves through the domain and has an associated momentum (at time $t^k$) given by $m_{sp}\mathbi{v_s}^k_p$ or $m_{vp}\mathbi{v_f}^k_p$ respectively with,
\begin{equation}
m_{sp} = v^k_{sp}\bar{\rho}^k_{sp}, \qquad m_{fp} = v^k_{fp}\bar{\rho}^k_{fp}
\label{Eqn: material_point_mass}
\end{equation}
and $\{m_{sp}\}$, $\{m_{fp}\}$ constant but not necessarily uniform.

\subsubsection{Definition of Background Grid Basis}\label{Sec: nodal_basis_definition}
In addition to the material point representation of the continuum bodies, we also use a grid to solve the weak form equations of motion and for approximating material fields (for post-processing and simplifying intermediate calculations). Since both bodies live within the same computational domain, $\Omega$, we let one discrete grid serve this purpose for the entire mixture. The grid is defined by a set of continuous nodal basis functions,
\begin{equation}
\sum_{i=1}^{[n]} \mathcal{N}_i(\mathbi{x}) = 1 \quad \forall \mathbi{x} \in \Omega
\label{Eqn: partition_of_unity}
\end{equation}
where $\mathcal{N}_i(\mathbi{x})$ is the $i$th nodal basis function and $[n]$ is the total number of nodes (or degrees of freedom if discontinuous shape functions are used). With this definition we can then define the nodal fields $\mathbi{a_s}(\mathbi{x})$, $\mathbi{v_s}(\mathbi{x})$, $\mathbi{a_f}(\mathbi{x})$, $\mathbi{v_f}(\mathbi{x})$, and $n(\mathbi{x})$ at time $t^k$ by the $[n]$ coefficients $\{\mathbi{a_s}_i^k\}$, $\{\mathbi{v_s}_i^k\}$, $\{\mathbi{a_f}_i^k\}$, $\{\mathbi{v_f}_i^k\}$, $\{n_i^k\}$ respectively.

In addition to the fields above, we also introduce a measure of the nodal basis weight, $V_i$,
\begin{equation}
V_i = \int_{\Omega}{\mathcal{N}_i(\mathbi{x}) dv}
\label{Eqn: nodal_volume}
\end{equation}

It is numerically convenient to let the background grid be composed of regular Cartesian elements. We therefore let the construction of the basis functions $\{\mathcal{N}_i(\mathbi{x})\}$ be the tensor product of 1D functions $\mathcal{N}_{1\text{D}}(\hat{x}_{ij})$ with $\hat{x}_{ij}$ a measure of the distance from the $i$th grid node to the spatial position $\mathbi{x}$ along the $j$th primary Cartesian direction, $\{\hat{x}_1,\hat{x}_2,\hat{x}_3\}$.
\begin{equation}
\mathcal{N}_i(\mathbi{x}) = \prod_{j=1}^{\text{DIM}} \mathcal{N}_{1\text{D}}(\hat{x}_{ij})
\end{equation}
where DIM is the dimension of the simulation. The choice of $\mathcal{N}_{1\text{D}}(\hat{x}_{ij})$ can have significant impact on the accuracy of the material point method, especially for reduction of `grid-crossing' error (see \cite{bardenhagen}) and quadrature error (see \cite{steffen}). In this work we use adjusted cubic splines based on those presented in \cite{steffen}.

\subsection{Time Marching Procedure}\label{Sec: time_marching}
\begin{figure}
	\centering
	\includegraphics[scale=0.35]{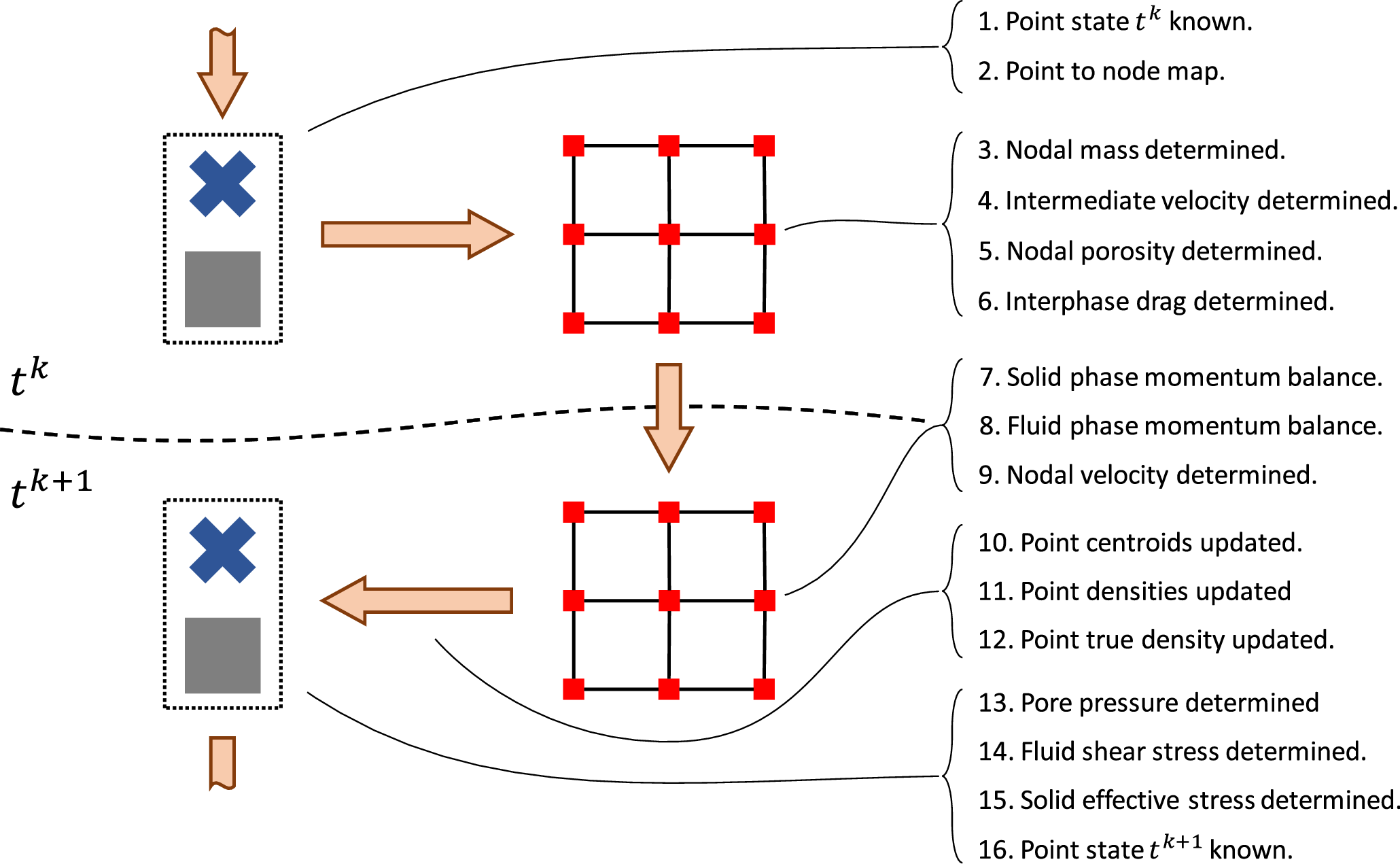}
	\caption{The explicit time integration procedure described in section \ref{Sec: time_marching} is shown. At the beginning of the step, the material points carry the full state of the mixture. The mixture state is then mapped from the points to the background grid nodes, where the equations of motion are solved according to the weak form of momentum balance. At the end of the step, the solved equations of motion are used to update the mixture state on the material points.}
	\label{Fig: time_step}
\end{figure}

The weak forms of the governing equations are solved according to the following explicit procedure (shown in figure \ref{Fig: time_step}) to step from time $t^k$ to time $t^{k+1}$ where,
\begin{equation}
t^{k+1} = t^k + \Delta t
\end{equation}

\begin{enumerate}
	\item The discrete material point states of the two phases are known at time $t^k$.
	$$\begin{aligned}
	\text{solid phase:\quad}&\{\bar{\rho}_{sp}^k, \boldsymbol{\tilde{\sigma}}_p^k, m_{sp}, \mathbi{x_s}_p^k, \mathbi{v_s}_p^k\} \\
	\text{fluid phase:\quad}&\{\bar{\rho}_{fp}^k, \rho_{fp}^k, \boldsymbol{\tau_f}_p^k, p_{fp}^k, m_{fp}, \mathbi{x_f}_p^k, \mathbi{v_f}_p^k\}
	\end{aligned}$$
	
	\item The material point centroids, $\{\mathbi{x_s}^k_p\}$ and $\{\mathbi{x_f}^k_p\}$ are used to generate the mapping coefficients $\{\mathcal{S}_{sip}^k\}$, $\{\mathcal{S}_{fip}^k\}$, $\{\nabla\mathcal{S}_{sip}^k\}$, and  $\{\nabla\mathcal{S}_{fip}^k\}$.
	\begin{equation}
	\begin{aligned}
	\mathcal{S}^k_{sip} = \mathcal{N}_i(\mathbi{x_s}^k_p),\qquad \nabla\mathcal{S}^k_{sip} = \grad\big(\mathcal{N}_i(\mathbi{x})\big)\big|_{\mathbi{x_s}^k_p}\\
	\mathcal{S}^k_{fip} = \mathcal{N}_i(\mathbi{x_f}^k_p),\qquad
	\nabla\mathcal{S}^k_{fip} = \grad\big(\mathcal{N}_i(\mathbi{x})\big)\big|_{\mathbi{x_f}^k_p}
	\end{aligned}
	\label{Eqn: numerical_mapping}
	\end{equation}
	
	\item The nodal mass coefficients, $\{m^k_{si}\}$ and $\{m^k_{fi}\}$, are determined.
	\begin{equation} 
	m^k_{si} = \sum_{p=1}^{N_s} m_{sp} \mathcal{S}^k_{sip}, \qquad m^k_{fi} = \sum_{p=1}^{N_f} m_{fp} \mathcal{S}^k_{fip}
	\label{Eqn: nodal_mass}
	\end{equation}
	
	\item An intermediate nodal representation of the phase velocity fields, given by the coefficients $\{\mathbi{v_s}_i^*\}$ and $\{\mathbi{v_f}_i^*\}$, is determined by approximating the material point velocity fields, given by the coefficients $\{\mathbi{v_s}^k_p\}$ and $\{\mathbi{v_f}^k_p\}$.
	\begin{equation}
	m^k_{i} \mathbi{v_s}^*_i = \sum_{p=1}^{N_s} m_{sp} \mathbi{v_s}^k_p \mathcal{S}^k_{sip}, \qquad
	m^k_{fi} \mathbi{v_f}^*_i = \sum_{p=1}^{N_f} m_{fp} \mathbi{v_f}^k_p \mathcal{S}^k_{sip}
	\end{equation}
	
	\item The nodal porosity coefficients, $\{n_i^k\}$, are determined.
	\begin{equation} 
	n^k_i = 1 - \frac{m^k_{si}}{V_i\rho_s}
	\label{Eqn: nodal_porosity}
	\end{equation}
	
	\item The nodal approximation of the inter-phase drag, given by $\{\mathbi{f_d}^*_i\}$, is determined.
	\begin{equation}
	\mathbi{f_d}^*_i = \frac{18 n_i^k (1-n_i^k)\eta_0}{d^2}\ \hat{F}((1-n_i^k), \Reyn_i^*)\ (\mathbi{v_s}_i^* - \mathbi{v_f}_i^*) \sum_{p=1}^{N_f}v^k_{pf}\mathcal{S}_{fip}^k
	\label{Eqn: nodal_drag_force}
	\end{equation}
	\begin{equation}
	\Reyn_i^* = \frac{n_i^k \|\mathbi{v_s}_i^*-\mathbi{v_f}_i^*\| d}{\eta_0}
	\label{Eqn: nodal_reynolds}
	\end{equation}
	
	\item The acceleration of the solid phase at time $t^{k+1}$, given by $\{\mathbi{a_s}_i^{k+1}\}$, is determined.
	\begin{equation} 
	\begin{matrix}
	m^k_{si}\mathbi{a_s}^{k+1}_i = m^k_{si}\mathbi{g} - \mathbi{f_d}^k_i - \sum_{p=1}^{N_s}\big(v^k_{sp} \boldsymbol{\tilde{\sigma}}^k_p \nabla\mathcal{S}^k_{sip}\big) + (1-n^k_i)\sum_{p=1}^{N_f}\big(v^k_{fp} p^k_{fp} \nabla\mathcal{S}^k_{sip}\big) + \mathbi{s_s}^k_i\\[5pt]
	\text{($\mathbi{s_s}_i^k$ is a boundary condition enforced on the $i$th node.)}
	\end{matrix}
	\label{Eqn: numerical_solid_momentum}
	\end{equation}
	
	\item The acceleration of the fluid phase at time $t^{k+1}$, given by $\{\mathbi{a_f}_i^{k+1}\}$, is determined.
	\begin{equation}
	\begin{matrix}
	m^k_{fi}\mathbi{a_f}^{k+1}_i = m^k_{fi}\mathbi{g} + \mathbi{f_d}^k_i - \sum_{p=1}^{N_f}\big(v^k_{fp} \boldsymbol{\tau_f}^k_p \nabla\mathcal{S}^k_{fip}\big) + n^k_i\sum_{p=1}^{N_f}\big(v^k_{fp} p^k_{fp} \nabla\mathcal{S}^k_{sip}\big) + \mathbi{s_f}^k_i\\[5pt]
	\text{($\mathbi{s_f}_i^k$ is a boundary condition enforced on the $i$th node.)}
	\end{matrix}
	\label{Eqn: numerical_fluid_momentum}
	\end{equation}
	
	\item The phase velocity fields at time $t^{k+1}$, given by $\{\mathbi{v_s}^{k+1}_i\}$ and $\{\mathbi{v_f}^{k+1}_i\}$, are determined explicitly according to,
	\begin{equation}
	\begin{aligned}
	\mathbi{v_s}^{k+1}_i = \mathbi{v_s}^*_i + \Delta t\  \mathbi{a_s}^{k+1}_i, \qquad 
	\mathbi{v_f}^{k+1}_i = \mathbi{v_f}^*_i + \Delta t\ \mathbi{a_f}^{k+1}_i
	\end{aligned}
	\end{equation}
	
	\item The material point centroid positions and velocities are updated explicitly as in \cite{brackbill1986} and \cite{brackbill1988},
	\begin{equation}
	\begin{aligned}
	\mathbi{x_s}_p^{k+1} = \mathbi{x_s}_p^k + \Delta t\ \sum_{i=1}^{[n]}\mathbi{v_s}_i^{k+1}\mathcal{S}^k_{sip} + (\boldsymbol{\delta_s}_p^k), &\qquad \mathbi{v_s}_p^{k+1} = \mathbi{v_s}_p^k + \Delta t\ \sum_{i=1}^{[n]}\mathbi{a_s}_i^{k+1}\mathcal{S}^k_{sip}\\
	\mathbi{x_f}_p^{k+1} = \mathbi{x_f}_p^k + \Delta t\ \sum_{i=1}^{[n]}\mathbi{v_f}_i^{k+1}\mathcal{S}^k_{fip}+ (\boldsymbol{\delta_f}_p^k), &\qquad \mathbi{v_f}_p^{k+1} = \mathbi{v_f}_p^k + \Delta t\ \sum_{i=1}^{[n]}\mathbi{a_f}_i^{k+1}\mathcal{S}^k_{fip}
	\end{aligned}
	\label{Eqn: discrete_centroid_update}
	\end{equation}
	where $\boldsymbol{\delta_s}_p^{k+1}$ and $\boldsymbol{\delta_f}_p^{k+1}$ are the \textit{$\delta$ position} correction described in section \ref{Sec: delta_position}.
	
	\item The material point densities at time $t^{k+1}$, $\{\bar{\rho}_s^{k+1}\}$ and $\{\bar{\rho}_s^{k+1}\}$, are updated.
	\begin{equation} 
	\begin{aligned}
	\bar{\rho}^{k+1}_{sp} &= \bar{\rho}^{k}_{sp} \exp\bigg(-(\Delta t) \tr\bigg(\sum_{i=1}^{[n]} \mathbi{v_s}^{k+1}_i \otimes \nabla\mathcal{S}_{sip}^k\bigg)\bigg)\\
	\bar{\rho}^{k+1}_{fp} &= \bar{\rho}^{k}_{fp} \exp\bigg(-(\Delta t) \tr\bigg(\sum_{i=1}^{[n]} \mathbi{v_f}^{k+1}_i \otimes \nabla\mathcal{S}_{fip}^k\bigg)\bigg)\end{aligned}
	\label{Eqn: numerical_mass_balance}
	\end{equation}
	where $\otimes$ is the tensor product operator.
	
	\item The fluid phase material point \textit{true} densities, $\{\rho_{fp}^{k+1}\}$, are determined. (Note that for numerical stability, we do not require that $\{n_p^{k+1}\}$, $\{\bar{\rho}_{fp}^{k+1}\}$, and $\{\rho_{fp}^{k+1}\}$ be consistent.)
	\begin{equation} 
	n_p^{k+1} = \sum_{i=1}^{[n]}n_i^k\mathcal{S}_{fip}^k
	\end{equation}
	\begin{equation} 
	\rho^{k+1}_{fp} = \rho^{k}_{fp} \exp\bigg(-\bigg(\frac{\Delta t}{n_p^{k+1}}\bigg) \tr\bigg(\sum_{i=1}^{[n]} [(1-n_i^k)\mathbi{v_s}^{k+1}_i + n_i^k \mathbi{v_f}^{k+1}_i] \otimes \nabla\mathcal{S}_{fip}^k\bigg)\bigg)
	\label{Eqn: numerical_fluid_density_evolution}
	\end{equation}
	
	\item The fluid phase material point pore pressure state is determined directly from the \textit{true} fluid density.
	\begin{equation} 
	p_{fp}^{k+1} = \kappa \ln\bigg(\frac{\rho_{fp}^{k+1}}{\rho_{0f}}\bigg)
	\label{Eqn: numerical_pore_pressur}
	\end{equation}
	
	\item The fluid phase material point shear stresses, $\{\boldsymbol{\tau_f}_p^{k+1}\}$, are determined directly from the fluid phase velocity gradient.
	\begin{equation} 
	\boldsymbol{\tau_f}_p^{k+1} = 2\eta_0\big(1 + \tfrac{5}{2}(1-n_p^{k+1})\big) \mathsfbi{D_{0f}}_p^{k+1}
	\label{Eqn: numerical_fluid_shear_stress}
	\end{equation}
	$$\mathsfbi{D_f}_p^{k+1} = \sym \bigg(\sum_{i=1}^{[n]} \mathbi{v_f}_i^{k+1} \otimes \nabla\mathcal{S}_{fip}^k\bigg)$$
	
	\item The solid phase material point effective stresses, $\{\boldsymbol{\tilde{\sigma}}_p^{k+1}\}$, are determined with a semi-implicit method described in section \ref{Sec: implicit_stress_update}. 
	\begin{equation}
	\boldsymbol{\tilde{\sigma}}_p^{k+1} = \boldsymbol{\tilde{\sigma}}_p^k + \Delta t \big[ 2G \big(\mathsfbi{D_{0s}}^{k+1}_p - (\mathsfbi{\tilde{D}^p_0})_p^{k+1}\big) + K \tr(\mathsfbi{D_s}^{k+1}_p - (\mathsfbi{\tilde{D}^p})_p^{k+1})\mathsfbi{1} + \mathsfbi{W_s}^{k+1}_p\boldsymbol{\tilde{\sigma}}^{k}_p - \boldsymbol{\tilde{\sigma}}^{k}_p\mathsfbi{W_s}^{k+1}_p\big]
	\label{Eqn: implicit_stress_update}
	\end{equation}
	
	\item The discrete material point states of the two phases are known for time $t^{k+1}$,
	$$\begin{aligned}
	\text{solid phase:\quad}&\{\bar{\rho}_{sp}^{k+1}, \boldsymbol{\tilde{\sigma}}_p^{k+1}, m_{sp}, \mathbi{x_s}_p^{k+1}, \mathbi{v_s}_p^{k+1}\} \\
	\text{fluid phase:\quad}&\{\bar{\rho}_{fp}^{k+1}, \rho_{fp}^{k+1}, \boldsymbol{\tau_f}_p^{k+1}, p_{fp}^{k+1}, m_{fp}, \mathbi{x_f}_p^{k+1}, \mathbi{v_f}_p^{k+1}\}
	\end{aligned}$$
	and the procedure is repeated for the $k+1$ time-step.
\end{enumerate}

\subsection{Semi-Implicit Effective Stress Algorithm}\label{Sec: implicit_stress_update}
The solid phase material point effective stress is updated at each time-step with the semi-implicit time integration scheme described in this section. Given the material point stress states at time $t^k$, $\{\boldsymbol{\tilde{\sigma}}_p^k\}$, and the total material point flow rates at time $t^{k+1}$,
\begin{equation}
\mathsfbi{D_s}_p^{k+1} = \sym(\mathsfbi{L_s}_p^{k+1}), \qquad\mathsfbi{W_s}_p^{k+1} = \skw(\mathsfbi{L_s}_p^{k+1}),\qquad \mathsfbi{L_s}_p^{k+1} =  \sum_{i=1}^{[n]} \mathbi{v_f}_i^{k+1} \otimes \nabla\mathcal{S}_{fip}^k
\label{Eqn: discrete_solid_strainrate}
\end{equation}
we solve for the plastic flow rates $\{(\mathsfbi{\tilde{D}^p})_p^{k+1}\}$ given by,
\begin{equation}
(\mathsfbi{\tilde{D}^p})_p^{k+1} = \frac{(\dot{\bar{\gamma}}^p)^{k+1}_p}{\sqrt{2}} \frac{\boldsymbol{\tilde{\sigma}_0}^{k+1}_p}{\|\boldsymbol{\tilde{\sigma}_0}^{k+1}_p\|} + \tfrac{1}{3} \big( \beta (\dot{\bar{\gamma}}^p)^{k+1}_p + (\dot{\xi}_1)_p^{k+1} + (\dot{\xi}_2)_p^{k+1} \big) \mathsfbi{1}
\label{Eqn: discrete_plastic_flow_rate}
\end{equation}
such that (with $(\dot{\bar{\gamma}}^p)^{k+1}_p$, $(\dot{\xi}_1)_p^{k+1}$, and $(\dot{\xi}_2)_p^{k+1}$ determined for each material point) the material point stress state at time $t^{k+1}$ is given by \eqref{Eqn: implicit_stress_update}.

\subsubsection{Definition of Trial Stress}
The update from \eqref{Eqn: implicit_stress_update} can be separated into a trial step,
\begin{equation}
\boldsymbol{\tilde{\sigma}}_p^{tr} = \boldsymbol{\tilde{\sigma}}_p^k + \Delta t \big[ 2G \mathsfbi{D_{0s}}^{k+1}_p + K \tr(\mathsfbi{D_s}^{k+1}_p)\mathsfbi{1} + \mathsfbi{W_s}^{k+1}_p\boldsymbol{\tilde{\sigma}}^{k}_p - \boldsymbol{\tilde{\sigma}}^{k}_p\mathsfbi{W_s}^{k+1}_p\big]
\label{Eqn: discrete_trial_stress}
\end{equation}
and a plastic step,
\begin{equation}
\boldsymbol{\tilde{\sigma}}_p^{k+1} = \boldsymbol{\tilde{\sigma}}_p^{tr} - \Delta t \big[ 2G (\mathsfbi{\tilde{D}^p_0})_p^{k+1} + K \tr((\mathsfbi{\tilde{D}^p})_p^{k+1})\mathsfbi{1}\big]
\label{Eqn: discrete_solid_stress_from_trial}
\end{equation}
where $\boldsymbol{\tilde{\sigma}}^{tr}_p$ is a \textit{trial stress} found between times $t^k$ and $t^{k+1}$. Since the trial stress given in \eqref{Eqn: discrete_trial_stress} is an explicit function of the strain-rates in Equations \eqref{Eqn: discrete_solid_strainrate}, we use it as the starting point of our implicit algorithm for solving \eqref{Eqn: discrete_solid_stress_from_trial}.

\subsubsection{Simplification to Scalar Relation}\label{Sec: scalar_stress_update}
The expression in \eqref{Eqn: discrete_solid_stress_from_trial} is separable into a deviatoric part and isotropic part,
\begin{equation}
\boldsymbol{\tilde{\sigma}_0}_p^{k+1} = \boldsymbol{\tilde{\sigma}_0}_p^{tr} - 2G \Delta t  (\mathsfbi{\tilde{D}^p_0})_p^{k+1}, \qquad
\tr(\boldsymbol{\tilde{\sigma}})^{k+1}_p = \tr(\boldsymbol{\tilde{\sigma}})^{tr}_p - 3K\Delta \tr((\mathsfbi{\tilde{D}^p})_p^{k+1})
\label{Eqn: algorithm_by_part}
\end{equation}

The following scalar stress measures reduce the implicit tensor relations above to a set of implicit scalar relations (which are much simpler to solve numerically),
\begin{equation}
\bar{\tau}_p^{tr} = \tfrac{\|\boldsymbol{\tilde{\sigma}_0}_p^{tr}\|}{\sqrt{2}}, \qquad \bar{\tau}_p^{k+1} = \tfrac{\|\boldsymbol{\tilde{\sigma}}_p^{k+1}\|}{\sqrt{2}}, \qquad
\tilde{p}_p^{tr} = -\tfrac{1}{3} \tr(\boldsymbol{\tilde{\sigma}_0}_p^{tr}), \qquad \tilde{p}_p^{k+1} = -\tfrac{1}{3} \tr(\boldsymbol{\tilde{\sigma}}_p^{k+1})
\label{Eqn: algortihm_tau_and_p}
\end{equation}
and therefore \eqref{Eqn: discrete_solid_stress_from_trial} becomes,
\begin{equation}
\bar{\tau}^{k+1}_p = \bar{\tau}^{tr}_p - G\Delta t\  (\dot{\bar{\gamma}}^p)^{k+1}_p
\label{Eqn: algorithm_tau_relation}
\end{equation}
\begin{equation}
\tilde{p}_p^{k+1} = \tilde{p}_p^{tr} + K \Delta t \big(\beta (\dot{\bar{\gamma}}^p)^{k+1}_p + (\dot{\xi}_1)_p^{k+1} + (\dot{\xi}_2)_p^{k+1}\big)
\label{Eqn: algorithm_p_relation}
\end{equation}

By solving the system of equations in \eqref{Eqn: algorithm_tau_relation} and \eqref{Eqn: algorithm_p_relation} subject to the following discrete yield conditions,
\begin{equation}
\begin{aligned}
&(f_1)_p^{k+1} = \bar{\tau}^{k+1}_p - \max\big((\mu_p + \beta)\tilde{p}^{k+1}_p,\ 0\big)\\
&(f_1)_p^{k+1} \leq 0, \qquad (\dot{\bar{\gamma}}^p)^{k+1}_p \geq 0, \qquad (f_1)_p^{k+1} (\dot{\bar{\gamma}}^p)^{k+1}_p = 0\\\\
&(f_2)_p^{k+1} = -\tilde{p}^{k+1}_p\\
&(f_2)_p^{k+1} \leq 0, \qquad (\dot{\xi}_1)_p^{k+1} \geq 0, \qquad (f_2)_p^{k+1} (\dot{\xi}_1)_p^{k+1} = 0\\\\
&(f_3)_p^{k+1} = g(\phi) \tilde{p}_p^{k+1} - (a\phi)^2 \big[ ((\dot{\bar{\gamma}}^p)^{k+1}_p - K_4 (\dot{\xi}_2)_p^{k+1})^2 d^2 \rho_s + 2 \eta_0 ((\dot{\bar{\gamma}}^p)^{k+1}_p - K_4 (\dot{\xi}_2)_p^{k+1}) \big]\\
&(f_3)_p^{k+1} \leq 0, \qquad (\dot{\xi}_2)_p^{k+1} \leq 0, \qquad (f_3)_p^{k+1} (\dot{\xi}_2)_p^{k+1} = 0
\end{aligned}
\label{Eqn: algorithm_yield_surfaces}
\end{equation}
we arrive at the final effective granular stresses at time $t^{k+1}$,
\begin{equation}
\boldsymbol{\tilde{\sigma}}_p^{k+1} = \frac{\bar{\tau}_p^{k+1}}{\bar{\tau}_p^{tr}} \boldsymbol{\tilde{\sigma}_0}_p^{tr} - \tilde{p}_p^{k+1} \mathsfbi{1}
\end{equation}

\subsubsection{Complete Algorithm for Stress Update}
\begin{figure}
	\centering
	\includegraphics[scale=0.32]{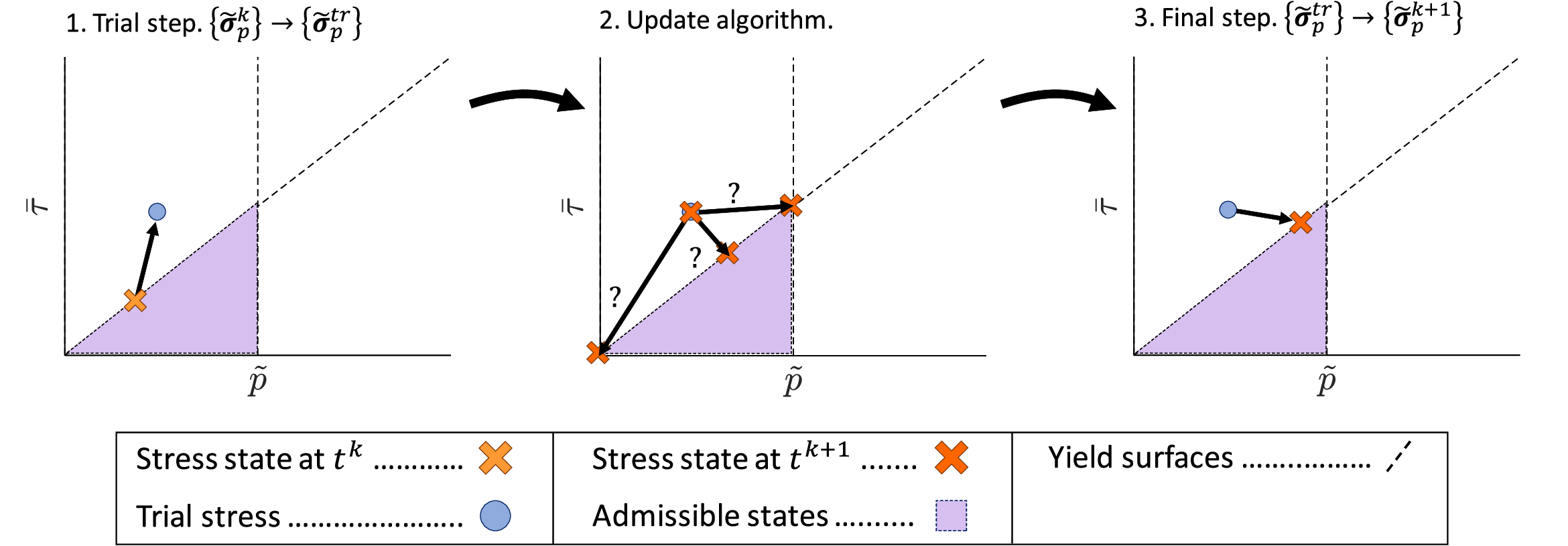}
	\caption{The basic solid phase effective stress update proceeds as follows. (1.) A trial step is taken from the stress state at time $t^k$ assuming that all flow is elastic. (2.) The method described in algorithm \ref{Alg: short_stress_update} is used to determine how to project the trial stress to an admissible stress state. (3.) The final stress state is updated from the trial stress state.}
	\label{Fig: stress_update_algorithm}
\end{figure}

To solve the system of equations from section \ref{Sec: scalar_stress_update}, we use the procedure described in algorithm \ref{Alg: short_stress_update} to successively project the trial stress state defined by $\{\tilde{p}_p^{tr}\}$ and $\{\bar{\tau}_p^{tr}\}$ to the yield surfaces given in \eqref{Eqn: algorithm_yield_surfaces}. As shown in figure \ref{Fig: stress_update_algorithm}, once an admissible stress update is found, the algorithm exits and proceeds to the next time-step. In our implementation of this procedure, we choose to use a simple Newton iteration scheme to solve for each of the projections.

\begin{algorithm}
	\caption{Outline of Stress Update Procedure}
	\label{Alg: short_stress_update}
	\begin{algorithmic}[1]
		\Procedure{Stress Update}{}
		\State Determine trial state.
		\State Check if trial stress is admissible (if so, \textbf{exit}).
		\State Solve assuming final stress is on $f_2$ yield surface
		\State Check if solved state is admissible (if so, \textbf{exit}).
		\State Solve assuming final stress is on $f_1$ yield surface only.
		\State Check if solved state is admissible (if so, \textbf{exit}).
		\State Solve assuming final stress is on $f_1$ and $f_3$ yield surfaces.
		\State This state must be admissible, so \textbf{exit}.
		\EndProcedure
	\end{algorithmic}
\end{algorithm}

\subsection{Specific Notes About Implementation}
In this section we briefly discuss the implementation of the boundary conditions, contact forces, partial saturation, and what we call the \textit{$\delta$ position} correction.

\subsubsection{Kinematic Boundary Conditions}
The kinematic boundary condition used in this work is inherited from that used by \cite{dunatunga}. In this method, the boundary force vectors, $\mathbi{s_s}_i^k$ and $\mathbi{s_f}_i^k$, on the boundary nodes are determined such that some prescribed velocity is achieved at the end of the explicit time-step.

\subsubsection{Mixed Boundary Conditions}\label{Sec: boundary_friction}
In some simulations, we implement a frictional boundary condition on the solid phase. For these simulations, only the component of $\mathbi{v_s}_p^{k+1}$ normal to the boundary is prescribed (and therefore, the normal component of $\mathbi{s_s}_i^k$ is also determined). We then let the tangential force component be given by either a no-slip condition or,
\begin{equation}
\mathbi{s_s}_i^k - (\mathbi{s_s}_i^k \cdot \mathbi{n_b}_i)\mathbi{n_b}_i = -\mu_1 \bigg\|\sum_{p=1}^{N_s}(v_{sp}\tilde{p}_p^{k} \nabla \mathcal{S}_{sip})\bigg\| \frac{\mathbi{v_s}_i^* - (\mathbi{v_s}_i^* \cdot \mathbi{n_b}_i)\mathbi{n_b}_i}{\|\mathbi{v_s}_i^* - (\mathbi{v_s}_i^* \cdot \mathbi{n_b}_i)\mathbi{n_b}_i\|}
\end{equation}
whichever is smaller, where $\mathbi{n_b}_i$ is the boundary normal at the $i$th node.

\subsubsection{Contact Algorithm}
In some of the qualitative results presented in this work, we implement the contact algorithm from \cite{huang}. This algorithm calculates an explicit inter-body force (when a third material body is introduced) which enforces a frictional, non-penetrating contact between the third body and each of the two phases presented in this work.

\subsubsection{Partial Immersion}
In the parts of the solid body where there is no fluid, we say that the viscosity, $\eta_0$, is zero. Numerically we accomplish this by constructing a nodal viscosity field at each time-step given by the coefficients $\{\eta_{0i}^k\}$. We then let the value of $\eta_0$ in section \ref{Sec: time_marching} be determined on each solid phase material point by, $\{\eta^k_{0p}\}$ where,
$$\eta_{0i}^k = \bigg\{ \begin{matrix}
\eta_0 &\text{if\quad} m_{fi}^k > 0\\
0 &\text{if\quad} m_{fi}^k = 0
\end{matrix} \quad \forall i\in[1,[n]], \qquad \eta^k_{0p} = \sum_{i=1}^{[n]}\eta^k_{0i}\mathcal{S}_{sip} \quad \forall p \in [1,N_s]$$

\subsubsection{Dynamic Quadrature Error Reduction} \label{Sec: delta_position}
Particle methods for simulating fluid flows have an inherent problem with (among other things) point clumping (see recent work by \cite{koh} and \cite{maljaars}). There are many physically admissible flows, such as those with stagnation points, which will result in material point tracers gathering together. By choosing the material point centroids as the quadrature points for our integral approximations, this clumping leads to significant quadrature error. In some fluid simulations, we see extremely spurious flows develop, which we attribute to this quadrature error.

To address this issue, we have developed a novel approach which `nudges' material point centroids as the material flows. This nudge is the $\boldsymbol{\delta_s}_p^k$ and $\boldsymbol{\delta_f}_p^k$ from \eqref{Eqn: discrete_centroid_update}. The method we introduce relies on the nodal weight measure from \eqref{Eqn: nodal_volume} (which is known \textit{a priori}). Since our material point characteristic functions are partitions of unity within the body (by \eqref{Eqn: point_characteristic_function}), we have,
\begin{equation}
V_i = \sum_{p=1}^{N_s}\int_\Omega{\mathcal{N}_i(\mathbi{x})U_{\alpha p}(\mathbi{x})dv} \qquad \text{if} \qquad (\mathcal{N}_i(\mathbi{x}) = 0 \quad \text{for} \quad \mathbi{x}\notin\mathcal{B}_\alpha^t)
\end{equation}
where $\alpha$ is a more general notation for either $s$ or $f$.

We determine how much quadrature error has accumulated by using the material point weights and centroids as quadrature points for the above integral and measure the relative \textit{overshoot error}, $e_{si}$ and $e_{fi}$, as follows,
\begin{equation}
v_{\alpha i} \equiv \sum_{p=1}^{N_\alpha} v_{\alpha p}\mathcal{N}_i(\mathbi{x_\alpha}_p), \qquad e_{\alpha i} = \max\bigg(0,\frac{v_{\alpha i} - V_i}{V_i}\bigg)
\end{equation}
We have attempted several methods of reducing this error (which will be explored in a future work); however the method used in this work is a strain-rate-dependent position correction given as follows,
\begin{equation}
\boldsymbol{\delta_\alpha}_p^k = -\lambda \Delta t (\Delta x)^2\|\mathbf{L_{\alpha0}}_p^{k+1}\| \sum_{i=1}^{[n]}e_{\alpha i}\nabla\mathcal{S}_{\alpha ip}
\end{equation}
with $\Delta x$ the grid spacing of the Cartesian grid and $\lambda$ an arbitrary scale factor.


\bibliographystyle{jfm}
\bibliography{ms}

\end{document}